%% file: main.tex
\documentclass[]{aastex631}

\newcommand\aastex{AAS\TeX}

\usepackage{graphicx}	
\usepackage{amsmath}	
\usepackage{amssymb}	
\usepackage{array}

\usepackage{color}

\usepackage[normalem]{ulem}

\definecolor{orange}{rgb}{1,0.5,0}
\definecolor{armygreen}{rgb}{0.29, 0.33, 0.13}
\definecolor{brightmaroon}{rgb}{0.76, 0.13, 0.28}
\definecolor{darkviolet}{rgb}{0.4,0.0,0.6}
\definecolor{cadmiumgreen}{rgb}{0.0, 0.42, 0.24}
\definecolor{indiagreen}{rgb}{0.07, 0.53, 0.03}

\begin{document}

\title{The Tianlai-WIYN North Celestial Cap Redshift Survey
\footnote{Template \aastex Article with Examples: 
v6.3.1Released on March, 1st, 2021}}

\correspondingauthor{Albert Stebbins}
\email{stebbins@fnal.gov}

\author{Reza Ansari}
\affiliation{Université Paris-Saclay,  \\
Université Paris Cité, CEA, CNRS, AIM, \\ 
91191 Gif sur Yvette, France}

\author{Gabriela A. Marques}
\affiliation{Fermi National Accelerator Laboratory \\
P.O. Box 500, \\
Batavia IL 60510, USA}

\author{John P. Marriner}
\affiliation{Fermi National Accelerator Laboratory \\
P.O. Box 500, \\
Batavia IL 60510, USA}

\author{Olivier Perdereau}
\affiliation{ IJCLab, CNRS/IN2P3, Universit\'e Paris-Saclay, \\
91405 Orsay, France}

\author{Elena Pinetti}
\affiliation{Fermi National Accelerator Laboratory \\
P.O. Box 500, \\
Batavia IL 60510, USA}

\author{Lily Robinthal}
\affiliation{Lunar and Planetary Laboratory, University of Arizona\\
Tucson AZ 85721, USA}

\author{Albert Stebbins}
\affiliation{Fermi National Accelerator Laboratory \\
P.O. Box 500, \\
Batavia IL 60510, USA}

\author{Haoxuan Sun}
\affiliation{Department of Physics, Brown University, \\
182 Hope St., Providence, \\
RI 02912, USA}

\author{Peter Timbie}
\affiliation{Department of Physics, University of Wisconsin Madison, \\
1150 University Ave,\\ 
Madison WI 53703, USA}

\author{Gregory S. Tucker}
\affiliation{Department of Physics, Brown University, \\
182 Hope St., Providence, \\
RI 02912, USA}

\author{Eli Doyle}
\affiliation{Department of Physics, Brown University, \\
182 Hope St., Providence, \\
RI 02912, USA}

\author{Jocelyn Chu}
\affiliation{Department of Physics, Brown University, \\
182 Hope St., Providence, \\
RI 02912, USA}

\author{E. Revsen Karaalp}
\affiliation{Department of Physics, Brown University, \\
182 Hope St., Providence, \\
RI 02912, USA}

\author{Xuelei Chen} 
\affiliation{National Astronomical Observatory, Chinese Academy of Sciences, \\
20A Datun Road, \\
Beijing 100101, P. R. China}

\author{Jixia Li} 
\affiliation{National Astronomical Observatory, Chinese Academy of Sciences, \\
20A Datun Road, \\
Beijing 100101, P. R. China}

\author{Fengquan Wu} 
\affiliation{National Astronomical Observatory, Chinese Academy of Sciences, \\
20A Datun Road, \\
Beijing 100101, P. R. China}

\collaboration{20}{(Tianlai collaboration)}

\begin{abstract}

We present the results of a small, low redshift 
spectroscopic survey of galaxies within $3^\circ$ of the North Celestial Pole (NCP) selected using V-band photometry obtained from the North Celestial Cap Survey (NCCS) \citep{Gorbikov_2014}.  The purpose of the current survey is to create a redshift space template for 21\,cm emission from neutral hydrogen with which to correlate radio line intensity observations by the Tianlai dish and cylinder interferometers.  A total of 898 redshifts were obtained from the 2102 
extended objects in the NCCS with $m_V<19$ in the survey area.  After accounting for extinction, the survey geometry and selection effects, the number density and clustering pattern of galaxies in the redshift catalog are consistent with other low redshift surveys. We were also able to identify 11 galaxy cluster candidates from this redshift catalog.

\end{abstract}


\input{sec_1_introduction} 
\input{sec_2_survey}

\input{sec_3_datareduction} 
\input{sec_4_spectrocatalog}

\input{sec_5_clustering} 
\input{sec_6_conclusions}

\section{Acknowledgments}
We thank Eric Hooper, Susan Ridgway, Heidi Schweiker, Dan Li, and Daryl Wilmarth for advice on using the WIYN telescope and Hydra spectrometer. This document was prepared by the Tianlai Collaboration and includes personnel and uses resources of the Fermi National Accelerator Laboratory (Fermilab).
This work was produced by Fermi Forward Discovery Group, LLC under Contract No. 89243024CSC000002 with the U.S. Department of Energy, Office of Science, Office of High Energy Physics. Publisher acknowledges the U.S. Government license to provide public access under the DOE Public Access Plan DOE Public Access Plan.
Work at UW-Madison and Fermilab was partially supported by NSF Award AST-1616554.    Work at UW-Madison was further supported by an NSF REU award, the Graduate School, and the Thomas G. Rosenmeyer Cosmology Fund.
Undergraduate students at Brown were supported by the Brown University SPRINT/UTRA program.
Authors affiliated with French institutions acknowledge partial support from CNRS (IN2P3 \& INSU), Observatoire de Paris and Irfu/CEA. They
 acknowledge financial support from “Programme National de Cosmologie and Galaxies” (PNCG) and from  the FCPPL (France-China Particle Physics Laboratory) of CNRS, France. 
The Tianlai Dish Array was built with the support of the CAS special fund for repair and purchase, and operated with the support of the NAOC Astronomical Technology Center.
The Tianlai Pathfinders are supported by the National SKA Program of China (Nos. 2022SKA0110100 and 2022SKA0110101), the National Natural Science Foundation of China (12361141814, 12203061, 12273070, 12303004). 

\appendix
\input{appendix_additional_spectra}

\bibliography{tnccsz_wiyn_ss}{}
\bibliographystyle{aasjournal}

\end{document}

%% file: sec_1_introduction.tex
\section{Introduction}
\label{sec:introduction}

In this paper we present the results of a spectroscopic optical galaxy redshift survey of the North Celestial Cap (NCC).  This survey was done in support of the Tianlai Pathfinder \citep{2021MNRAS.506.3455W, 2020SCPMA..6329862L}, a radio telescope array being used to develop the 21\,cm intensity mapping technique \citep{2008PhRvL.100i1303C, 2008PhRvL.100p1301L, 2020PASP..132f2001L}.    
For decades, the 21\,cm radio emission line has been used to construct redshift catalogs of individual galaxies with depths limited by the sensitivity to individual galaxies \citep{2019MNRAS.486.5124O}. 21\,cm intensity mapping can function at greater depths where individual galaxies are unresolved, determining 
the large scale spatial distribution of the neutral hydrogen gas.  Indeed, 21\,cm intensity mapping observations have set upper limits on the the spatial distribution of hydrogen gas at redshifts spanning the Epoch of Reionization and Cosmic Dawn.  They may eventually measure the distribution of hydrogen gas during the cosmic Dark Ages, before galaxies formed, where most of the gas is neutral hydrogen.  Observations in all redshift ranges suffer from common challenges, the most significant being radio foregrounds orders of magnitude brighter than the hydrogen signal. Until recently, measurements by individual instruments have resulted only in upper limits on the hydrogen signal, with the exception of measurements at $z = 0.33$ and $z = 0.44$ using the MeerKAT \citep{Paul2023} interferometer and at $z \sim 1$ with the CHIME interferometer \citep{CHIME2025}.   In addition, successful detections have been achieved by cross-correlations with optical surveys.   The CHIME interferometer has also successfully detected HI emission by cross-correlating with large-scale structure in the eBOSS survey \citep{2023ApJ...947...16A, 2024ApJ...963...23A} as has the MeerKAT interferometer (operated as multiple single dishes) \citep{2023MNRAS.518.6262C} with the Wiggle-Z galaxy redshift survey. Single dish instruments (e.g. \cite{Wolz_2021} with GBT data,  \cite{Anderson_2018} with Parkes data) have done so as well. Observing cross-correlation signals at levels on par with detailed simulations validates the understanding both of systematic effects and the analysis pipelines.  So far, in redshift ranges where galaxy redshift surveys exist,  21\,cm intensity mapping results are far from being competitive with them.  This is largely due to a variety of systematic effects which it is hoped can be ameliorated.  

Traditional 21\,cm galaxy redshift surveys operate in a regime where individual galaxies can be differentiated.  At somewhat larger $z$'s there is an intermediate (``hybrid") regime where one can identify individual large galaxies or groups of galaxies but where most of the 21\,cm signal comes from unresolved galaxies which can be studied using 21\,cm intensity mapping.  This is the regime in which the Tianlai Dish Array (TDA) low redshift survey will operate.  It is expected to detect the $21\,$cm signal, including a few HI ``clumps" \citep{2022MNRAS.517.4637P}.  Currently, the TDA correlator has an instantaneous bandwidth of $\sim100\,\mathrm{MHz}$ that can be tuned within the range from $400 - 1430\,$MHz ($2.55 > z > -0.01$). The nearest volume that can be surveyed with a single tuning of the array includes $0\lesssim z\lesssim0.07$.  Just as other 21\,cm intensity mapping surveys at higher redshift have benefited from cross-correlation with galaxy redshift surveys, having a spectroscopic optical galaxy redshift survey overlapping the TDA survey volume can help to disentangle signal from systematics. This is the purpose of the current survey. 

 One systematic uncertainty is the shape of the radio beam, particularly the sidelobes.  Another is electromagnetic interactions  between different elements of the array \citep{Kern2019,Kern2020}.  Both of these are sensitive to the array's configuration. To minimize the complexity of these uncertainties, arrays designed for 21\,cm intensity mapping generally operate as transit telescopes pointing at a fixed declination and scanning the sky over right ascension as the Earth rotates.    
Since there is a much larger solid angle at low declination (near the celestial equator) than at high declination (near the NCP), the telescope will spend a smaller fraction of time observing any given patch of the sky at low declination than at high. This results in poorer brightness temperature (or intensity) sensitivity at low declination for a given observing time.  At the other extreme, if one points at the Celestial Poles (North NCP or South SCP) then one is always observing the same patch of the sky, which leads to far better sensitivity. For this reason, most of the TDA observations to date (near $z\sim 1$) were performed with the array pointing directly at the NCP. We chose the same observing strategy for the low redshift survey.  At the lowest $z$'s the observational wavelength is $\lambda\sim21\,\mathrm{cm}$ and the TDA angular resolution is $\sim2.3^\circ$.  Thus, to complement the low-$z$ TDA survey, one wants an optical galaxy spectroscopic redshift survey of the $z\le0.07$ volume within $\sim2^\circ$ of the NCP.  We refer to our survey area around the NCP as the North Celestial Cap or NCC and use tNCCSz as an acronym for our redshift survey.

\begin{figure}[!htp]
    \centering
    \includegraphics[width=0.4\textwidth]{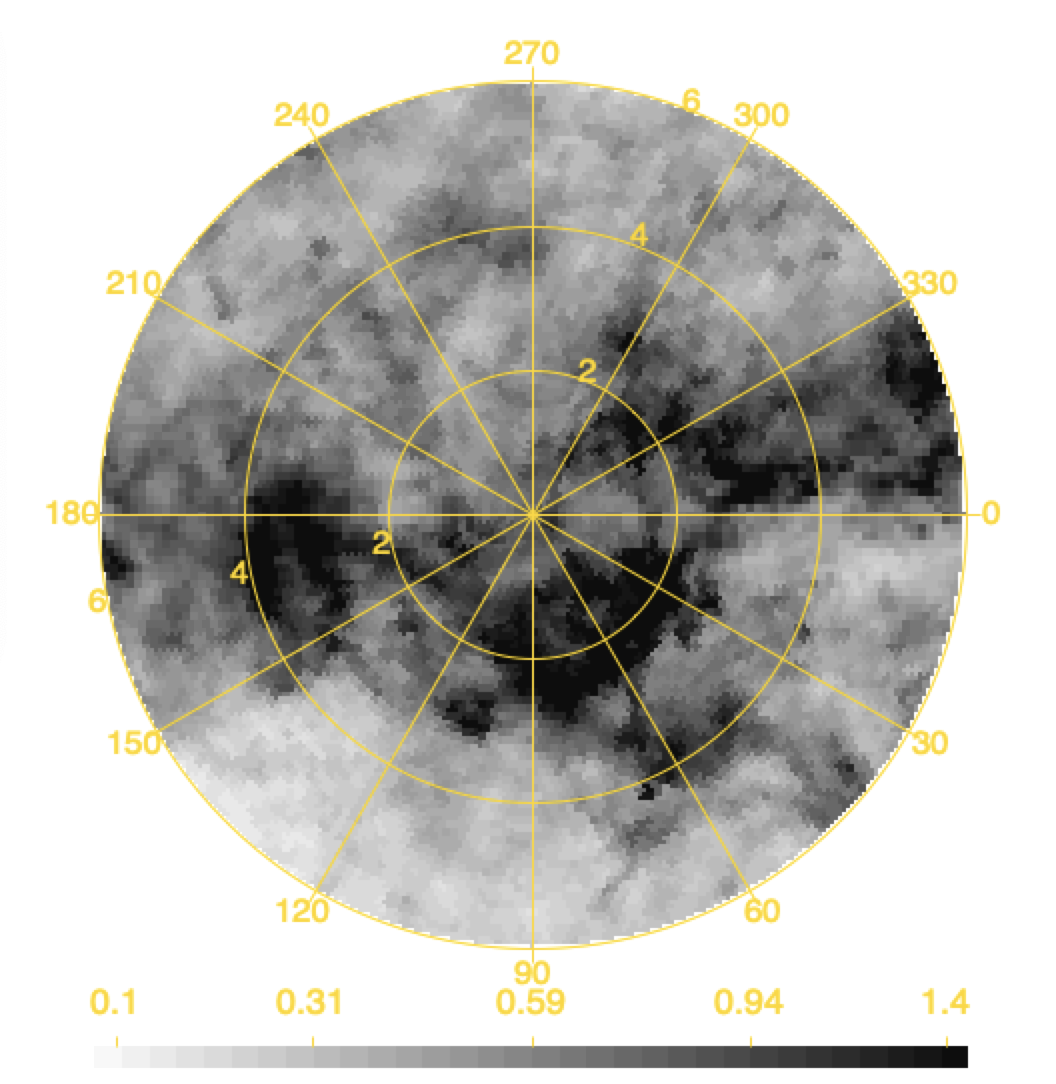}
      \includegraphics [width=0.5\textwidth]{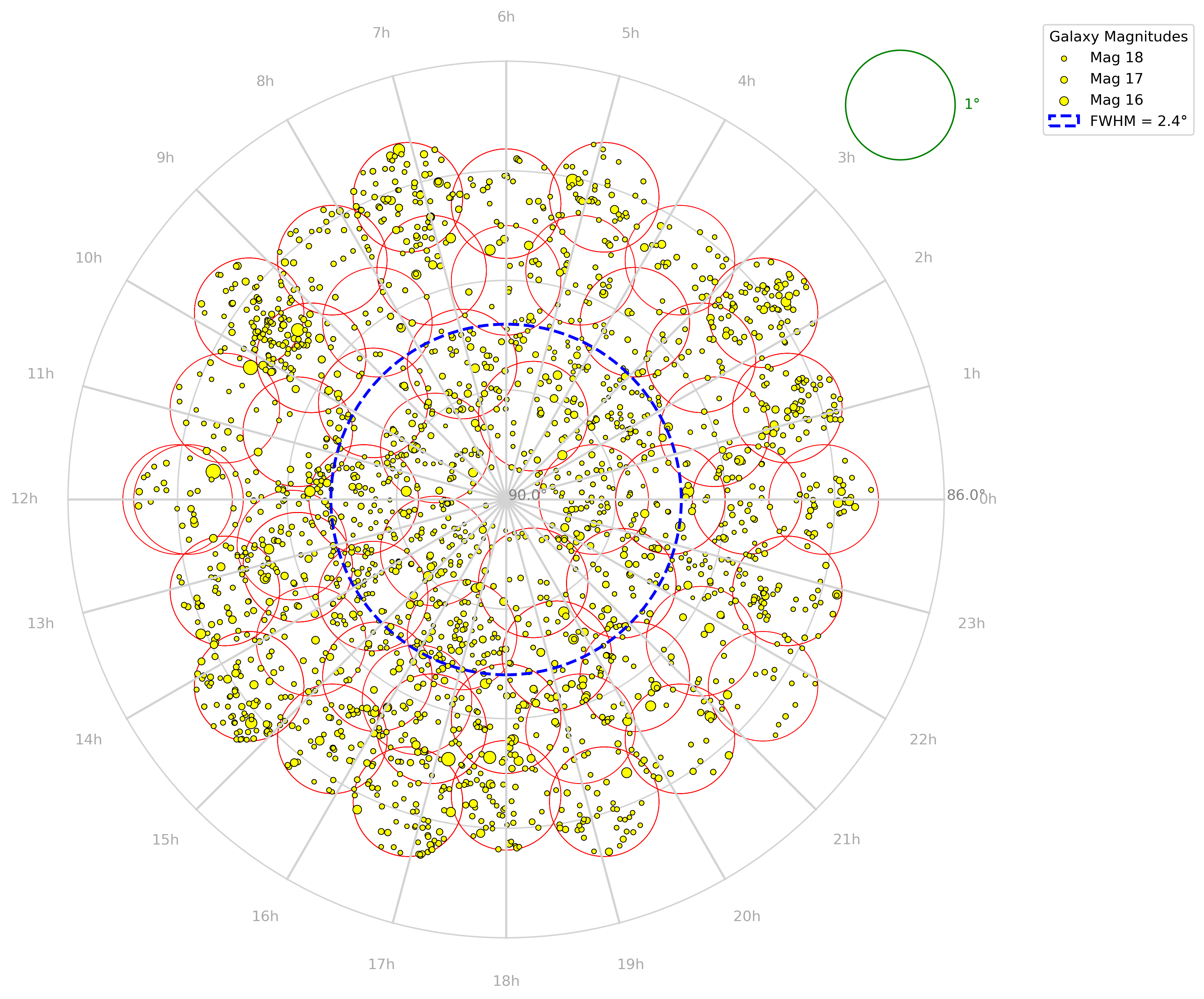}
     
   \caption{Left panel: Planck V-band extinction map showing $A_v$ (in magnitudes) in a 6$^\circ$ radius region around the NCP. We have used the Planck dust extinction map \citep{2016A&A...586A.132P} to model the Galactic extinction.   Note that the tNCCSz survey covers the central region to a radius of 3$^\circ$.   One can notice that this area is subject to significant absorption by dust, reaching 1.7 magnitude in some directions covered by the survey. Right panel: Distribution of targets and associated tiling map for the tNCCSz survey. 
   Color-coded by magnitude, the map highlights galaxies in a specific magnitude range for redshift analysis. The circular red outlines represent the 56 observational tiles, each corresponding to a unique pointing of the WIYN telescope to cover the declination range above $87^\circ$ and parts of the 86.6$^\circ$ to 87.0$^\circ$ range. The blued dashed circle represents the full width at half maximum of the primary beams of the TDA antennas at 1380 MHz ($z \sim 0.035$).  
   \label{fig:extinction-tiling}}
\end{figure}

The Tianlai collaboration carried out the spectroscopic survey prescribed above 
because no such survey existed. (Fortunately, the NCCS photometric survey \citep{Gorbikov_2014} had already been performed for other reasons.) There are many reasons why this survey has not been done before: 
\begin{enumerate}
    \item The NCC is near the Galactic plane, it has high reddening (or extinction), $0.05\lesssim\mathrm{E}(B-V)\lesssim 0.55$, which makes it an ill-suited area for extra-Galactic astronomy (Fig. \ref{fig:extinction-tiling} Left); 
    \item The NCC can only be observed  at high airmass, $\gtrsim2$, for any existing, large, ground-based telescope;
    \item Most equatorial mount telescopes are unable to observe very close to the NCP. 
\end{enumerate}
Although the survey described here is suitable for this purpose, it adds little to our knowledge of cosmological structure because of the small area ($\sim30\,\mathrm{sq\,deg}$) and volume ($2.5\times10^{-4}\,\mathrm{Gpc}^3$).  It does, however, map a part of the nearby Universe that has been neglected heretofore.  It will also aid the Tianlai project and perhaps other observations in this volume.

This paper is organized as follows.    Section \ref{sec:survey} describes the survey design and  observations.  Section \ref{sec:datareduction} describes the process of reducing the raw spectroscopic data to redshifts. Section \ref{sec:spectrocatalog} gives statistics for the catalog, provides verification of the survey by cross-matches with other surveys in the same region, and provides the selection function and the redshift distribution. Section \ref{sec:clustering} further validates the survey by comparing the observed clustering statistics with those of comparable surveys in other regions of the sky. We summarize the results in Section \ref{sec:conc}.

%% file: sec_2_survey.tex
\section{The Survey}
\label{sec:survey}

To conduct an optical galaxy redshift survey, one must first identify the angular positions of the galaxies using a photometric survey and then perform spectroscopic observations. The photometry for our study was previously done in the North Celestial Cap Survey (NCCS \cite{Gorbikov_2014}). The NCCS identifies approximately $4 \times 10^6$ objects within $10^\circ$ of the NCP using V-R-I photometry with magnitude limits $m_V<20.3$, $m_R<21.0$, and $m_I<19.2$. Some of these objects are stars, and some galaxies, particularly at the magnitude limit, are likely beyond our primary interest in redshifts with $z<=0.07$. Thus, only a  fraction of the NCCS survey is useful for our purposes. The catalog also includes a PESS (point-extended-source-separation) index indicating whether an object is point-like (PESS $<=$ 1, most likely a star) or extended (PESS = 2 or 3, most likely a galaxy). Selecting spectroscopic targets involved cutting on declination and magnitude and excluding point-like objects.

\subsection{NCCS Target Selection}
\label{sec:targetselection}

Target selection was 
based on the NCCS catalog, focusing on identifying galaxies within the V-band magnitude range  $13 < m_V < 18.995$.  (We originally set our magnitude threshold at $m_V < 20$, but backed off after taking the first night of data.) 
We chose only extended objects in the NCCS catalog, with PESS = 2 or 3.  The spatial coverage concentrated on regions with declinations greater than $87^\circ$. Although the radio beam of the Tianlai Pathfinder falls off with angle, it does not go to zero, and there is some benefit to surveying larger areas
for the tNCCSz. We may extend this survey to lower declinations in the future.
This region was divided into 56 observational tiles. The center of each tile corresponds to a specific pointing of the WIYN telescope and the diameter corresponds to the  $1^\circ$ field of view. (Figure~\ref{fig:extinction-tiling} Right). The tiling ensured that each target was assigned to a unique tile, minimizing redundancy and maximizing coverage. Due to limitations in fiber positioning within the WIYN 3.5~m telescope's focal plane, each tile required multiple observations, typically two or more, to observe nearly all of the targets. The observation strategy also yielded partial coverage of the declination range from 86.6$^\circ$ to 87.0$^\circ$.

\subsection{Observations}  
\label{sec:observations}
Between February 2020 and May 2023, we obtained 15 nights of gray and dark time on the WIYN 3.5~m telescope to use the Hydra multifiber positioner and bench spectrograph \citep{1995SPIE.2476...56B} to survey the NCC. 
We used the bench spectrograph with the blue-light optimized fibers and the 400@4.2 (WIYN Hydra and Bench spectrograph users manual - v5) grating.  These choices resulted in wavelength coverage from below 3800 {\AA} to above 7600 \AA.  However, our sensitivity below about 4600 {\AA} was typically insufficient to identify the common absorption features in galaxy spectra, and the $\textrm{H}_{\alpha}$ emission line could be confused with prominent sky lines above 7200 {\AA}.

%% file: sec_3_datareduction.tex
\section{Data Reduction}
\label{sec:datareduction}

\subsection{Processing}
Each fiber in the Hydra spectrograph produces a linear image on the charge-coupled device (CCD). The images are reduced to spectra using a Hydra-specific computer code written by one of the authors (JPM).  Each night a number of calibration images are taken.  These usually comprise 21 bias frames, 3 dark frames, and 5 dome flats.  The exposures are averaged and used in the processing of the science images.  In addition, the spectrum of a standard star ($\delta$ Ursa Minoris) is measured and used to obtain the ratio of measured analog-to-digital converter (ADC) counts to physical flux units. The steps for processing the science images are summarized below.
\begin{enumerate}
\item{Subtract bias images and mask pixels that are flagged as bad because the bias levels are too low or too high.}
\item{Subtract dark images (normalized to the exposure time) and mask pixels that are flagged for excessive dark current.}
\item{Estimate pixel value errors as a combination of read noise and photon statistics.}
\item{Remove cosmic rays by looking for unusually large gradients in the pixel map.  Pixels contaminated by cosmic rays are masked.}
\item{Extract the spectra for each fiber.    The extraction follows the peak of each line and adds the pixels at each wavelength that lie within $\pm7$ pixels of the peak.}
\item{Normalize the spectra in the previous step using the dome flat.  Assuming that a dome flat illuminates the fibers uniformly, this step will equalize the responses of all the fibers.}
\item{Subtract a sky spectrum from each of the spectra in the previous step.  The sky background is determined from the average of the sky fibers, which are placed in the focal plane without regard to the positions of the targets; most often they measure positions with negligible light from celestial sources.   An average spectrum is determined and subtracted from each individual spectrum.}
\item{Correct the spectra for atmospheric extinction.}
\item{Normalize the spectra to physical flux units using the observation of the standard star.  The normalization is discussed in more detail below.}
\end{enumerate}

\begin{figure}
    \centering
   \includegraphics[width=5 in,trim= 0 50 0 0]{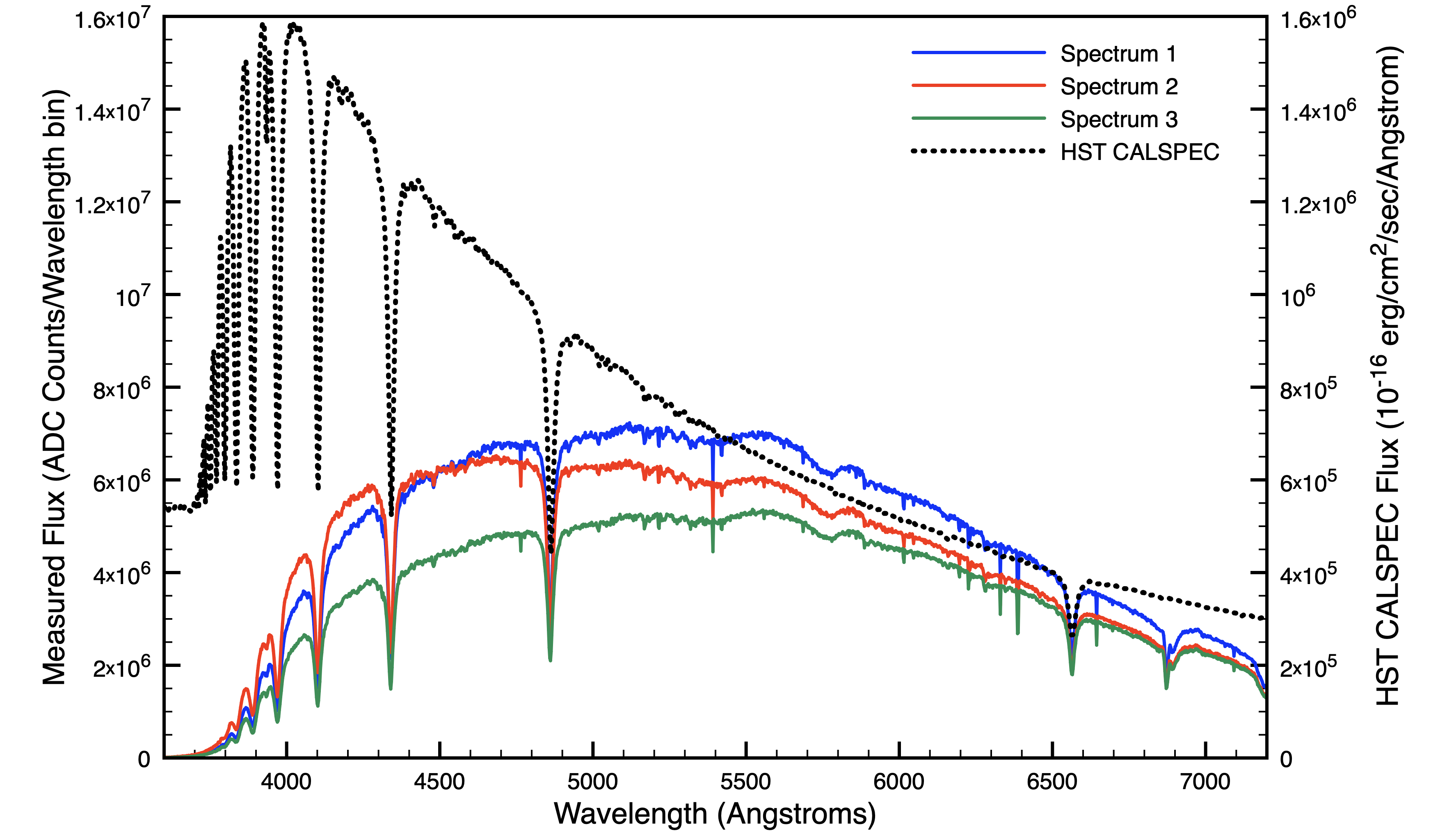}
   \caption{Three spectra of $\delta$ Ursae Minoris obtained on March 7, 2022 are plotted in raw ADC counts per wavelength bin (scale on the left axis).  For our configuration the wavelength bins are about 2.1~\AA. The {\sl HST CALSPEC} standard spectrum is plotted in physical units (scale on the right axis).  The ratio of the average of the measured spectra to the standard establishes the normalization factor.}
   \label{fig:deltaUM}
\end{figure}

A standard star was measured to normalize the spectra. On most nights at least three spectra were taken 
of a reference star, $\delta$ Ursae Minoris, to compare with the standard star spectrum obtained by the {\sl Hubble Space Telescope} (HST).
Figure \ref{fig:deltaUM} shows three consecutive spectra.  The spectra are all high quality but the variations in amplitude and shape affect the overall normalization.  The normalization curve is derived from the ratio of the average of the measured spectra to the standard flux.

 The sensitivity of the Hydra instrument is limited by the brightness of the night sky.  The night sky spectra exhibit prominent atmospheric emission lines and a broad continuum.
 For most of our data, the moon was at $\sim\!\!50\%$ illumination and above the horizon for about half of the night.
 The moonlight is particularly bright towards the blue part of the spectrum.  The observed night sky brightness is reduced by $\sim 3.5$ times after moonset.

\subsection{Sample Spectra}
\begin{figure}[ht]
    \centering
\vglue -0.1 in
  \includegraphics[width=7 in, trim= 0 50 0 0]{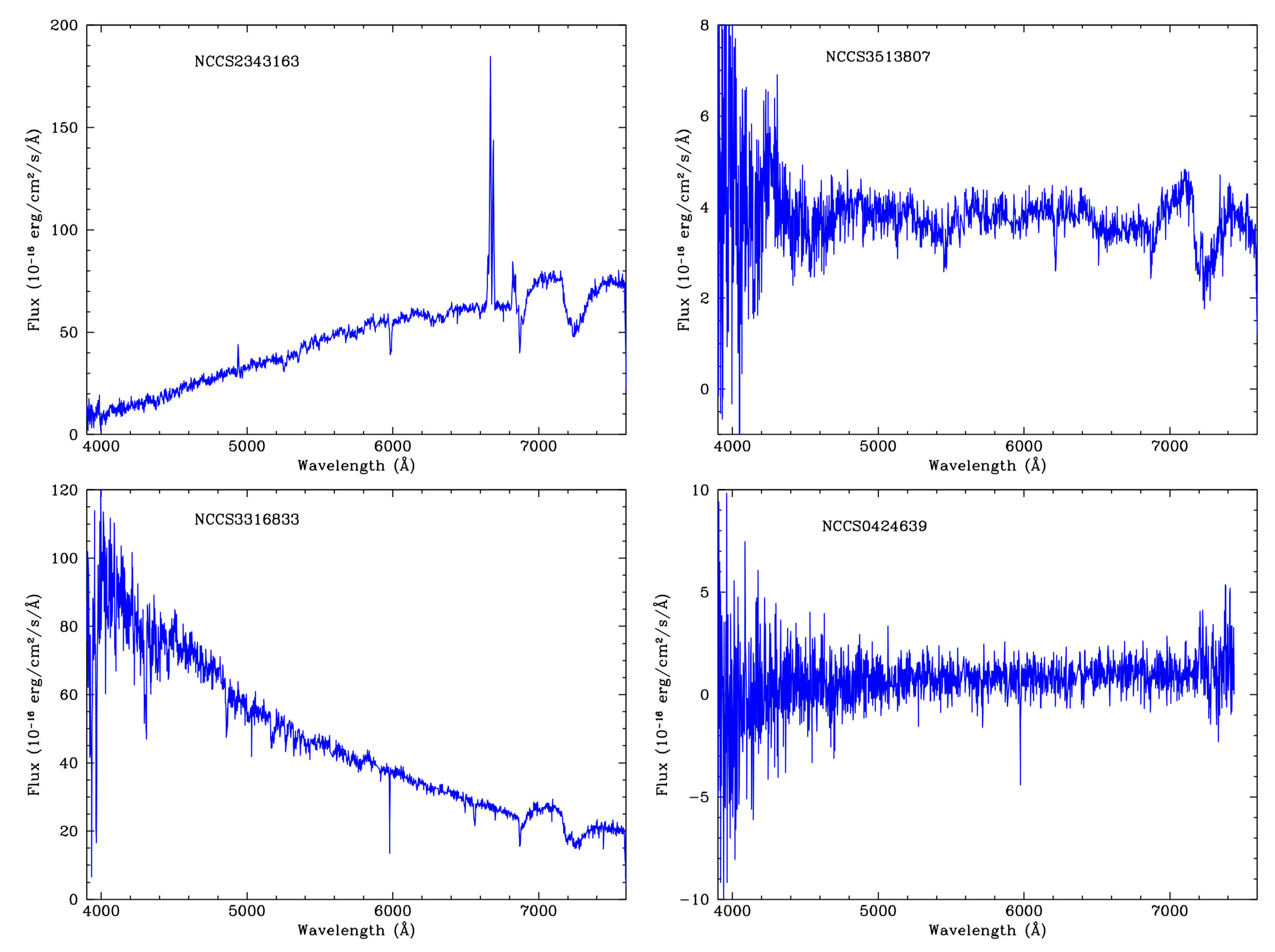}
  \vglue 0.4 in
 \caption{Upper left: The spectrum of NCCS2343163 with a catalog magnitude of $m_V=14.56$ is the spectrum of a bright emission line galaxy. The clear $H_\alpha$ (6565 \AA) emission line flanked by N[II] (6550, 6585 \AA) dominates the spectrum.  The measured redshift is $z=0.0158$. Upper right: The spectrum of NCCS3513807 with a catalog magnitude of $m_V=16.05$ is one of the brightest absorption line spectra that were obtained. The absorption lines are seen clearly at redshifted G-band (4307~\AA),  H$_\beta$ (4863~\AA),  Mg (5177~\AA) and Na (5896~\AA).  The Ca-K (3935 \AA) is present but difficult to discern from the noise until the wavelength scale is expanded while there is but a hint of the Ca-H (3970~\AA) line.  The measured redshift is $z=0.0543$. Lower left: The spectrum of NCCS3316833 with a catalog magnitude of $m_V=15.26$ is a typical spectrum of a bright star. There are clear hydrogen Balmer series lines as well as G-band (4307 \AA) and Mg (5177~\AA). Lower right: The spectrum of NCCS0424639 with a catalog magnitude of $m_V=17.91$ is a typical spectrum without any obvious features.}
  \label{fig:sample_spectra}
\end{figure}

Four representative spectra are shown in Figure \ref{fig:sample_spectra}.  The spectrum of NCCS2343163 has a spectrum with prominent emission lines, characteristic of a spiral galaxy with significant star formation. This spectrum is among the brightest galaxy spectra in our sample.
Emission lines, when present, can often be easily recognized even when absorption features are completely washed out by the noise in the spectrum.  In fact, the dimmest spectrum in our catalog that has a well determined redshift has $m_V=18.989$, nearly at our target selection criterion of $m_V\le19$.

The spectrum of NCCS3513807 has no prominent emission lines but does have some clear absorption lines, characteristic of elliptical or early type galaxies.  While the absorption lines are very clear in this relatively bright galaxy, they are typically more subtle.  Emission lines have relatively small errors from photon statistics, but absorption lines result from the absence of light and have relatively large errors from photon statistics.  Absorption lines also tend to be broader than emission lines.  For these reasons, it is difficult to identify absorption lines in the dimmer galaxies.  In fact, the overwhelming majority of identified redshifts come from emission line galaxies.

The spectrum of NCCS3316833 is that of a star.  The brightest targets in our sample are often not galaxies but stars.  While the targets were selected because the measured size exceeds the size of the point spread function, our NCCS galaxy catalog has substantial stellar contamination at the brighter end of the catalog.  Stars can generally be identified by absorption at the rest-frame wavelengths of the hydrogen Balmer series.  Mature stars often have prominent absorption at G-band, Mg, and Na as observed in this example.  Unfortunately the Na line is also present in the sky spectrum, making it an unreliable indicator of stellar spectra.
Cooler stars, like M-stars, have only broad absorption bands, making them more difficult to identify when the signal-to-noise ratio is poor.

The spectrum of NCCS0424639 has no identifiable features.  The NCCS magnitude of this target is $m_V=17.91$, significantly brighter than the faintest targets at $m_V=19$.  The spectrum has positive flux except possibly at the very blue portion, but there are examples with negative flux and significant positive and negative slopes.  The variation in the level of the continuum is thought to be caused by errors in the sky subtraction. No redshift was assigned to this object.

\subsection{Redshift determination}\label{sec:redshift}

We used two approaches to determine redshifts from the reduced spectra.  In the first method, the spectra are analyzed visually.  The second method is largely automated. 

\subsubsection{Visual Method}
Each reduced spectrum is examined for characteristic emission and absorption lines that would reveal the galaxy redshift.  In most cases, no significant features are found.
In the case where a spectral feature is potentially identified, the redshift determination is assigned a quality factor ($Q$) of 1 to 5.  A quality factor of 5 generally indicates a certain redshift with a clear feature (usually $H_\alpha$) and at least one other confirming feature, for example, O[III](5008).  A quality factor of $Q=1$ indicates a possible but unconfirmed feature or more than one feature with low significance.  A spectrum is assigned a quality factor $Q=1$ if it is judged that the feature(s) are more likely than not to have been correctly identified.  A quality factor of $Q=0$ is assigned for spectra where no redshift is determined.  Other quality factors describe intermediate levels of confidence in the redshift determination.  Of the 1874 spectra examined, it is determined that 1086, 160, 77, 173, 126, 252 spectra have quality factors of 0, 1, 2, 3, 4 and 5 respectively. 
It should be emphasized that the quality factor is a subjective assessment that has not been checked or calibrated.
 
Spiral galaxies, especially those with significant star formation, exhibit strong emission lines that are often several times the continuum level.
Redshift was determined using the most prominent line, which was usually the $H_\alpha$(6562) line.
Most of the galaxies whose redshifts are determined are of this type.  Elliptical galaxies typically have absorption lines that are a fraction of the continuum level, and these lines are difficult to distinguish from fluctuations in the continuum level.  Thus, it can be easy to determine redshifts for star-forming galaxies at $m_V=19$ but impossible to determine redshifts for elliptical galaxies at $m_V=16$.  The distribution of target magnitudes is shown in Figure \ref{fig:magdist-f-quality}.  Not surprisingly, we more often obtain a redshift when the target is bright.

\subsubsection{Semi-automatic Method}
\label{subsec:semiauto}

This method aimed at determining a robust redshift estimation from the computation of a correlation coefficient between the measured spectra and galaxy template spectra picked from the literature.  On average, the WIYN spectra consist in about 1720 data points, with a 2 Angstroms sampling step covering the wavelength range $3900 \lesssim \lambda \lesssim 7560$~\AA.

\begin{figure}
    \centering
    \includegraphics[width=0.9\linewidth]{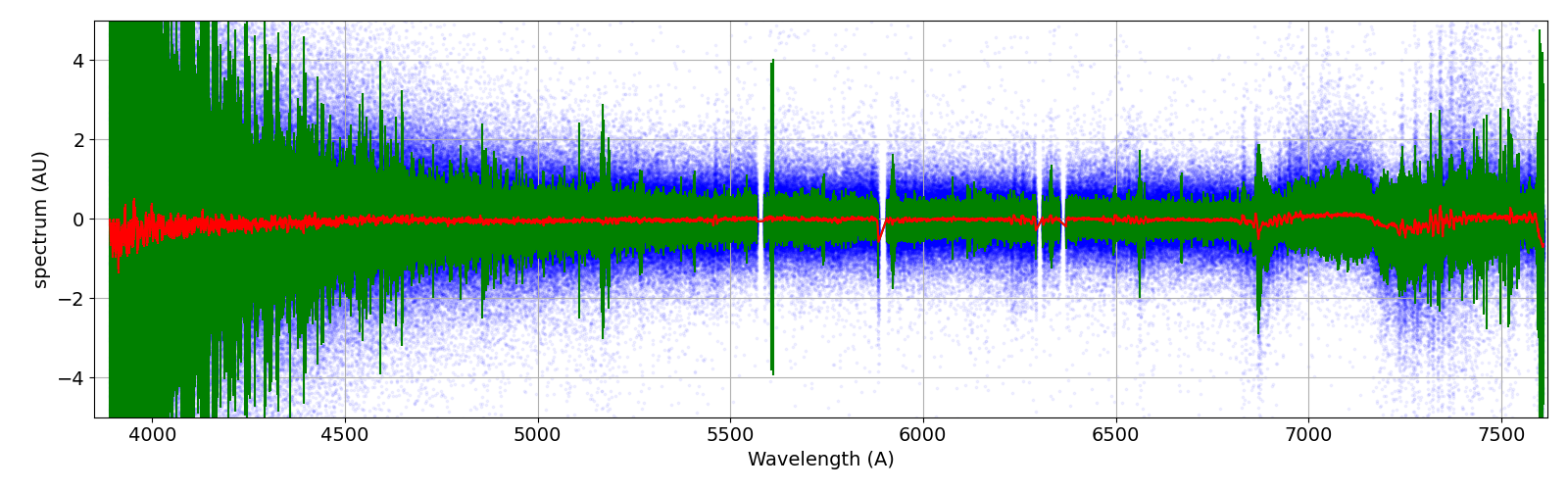}
    \caption{Systematics template in the WIYN spectra. All spectra, after subtraction of a low-order polynomial, have been superimposed (blue dots). The red curve shows their median in  wavelength bins of width 1~\AA. The root-mean-square of the measurements distribution in each bin is shown in green. Areas where atmospheric or instrumental effects either cause large dispersion or prevent any measurements are clearly apparent. Note the ``wiggles" at the long-wavelength end of the spectra.}
    \label{fig:syste_template}
\end{figure}

To perform such evaluation, we first subtracted a low-order polynomial from our spectra, as a similar operation is usually performed for the literature templates.  We determined a ``systematics" template from the resulting spectra, by computing their median with respect to frequency, as illustrated by Figure \ref{fig:syste_template}.
The structures in this median were caused e.g., by atmospheric features and/or specific features in the fiber system. To each reduced spectrum we then fitted an amplitude for this systematics shape and subtracted it. 

We chose to use two (for robustness) galaxy spectra templates with emission features and two with absorption features. 
Emission line galaxy spectra  have been chosen from the SDSS 
DR5 spectral cross-correlation templates\footnote{https://classic.sdss.org/dr5/algorithms/spectemplates/}. Specifically, we used templates  with IDs 24 and 25 in our processing. For absorption line galaxies, on the one hand, we used the NGC 4125 spectrum listed by  \cite{Brown_2014} in their representative subsample. We also used template no. 41 in the collection used in the {\tt AUTOZ} code developed within the GAMA collaboration \citep{Baldry_2014}\footnote{code and data can be found in I. Baldry's website: https://www.astro.ljmu.ac.uk/~ikb/research/}. Finally, two redshift measurements have been derived using template ID 30 in the SDSS DR5 collection, corresponding to objects with broad emission lines (like QSOs or AGN). Prior to their use in redshift determination, all templates are pre-processed by fitting and subtracting a low order polynomial. 

Given the limited signal-to-noise ratio (SNR) of our reduced spectra, we chose to single out the most salient features of the templates to evaluate a redshift. 
For each tested redshift value $z$, template spectra were redshifted, after which the procedure computes: 
$$c(z) = 1/N_{meas.}\sum_{f} \sum_{i}(d_i-\langle d\rangle )(f(z)_i-\langle f(z)\rangle ), $$
where $f(z)$ denotes the selected parts of the template (e.g. the $H_\alpha$ emission line region) translated according to $z$ and re-interpolated at the wavelengths of the tested reduced spectrum  points denoted $d_i$.
From the errors on each data point evaluated by our spectrum extraction pipeline, we also derived the uncertainty on $c(z)$, denoted by $\sigma_c(z)$, for each tested redshift. The median of the $c(z)$ values, denoted by $m_c$, is in general close to 0. We compute the clipped r.m.s.  of these values, denoted by $\beta$. 
The procedure selects the value of $z$ which corresponds to the maximum of $c(z)-m_c$ for each template. From the width of the peak around this maximum, we estimate in general an accuracy of $\sim 5\times10^{-4}$ on our redshift estimation, with 
a large dispersion (in particular, the accuracy is in general better for emission line spectra). Our procedure also provides an estimation of 
the ``significance" of this excursion: 
$$SN=\frac{  c(z)-m_c}{\sqrt{\sigma_c(z)^2 +\beta^2}}.    $$
We finally examine by eye the results of this determination to keep only those spectra for which the redshift determination seems reliable.  In general, only one of the two template categories leads to a redshift estimation, but in a few cases a compatible redshift is estimated  with both, as in one of the emission line galaxy templates some absorption features are also present. For 2 spectra, the redshift was estimated using the SDSS  ID 30 QSO template as both present broad emission lines characteristic of these objects. 
\begin{figure}
    \centering
    \includegraphics[width=0.8\linewidth]{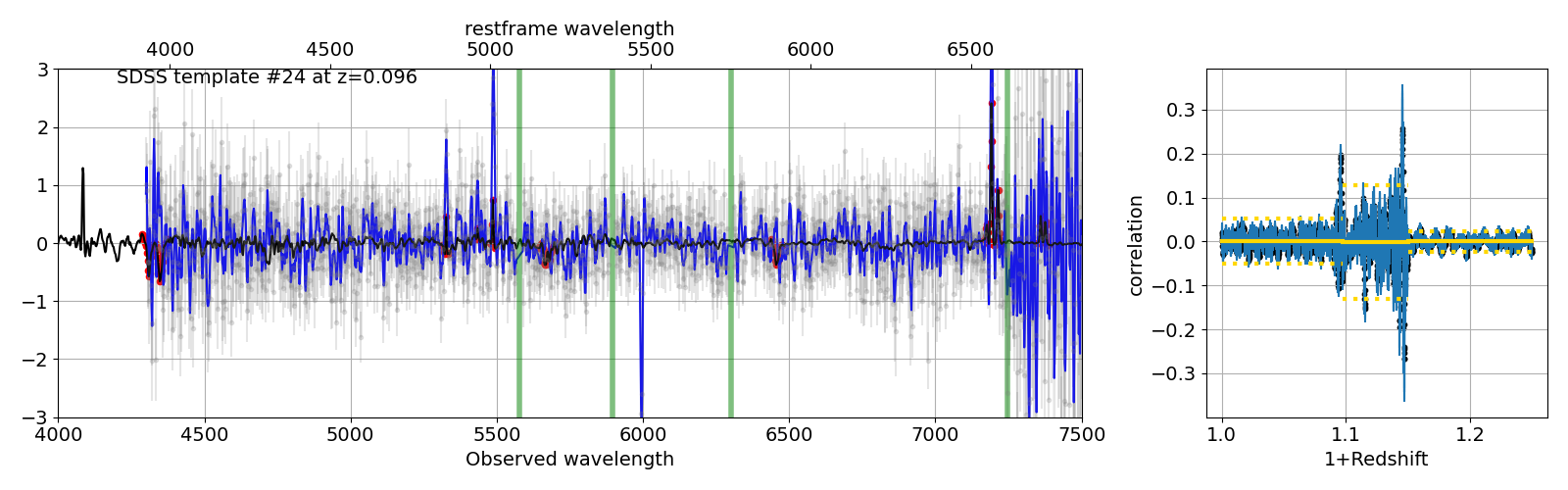}\\
    \includegraphics[width=0.8\linewidth]{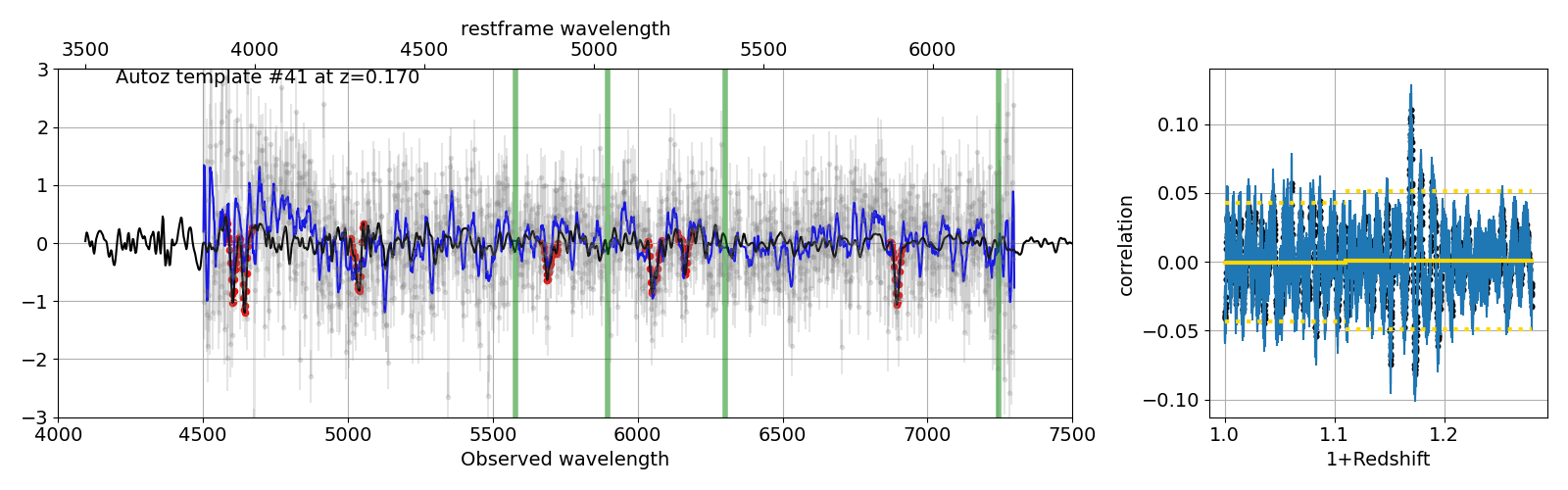}
    \caption{Results of the semi-automatic redshift determination procedures. The top panel shows the reduced spectrum of object NCCS426395, for which we measure $z=0.0956$ with template no. 24 from the SDSS collection. The bottom panel shows the reduced spectrum of object NCCS452010 for which we measure $z=0.1697$ using the template no. 41 from the {\tt AUTOZ} collection, corresponding to an elliptical galaxy. See text for a detailed explanation. }
    \label{fig:semiautoz}
\end{figure}

We show two examples of results we obtained with our procedure in Figure \ref{fig:semiautoz}. In both examples we show on the right the variations of $c(z)$
as a function of the tested redshifts. The value we compute is shown in black; we superimpose error bars obtained by propagating the spectrum errors. The yellow lines outline the median and dispersions computed in broad $z$ intervals. On the left hand side, we show the reduced spectra and template, after subtraction of a low order polynomial.
The WIYN spectra are shown by the gray dots and error bars. We also indicate, in blue, a smoothed rendition of the same data. The template, redshifted to the best redshift value, is shown in black. We indicate by  red dots the parts of the templates we used in the calculation of $c(z)$. They correspond to the most salient features of the templates. 

The NCCS426395  presents a few narrow emission lines. The most prominent is the H$_{\alpha}$ line  at 6564.6\AA. Several other lines are visible, such as $H_{\beta}$ (4862.7 \AA) and O III lines near 5000\AA.  We also indicate with green bands the location of usual atmospheric lines, used in the SDSS analyses\footnote{https://classic.sdss.org/dr6/algorithms/linestable.php}. On the $c(z)$ variations we clearly see the impact of the non-uniformity of noise and systematics: Above $z=0.1$, the $H_{\alpha}$ line is redshifted in the high end of the spectrum, where residual systematics show up. Hence, the correlation estimation is more dispersed around 0. Above $z\sim 0.14$ this line gets  shifted outside the spectrum's range. This directly impacts the efficiency of our redshift determination, which for emission line galaxies falls above $z\sim 0.1$.

The NCCS452010 shown below presents typical broader absorption lines from  Na (around 5895 \AA), Mg (around 5176.7 \AA), the H and K doublet from ionized Ca (3935 and  3970 \AA), the G band near 4310\AA. In general, due to the broad nature of the features used in the correlation analysis, the accuracy is lower than for emission line spectra. The rather low SNR of our data imposed us to require that more than two of the most prominent features are identified in the spectra. We also restricted the data used in a narrower wavelength interval to avoid results biased by the systematics at both ends of our spectra.  Finally, thanks to the marked HK doublet absorption lines, the efficiency of the procedure is better than for the emission line spectra for redshifts  above 0.15--0.20.  

%% file: sec_4_spectrocatalog.tex
\section{The spectroscopic catalog}
\label{sec:spectrocatalog}

The statistics of the redshift determinations for tNCCSz are given in Table \ref{tab:stats}.  The number of targets that were candidates for observation according to the criteria outlined in Section \S\ref{sec:targetselection} is 2102.  We obtained spectra for 1874 of those targets. More than one spectrum was obtained for some targets where the first results were not satisfactory. 
The visual method was able to extract information from the spectra for 789 targets ($Q_\mathrm{vis} \geq 1$), out of which 669 were classified as galaxies with a redshift, and the remaining 120 were classified as stars. In most cases, thetae failure to determine a redshift was due to 
a low SNR spectrum, but the semi-automatic analysis indicates that some redshifts were missed by the visual analysis.

The semi-automatic analysis was able to determine redshifts for 786 targets in total ($Q_\mathrm{auto} \geq 1$), out of which 553 were also identified as galaxies by the visual method. The two redshift determinations agree well for the vast majority of these objects, with $|Z_\mathrm{vis} - Z_\mathrm{auto} | < 0.005$ for 510 galaxies out of 553. In addition, there are 9 objects identified as stars by the visual method for which a redshift is obtained by the semi-automatic procedure. These 9 objects have been subjected to a complementary analysis, where the information from the GAIA catalog has also been used. We have classified as stars 4 of these 9 objects, and have classified the remaining 5 as galaxies with redshifts assigned by the semi-automatic method. In summary, combining the two methods, we have been able to extract information for 1013 targets, of which 115 have been classified as stars and 898 as galaxies with a redshift estimate.

\begin{table}[!ht]
  \centering
    \caption{Target Statistics.}
    \label{tab:stats}
\begin{tabular}{l@{\hspace {1.4in}}r@{\hspace {0.1in}}} 
      \hline
      \hline
      Targets & 2102 \\
      Targets with spectra  & $1874^{*}$ \\
      Targets with $\ge2$ attempted  spectra & $105^{\dag}$\\
      Targets with $\ge3$ attempted spectra & 7 \\
      \hline
      {\bf Visual method results} \\
      Objects with $Q_\mathrm{vis} \ge 1$ (spectrum analysis successful) & 789 \\
      Objects with $Q_\mathrm{vis} \ge 1$ \& redshift = 0, identified as stars & 120 \\
      Objects with $Q_\mathrm{vis} \ge 1$ \& redshift $>0$, identified as galaxies & 669 \\
      \hline
      {\bf Semi-automatic method results} \\
      Objects  with redshifts, identified as galaxies $Q_\mathrm{auto} \ge 1$ & 786 \\ 
      Objects  with redshifts, identified as galaxies $Q_\mathrm{auto} \ge 3$ & 699 \\
      \hline
      {\bf Total number of galaxies with redshifts (used in analyses below) } & 898 \\
     \hline
     \hline
\end{tabular}
\begin{itemize}
\item[*] {\footnotesize The main reason we were unable to obtain spectra for all the targets was due to the constraints in placing fibers in the focal plane and the lack of observing time to make additional observations of each tile.  In addition, we lost targets because the telescope software did not configure all the fibers according to the pre-observational simulation (49 targets) and a broken fiber that was not recognized during the early observations (36 targets.}

\item[\dag] {\footnotesize More than one spectrum indicates multiple attempts to obtain a spectrum--- normally because prior attempts did not actually produce a spectrum or because a redshift could not be determined from the spectrum.}
\end{itemize}
\end{table}

\begin{figure}[htp]
    \centering
   \includegraphics[width=0.6\textwidth]{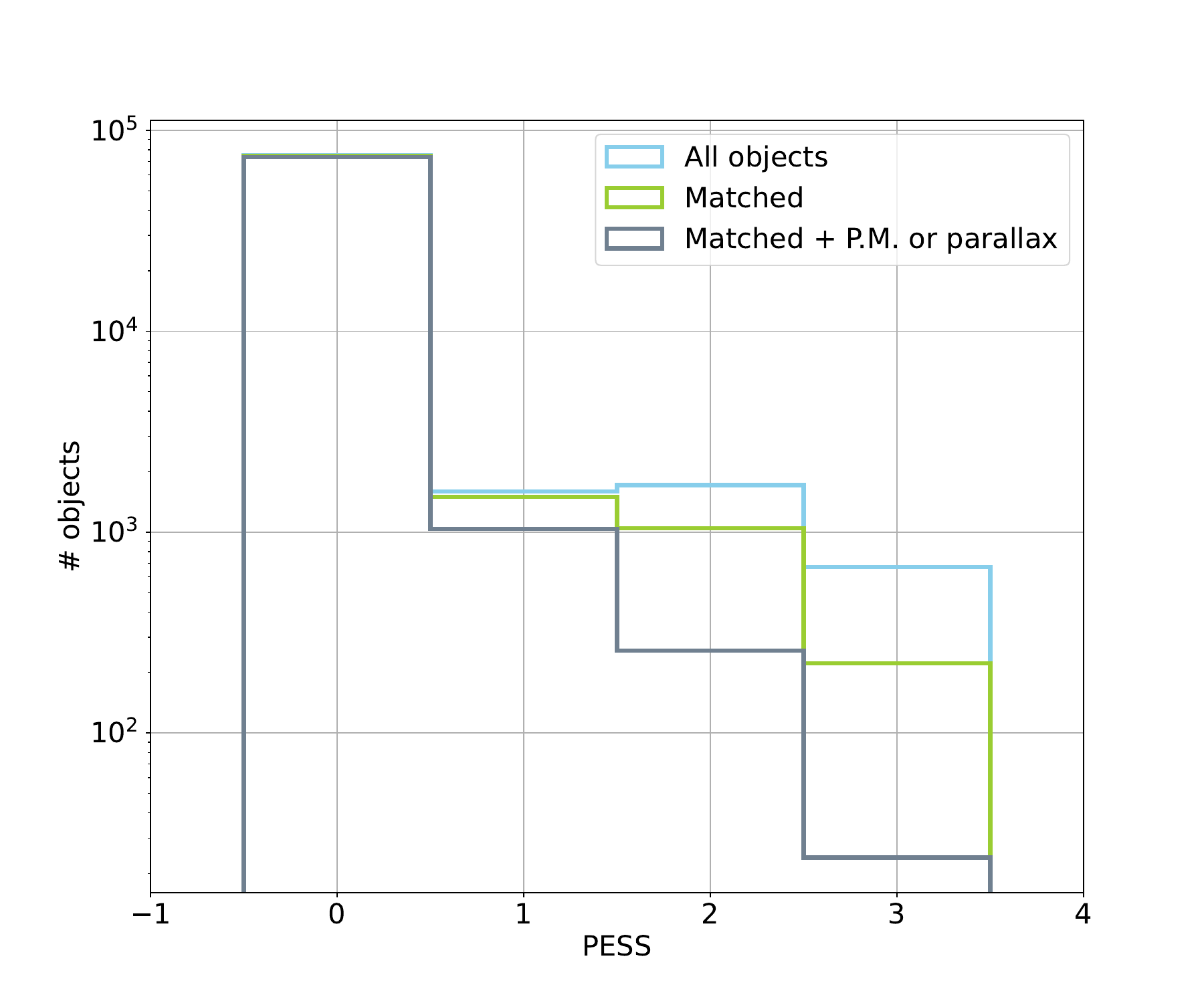}
   \caption{Distribution of the PESS flag (NCCS source identification flag) for various samples: all NCCS objects  with $\delta\geq 86.6^\circ$ and $13<m_V<18.995$ (light blue), objects with a match in Gaia DR3 (light green), objects with a match in Gaia DR3 (orange), and objects with a match in Gaia DR3 with a significant proper motion or a significant parallax  (gray). Our targets, which are restricted to have PESS = 2 or PESS = 3, are likely to be contaminated by stars at $ \lesssim 10 \%$. 
   \label{fig:PESS_NCCS_Gaia}} 
\end{figure}

\subsection{Cross-match with other catalogs}
\subsubsection{A posteriori check of target selection with Gaia}
\label{sec:targselgaia}
In order to estimate the fraction of proposed targets that might actually be stars, rather than galaxies, 
 we matched the NCCS catalog with the Gaia DR3 catalog \citep{gaia_collaboration_gaia_2023}.  As this catalog was released well after the beginning of our survey, this is only an a posteriori check.  First, we looked at the distribution of the PESS flag for NCCS objects with $\delta\geq 86.6^\circ$ and $13<m_V<18.995$ - essentially the same as the criteria used for our target selection - shown on Figure \ref{fig:PESS_NCCS_Gaia}.  Within this sample, 1714 and 669 objects have PESS flag equal to 2 and 3, respectively. Of these, 
1046 objects with PESS = 2, and 222 objects with PESS = 3, are closer than 2 arcsec from an object in Gaia DR3 (a ``match'' in the following). In contrast, more than 99.8\% of NCCS objects with PESS $\leq$ 1 (point-like) have a match in Gaia DR3.

Gaia DR3 provides parallax and proper motion measurements. An object in the Gaia catalog with a significant parallax or proper motion is most probably a star. For 258 of the 1046 PESS = 2 objects with a match in Gaia, the Gaia source  has a parallax greater than twice the reported error in the Gaia catalog, or a significant proper motion. (Note that $\sim$ 99.8\% of objects with a significant proper motion also have a significant  parallax.)
For the 222 NCCS objects with PESS = 3 and with a match in Gaia, 
the above category (with significant parallax or proper motion) amounts to 24 objects.  As Gaia is targeting stars, PESS=2 and 3 NCCS objects without a match are most probably not stars, but mostly galaxies. Assuming these have the same properties as the matched ones (a worst case scenario),  the fraction of our selected targets that could be stars would amount to $ \lesssim 10 \%$.  As noted above, we did not use the Gaia catalog to remove stars from our target list due to its late availability. 

\subsubsection{Comparison of our catalog with Gaia}

Based on our (visual) analysis of the spectra, we identified 120 stars in our NCCS target sample ($Q_\mathrm{vis} \ge 1$, $Z_\mathrm{vis} = 0$). All of these objects have a Gaia counterpart, all with measured parallax and proper motion.   For more than 90\% of these objects, the measured parallax and/or proper motion is significantly ($> 3 \sigma$) different from 0,  indicating that the vast majority of objects spectroscopically identified as stars are indeed stars. The fraction of objects identified as stars, 120 out of 789 objects with $Q_\mathrm{vis} \ge 1$, is $\sim 15 \%$ of the total.  This fraction is slightly higher than the overall star contamination of $ \lesssim 10 \%$ expected among targets, as described in \S\ref{sec:targselgaia}. This higher rate may be explained by the biased magnitude distribution; stars are brighter and hence have a larger spectrum based identification efficiency.%

\subsubsection{Comparison with the 2MASS Redshift Survey}
The 2MASS redshift survey catalog \citep{huchra_2mass_2012} contains a number of galaxies that overlap our survey area.  As a means for testing the completeness of the tNCCSz, we searched the 2MASS catalog for galaxies with $\delta\ge 86.5^\circ$ and $m_K > 13.5$.  We found 26 such galaxies and  

our tNCCSz survey obtained spectra for 14 of them.  The breakdown of the reasons the remaining 12 were not in our sample is summarized in Table \ref{tab:2MASS}.  The two targets that were ``Not in the NCCS Catalog'' were detected in V-band by NCCS but without corresponding measurements in R or I bands.  These objects appear in the full NCCS catalog but not in the ``cleaned'' catalog that we used for target selection. 
Somewhat surprisingly, 10 galaxies were not included in our target list because the NCCS catalog PESS score indicated that they were more likely stars.  This result appears to be inconsistent with our understanding that the NCCS catalog separation between stars and galaxies is $>92\%$ accurate, based on a comparison of NCCS with SDSS \citep{Gorbikov_2014}.

The overall star-galaxy separation is poorer (10 stars/24 objects = 42\%) for this particular small sample.  It may be that the star-galaxy separation for the 2MASS redshift sample is less accurate for the relatively bright galaxies in the 2MASS sample --- although naively one would expect a problem for fainter objects.
 
\begin{table}[!ht]
   \begin{center}
    \caption{ Comparison of 2MASS redshifts with the tNCCSz ones}
     \label{tab:2MASS}
     \begin{tabular}{lrr}
          \hline
          \hline
          Category & Number & Percentage (\%)\\
          \hline
Not in NCCS Catalog & 2	& 8 \\
NCCS Star (PESS 0 or 1)	& 10 & 38 \\
Agreement (roughly)	& 14 & 54 \\
Total	& 26 & 100 \\
\hline
\hline
     \end{tabular}
   \end{center}
\end{table}

\begin{figure}
    \centering
   \includegraphics[width=0.6\textwidth, trim= 0 50 0 0]{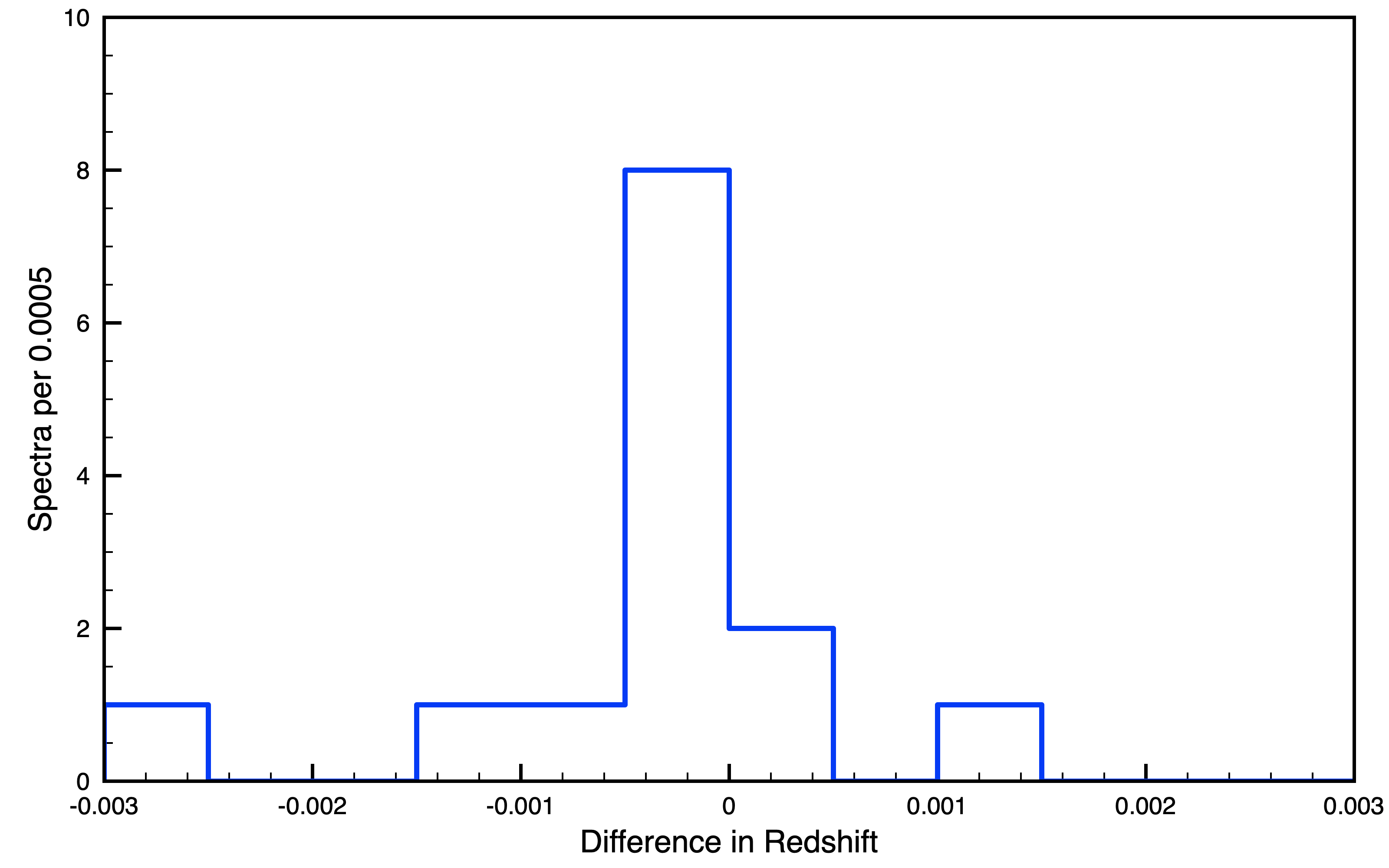}
   \caption{The distribution of differences in redshifts between the 2MASS redshift catalog and the tNCCSz catalog (this work):  $\Delta z = z_{\rm tNCCSz} - z_{\rm 2MASS}$. } 
   \label{fig:zDiff}
\end{figure}

The differences in redshift between the 2MASS catalog objects and those from this work for the 14 common objects are plotted in Figure \ref{fig:zDiff}.  All the WIYN spectra were judged to be $Q_\mathrm{vis} \geq 3$.   The differences are typically less than 0.0005, although there is an indication that the WIYN redshifts may be slightly smaller on average.  The WIYN redshift of the outlier at -0.0029 was determined from a putative H$_\beta$ emission line despite several absorption features which agree well with the 2MASS redshift.
In retrospect, using the H$_\beta$ emission line was probably an error in judgment.  In this case, the WIYN spectrum was judged to be $Q_\mathrm{vis}=3$.  The outlier at 0.0015 appears to have an H$_\alpha$ and an NII(6583) emission line and was also judged to be a $Q_\mathrm{vis}=3$ identification.  However, the difference in redshifts is inconsistent with the expected uncertainty in the WIYN spectra.  The WIYN spectrum for the outlier at -0.0013 was judged to be a $Q_\mathrm{vis}=5$ spectrum, but had only absorption lines, which are more difficult to accurately measure.  
 
\subsection{Selection Function}
\label{sec:selfuncs}

\begin{figure}
    \centering
   \includegraphics[width=0.65\textwidth]{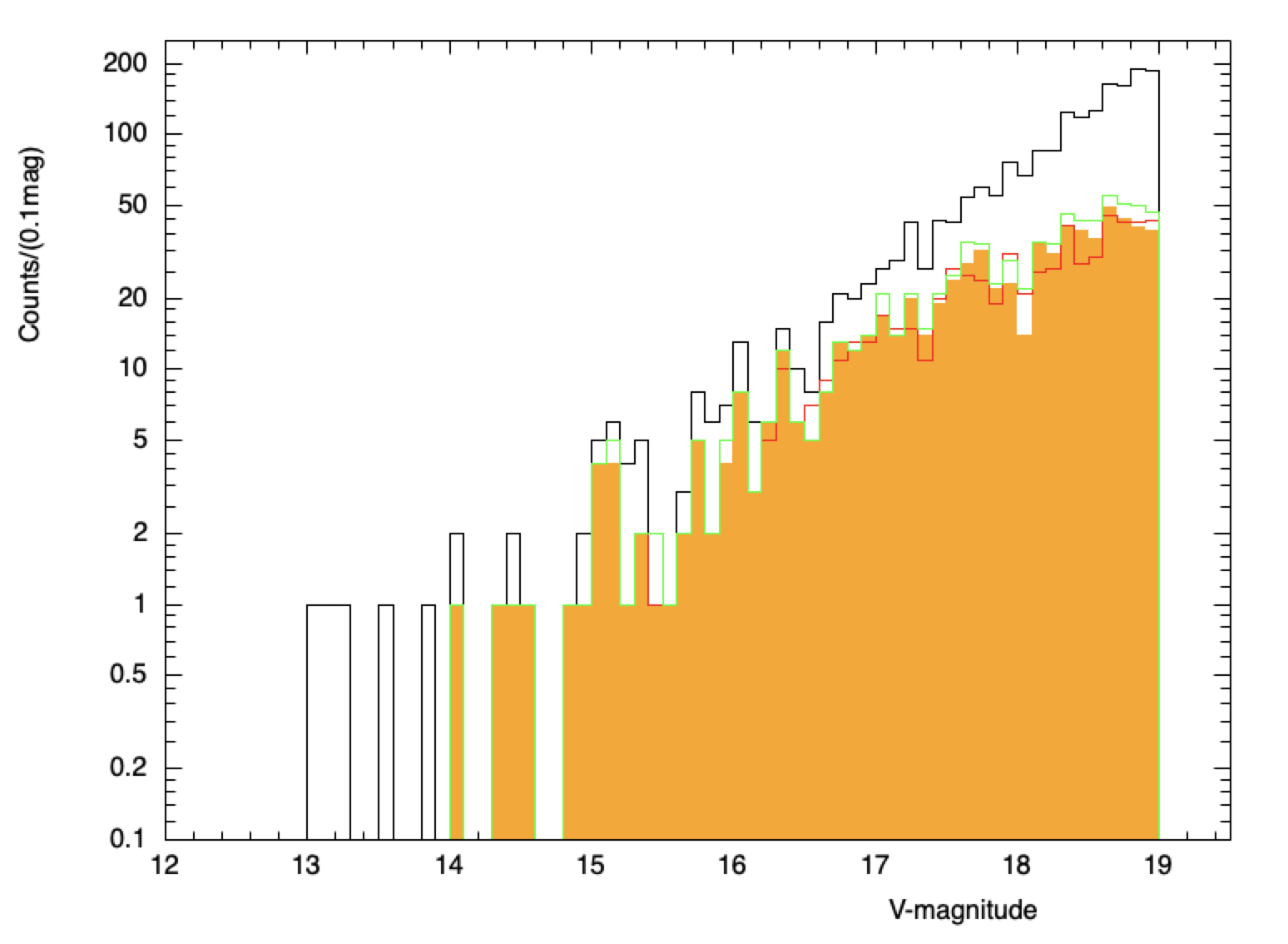}
   \caption{Magnitude distribution of observed targets is shown in black, while the filled orange histogram corresponds to the objects for which a redshift has been obtained by the automatic procedure and an associated quality $Q_{\rm auto} \ge 3$. The magnitude distribution for objects with $Q_{\rm auto} \ge 1$ is represented as the green histogram, while the red histogram shows the distribution for those 
   with a redshift assigned by the visual procedure with $Q_{\rm vis} \ge 1$.
   (Notice the logarithmic vertical scale.)
    \label{fig:magdist-f-quality} }  
\end{figure}

Figure \ref{fig:magdist-f-quality} shows the $V_{\rm mag}$ distribution of all observed NCCS targets, 
together with the distribution of objects identified as galaxies with an assigned redshift, either by the visual procedure or by the automatic procedure. 
There are a total of $N=669$ galaxies with visual redshifts $Q_\mathrm{vis} \ge 1$, and 
$N=786$ galaxies with redshifts assigned by the automatic procedure $Q_\mathrm{auto} \ge 1$. 
This number drops to $N=699$ for a more stringent quality cut $Q_\mathrm{auto} \ge 3$, corresponding to a higher SNR, dropping the $\sim 10\%$  galaxies with the lowest SNR.  

We have modeled the selection effects for the tNCCSz spectroscopic catalog $\mathcal{E} \left( V_{\rm mag}, \delta  \right)$ as a product of three selection functions which are described below:
\begin{eqnarray}
\mathcal{E} \left( V_{\rm mag}, \delta  \right) & = & 
\mathcal{E}_{\rm NCCS}(V_{\rm mag}) \times \mathcal{E}_{\delta}(\delta) \times \mathcal{E}_{z}(V_{\rm mag})
\end{eqnarray}
\begin{enumerate}
\item The NCCS catalog selection effect, modeled as a V-magnitude dependent effect:
\begin{eqnarray}
\mathcal{E}_{\rm NCCS}(V_{\rm mag}) & = & \frac{1}{1+\exp\left(c_1 \times (V_{\rm mag} - m_1^*) \right) } \hspace{10mm} c_1 = 2.5 \hspace{2mm} , \hspace{2mm} m_1^*=19.25 \label{eq:effnccs}
\end{eqnarray}
\item A significant fraction of the targets located beyond 3$^\circ$ of the NCP were dropped during the tiling for fiber assignment. We model this as a declination-dependent fiber assignment efficiency function:
\begin{eqnarray}
\mathcal{E}_{\delta}(\delta) & = & \frac{1}{1+\exp\left(c_\delta \times (\delta - \delta^*) \right) } \hspace{10mm} c_\delta = 15 \hspace{2mm} , \hspace{2mm} \delta^*=86.78^\circ
\label{eq:efffibass}
\end{eqnarray}
\item Finally, the efficiency of obtaining a usable spectra, leading to identifying a galaxy with a redshift is modeled as a second magnitude dependent selection function. 
\begin{eqnarray}
\mathcal{E}_{z}(V_{\rm mag}) & = & \frac{1}{1+\exp\left(c_2 \times (V_{\rm mag} - m_2^*) \right) } \hspace{10mm} c_2 = 1 \hspace{2mm} , \hspace{2mm} m_2^*=17.4 \label{eq:effzwiyn}
\end{eqnarray}
\end{enumerate}

We have used the sample of galaxies with redshifts assigned by the automatic procedure with $Q_{\rm auto} \ge 3$ in the clustering analysis presented below, although we obtain fully consistent results by using either the redshifts obtained by the visual procedure, or the larger sample $Q_{\rm auto} \ge 1$.
The parameters of the third selection function $\mathcal{E}_{z}(V_{\rm mag}) $ given above is consistent with the efficiency function describing either the sample of $N=669$ objects with redshifts assigned by the visual procedure or the sample $Q_{\rm auto} \ge 3$ of the semi-automatic procedure with $N=700$ objects. 

Using the above efficiency functions, we are able to reproduce the magnitude distribution of our final list of galaxies with redshifts, the NCCS catalog itself, and the selected targets for this survey. To do so, we start from the galaxy absolute luminosity distribution function, and we take into account the absorption by the Galactic dust, with its variations over the $\sim 35\, \mathrm{deg}^2$ survey area (Figure \ref{fig:extinction-tiling}, right panel). The  galaxy luminosity distribution is modeled as a Schechter function (see Eq. \ref{eq-lum-func}) with the parameters  given by
\citet{2009A&A...508.1217Z}. More precisely, we have used a characteristic magnitude $M^* = -20.5$, normalisation $\Phi^* = 0.0063 \mathrm{Mpc}^{-3}$ 
and a faint end slope $\alpha=-1.07$ as the luminosity function parameters. More details can be found in paragraph \ref{sec-rand-cat}.

\begin{figure}
    \centering
   \includegraphics[width=0.48\textwidth]{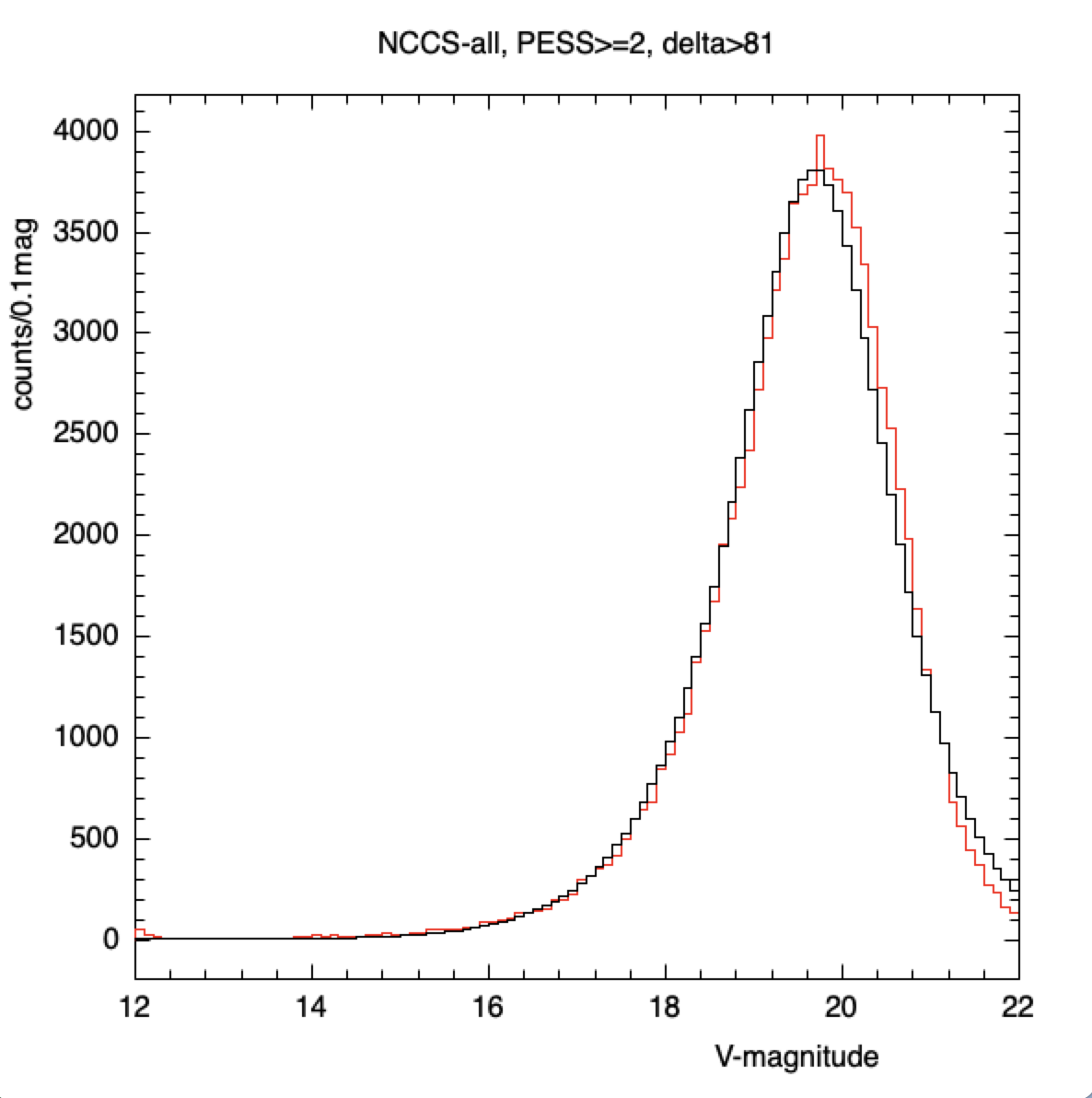}
    \includegraphics[width=0.48\textwidth]{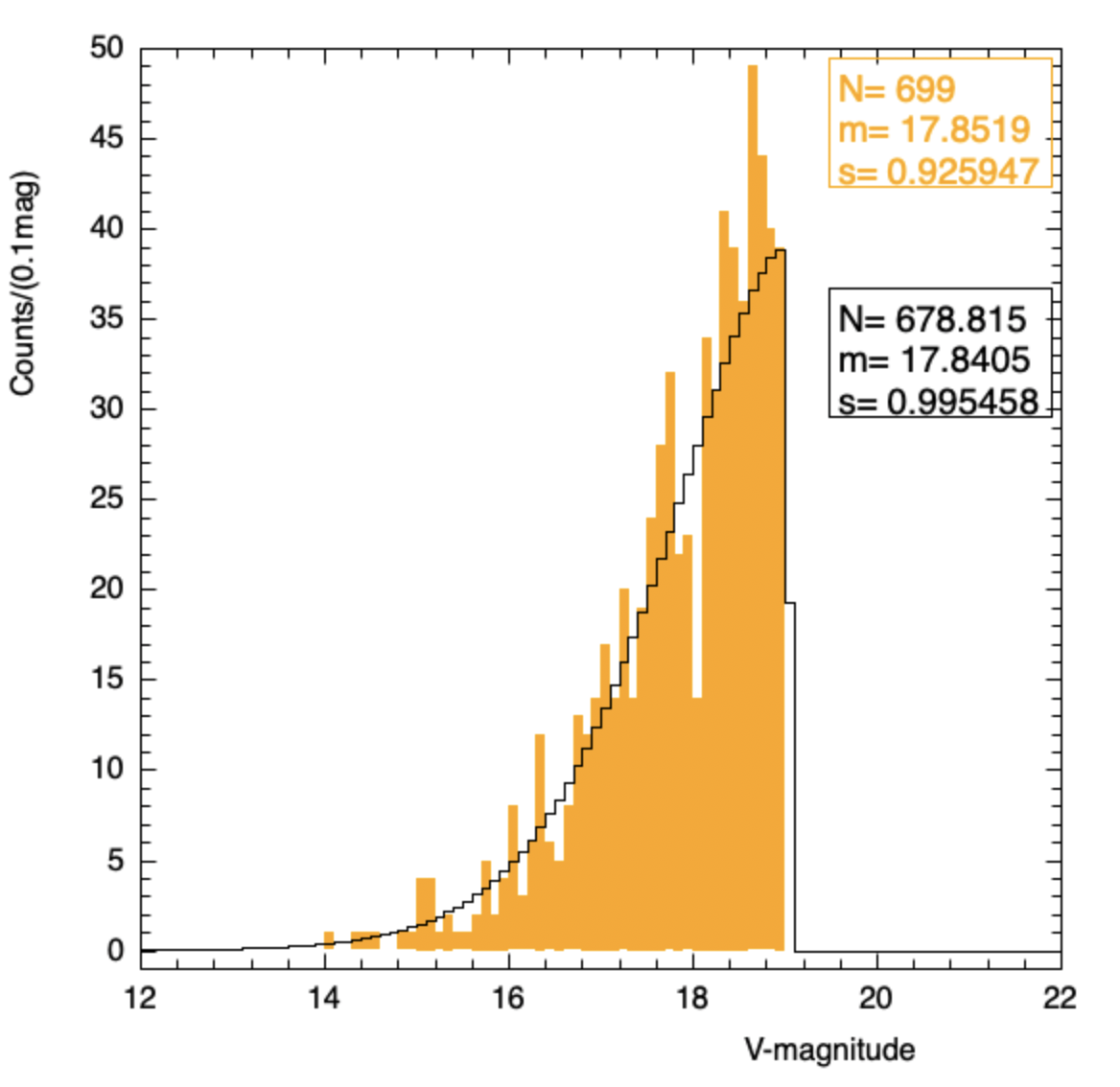}
\caption{Left: V-band magnitude distribution for all NCCS sources with PESS $\ge 2$ and $\delta > 81^\circ$ represented as the red histogram, and the expected magnitude distribution starting from the galaxy luminosity function in black. Right: V-band magnitude distribution for all targets for which a redshift has been obtained by the automatic procedure with associated quality $Q_\mathrm{auto} \ge 3$, represented as the filled orange histogram. The expected distribution using the selection functions and starting from galaxy luminosity function is shown as the black histogram. 
The values of N,m,s shown on the histograms on the right hand side correspond to the number of entries or the integral, the mean and standard deviation of V-magnitude distributions respectively.  \label{fig:magdistallsim}}
\end{figure}

Using the NCCS selection function defined in equation 
\ref{eq:effnccs}, we are able to reproduce well the apparent magnitude distribution of the NCCS catalog.  
The right panel of Figure \ref{fig:magdistallsim} shows 
the expected distribution of the magnitude of the galaxies with a redshift from the tNCCSz survey, obtained using the efficiency functions defined in equations 
\ref{eq:effnccs}, \ref{eq:efffibass} and \ref{eq:effzwiyn}
superimposed on the actual distribution (filled orange histogram). The agreement is quite satisfactory, including the total number of expected objects. 

\begin{figure}
    \centering
   \includegraphics[width=0.75\textwidth]{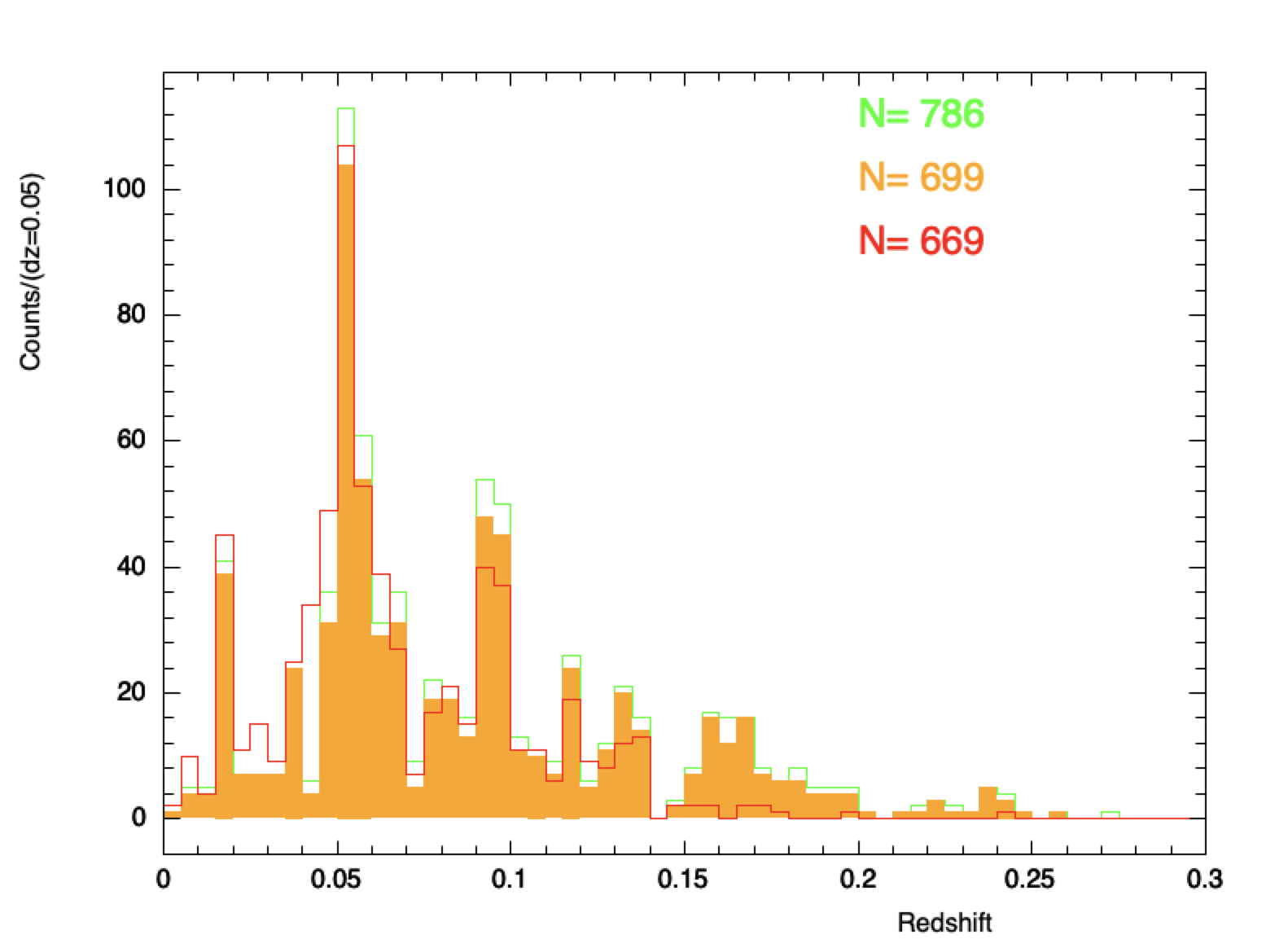}
\caption{Redshift distribution of tNCCSz galaxies, all galaxies with an estimated redshift by the automatic procedure with $Q_{\rm auto} \ge 3$, represented as the filled orange histogram, while the green histogram corresponds to the ones with $Q_{auto} \ge 1$. The distribution of redshifts obtained with the visual procedure
with $Q_{\rm vis} \ge 1$ is shown as the red histogram. The total number of galaxies in each sample $N$ is also shown.  \label{fig:redshiftdist}}
\end{figure}

\subsection{Catalog Description}

The tNCCSz redshift catalog contains a total of 23 columns and 1874 rows, with each row corresponding to an NCCS \citep{Gorbikov_2014} source, with the associated redshift information extracted from the current survey's observations. Eleven columns correspond to information in the original NCCS catalog, and an additional 12 columns contain redshift information derived from this survey. The full table in electronic form will be available from the the Centre de Donn\'ees astronomiques de Strasbourg (CDS)\footnote{CDS, Strasbourg astronomical Data Center: \href{https://cds.unistra.fr}{https://cds.unistra.fr}}.

The processed, normalized spectra will also be available as a set of plain text (ascii) files. These files are organized in 14 folders, each corresponding to an observation night.  Each line in the spectra text files contains three comma separated values: the wavelength in angstroms, the flux, and the estimated flux error. Wavelengths with negative flux errors should be masked. The NCCS\_ID in the filename can be used to link the file to the NCCS source. For example, the {\tt NCCS0036126mag1898\_33\_norm.txt} file is the spectrum file for the NCCS source with {\tt ID=36126}.  All NCCS\_ID are zero padded to represent 7 digits after the initial four letters {\tt NCCS}.

Table \ref{tab-catalog-extract} shows the first few rows of the full table and a subset of the columns. The columns in the full table are as follows:

\begin{itemize}
\item[-] Column 1, labeled {\bf Tile}, identifies the observational tile corresponding to the source spectra (See the right panel of Figure \ref{fig:extinction-tiling}).
\item[-] Column 2, labeled {\bf NCCS\_ID}, contains the source identification in the original NCSS catalog.
\item[-] Columns 3 and 4, labeled {\bf RA} and {\bf Dec}, contain the right ascension and declination, copied from the NCCS catalog.
\item[-] Columns 5, 6 (not shown) and 7 (not shown), labeled {\bf Vmag, Rmag, Imag}, are the source magnitudes in the V, R and I bands copied from the NCCS catalog.
\item[-] Columns 8--10 (not shown), labeled {\bf VFWHM, RFWHM} and {\bf IFWHM}, are the source sizes in arcsec in the three photometric bands copied from the NCCS catalog. 
\item[-] Column 11 (not shown), labeled {\bf errVmag}, contains the V-band photometric error {\bf errVmag}, and column 12 (not shown), labeled {\bf PESS}, corresponds to the extended source score from the NCCS catalog.
\item[-] Columns 13 and 14, labeled {\bf Zvis} and {\bf Qvis}, contain the redshift determined by the visual method and the associated quality. $\mathrm{Z_{vis} = 0}$ corresponds to sources identified as stars. Usable redshift values have quality factors $1 \leq \mathrm{Q_\mathrm{vis}} \leq 5$. $\mathrm{Q_{vis}} \leq 0$ indicates that the visual analysis of the spectrum was unsuccessful.
\item[-] Columns 15 and 16, labeled {\bf Zeml} and {SNReml}, contain the redshift determined by the semi-automatic method, using emission line template spectra and the corresponding SNR.
\item[-] Columns 17 and 18, labeled {\bf Zabs} and {\bf SNRabs}, contain the redshift determined by cross-correlating the spectrum with absorption line templates and the associated SNR.
\item[-] Columns 19--21, labeled {\bf Zauto}, {\bf Qauto} and {\bf SNRauto}, contain the best estimate of the redshift by the semi-automatic method, the associated quality factor and SNR.  One of the two redshifts Zeml or Zabs have been selected, according to the SNR value. Note, however, that SNRabs has been rescaled to make the SNR distribution more similar to the one from the emission line templates. $\mathrm{Q_{auto}}=0$ indicates that no redshift could be determined. Each time the quality factor is increased by a step of one from $Q_{\rm auto}=1$ to $Q_{\rm auto}=5$, 5\%--7\% of the objects with the lowest SNR are dropped.
\item[-] Column 22, labeled {\bf SFauto}, provides some more information about Zauto. $\mathrm{SF_{auto}} = 11,13$ indicates that redshift has been obtained with one of the two emission line templates, while  $\mathrm{SF_{\rm auto}} = 16$ corresponds to the QSO/AGN template with broad emission lines. Redshifts obtained from absorption line templates are identified as $\mathrm{SF_{\rm auto}} = 22,24$. $\mathrm{SF_{\rm auto}} = 35,45$ indicates that a redshift has been obtained with both emission and absorption line templates. The selected redshift is for the emission line template for $\mathrm{SF_{\rm auto}} = 35$ and for the absorption line template for  $\mathrm{SF_{\rm auto}} = 45$.
\item[-] Column 23, labeled {\bf Type}, is the best guess of the object type, either a star or a galaxy. The field is left blank if none of the redshift identification methods was successful. {\tt Type=STAR} indicates that the object is likely to be a star, based on the visual method and with no reliable redshift from the automatic procedure. If a redshift has been obtained by either of the two methods, then {\tt Type=GALAXY}. The two objects identified as QSO/AGN are tagged with {\tt Type=GALAXY/AGN}.
\end{itemize}

\begin{table}[tbp]
\begin{center}
\begin{tabular}{c|c|cccc|cc}
\hline 
Tile & NCCS\_ID & RA & Dec & Vmag & Col6-12 & Zvis & Qvis \\
\hline 
tile09h36m+87d06m & 19 & 139.831 & 86.9996 & 18.527 & \ldots & -1 & 0  \\
tile07h12m+87d06m & 259 & 102.061 & 87.0383 & 15.322 & \ldots & 0.0529 & 5  \\
tile07h12m+87d06m & 269 & 102.089 & 87.034 & 17.271 & \ldots & 0.0531 & 4 \\ 
tile08h24m+87d18m & 375 & 119.297 & 87.0247 & 18.768 & \ldots & 0.1479 & 1 \\ 
tile07h12m+87d06m & 443 & 117.087 & 86.9937 & 17.638 & \ldots & 0.0572 & 3 \\ 
tile07h12m+87d06m & 489 & 116.6 & 87.0417 & 17.975 & \ldots & -1 & 0  \\
tile07h12m+87d06m & 532 & 111.961 & 87.0558 & 16.729 & \ldots & 0 & 5 \\
tile07h12m+87d06m & 572 & 112.224 & 87.0358 & 17.648 & \ldots & 0.0533 & 3 \\ 
\hline 
\end{tabular} \\[2mm]
\begin{tabular}{c|cccc|cccc|c}
\hline 
NCCS\_ID & Zeml & SNReml & Zabs & SNRabs & Zauto & Qauto & SNRauto & SFauto & Type \\
\hline 
19 & -1 & -1 & -1 & -1 & -1 & 0 & -1 & 0 & \\
259 &  -1 & -1 & 0.0527 & 3.61982 & 0.0527 & 5 & 6.89125 & 22 & GALAXY \\
269 &  -1 & -1 & 0.0542 & 3.32445 & 0.0542 & 5 & 6.11282 & 22 & GALAXY \\
375 & 0.1768 & 7.41233 & -1 & -1 & 0.1768 & 5 & 7.41233 & 13 & GALAXY \\
443 & -1 & -1 & -1 & -1 & -1 & 0 & -1 & 0 & GALAXY \\
489 &  -1 & -1 & -1 & -1 & -1 & 0 & -1 & 0 & \\
532 & -1 & -1 & -1 & -1 & -1 & 0 & -1 & 0 & STAR \\
572 & 0.0528 & 3.39454 & 0.052 & 1.87262 & 0.0528 & 5 & 3.39454 & 35 & GALAXY \\
\hline 
\end{tabular}
\end{center}
\caption{The first few rows from the tNCCSz redshift catalog. Note that the seven columns {\tt Rmag, Imag, VFWHM, RFWHM, IFWHM, errVmag, PESS} are not shown, and the table is shown split in two parts, with the first part (top) showing 7 columns, the first 5 and {\tt Zvis, Qvis}, while the last 9 columns are shown in the second (bottom) part. The column {\tt NCCS\_ID} is included in both parts. \label{tab-catalog-extract}}
\end{table}

%% file: sec_5_clustering.tex
\section{Clustering analysis}
\label{sec:clustering}

One way to validate tNCCSz is to analyze the redshift space clustering of our galaxies and compare this with other surveys since the clustering statistics in different cosmological volumes are expected to be the same.  Due to its small solid angle tNCCSz can add little new to our knowledge of galaxy clustering, but this is not why this survey was made.  The comparison we make is with the Sloan Digital Sky Survey (SDSS).  SDSS is chosen because of its similar depth, excellent control of systematics, and its very large area, which provides a large statistical sample.  More specifically, we compare tNCCSz with mock tNCCSz catalogs derived from the SDSS spectroscopic catalog \citep{2017AJ....154...28B}.  The procedure for generating these mocks is described in \S~\ref{sec:mocks}.

The comparisons with SDSS clustering we will make are 1) a qualitative visual comparison, 2) a comparison of estimators of the 2-point correlation function, and 3) galaxy cluster identification.  While the correlation function is the primary quantity used to characterize galaxy clustering, it is insensitive to non-Gaussianities.  Galaxy clustering is intrinsically non-Gaussian on small scales and, in addition, sample incompleteness can introduce artificial non-Gaussianities. For example, the inherent limitations on fiber placement by Hydra both is difficult to model (it is not simply a matter of fiber collisions) and is a potential source of artificial clustering in an incomplete survey since the spectra not taken were partly driven by this.  A visual comparison between tNCCSz and mocks may reveal artifactual  clustering caused by such systematics.  Similarly, the properties of 3D concentrations of galaxies is a powerful probe of non-Gaussianity.

\subsection{SDSS extracted Mock Catalogs}
\label{sec:mocks}

While of similar depth, the SDSS and NCCS photometric surveys have significant differences.  These include  
\begin{itemize}
    \item SDSS uses $ugriz$ photometry to select spectroscopic targets while the NCCS measures only $VRI$ photometry.
    \item The SDSS main sample is a nearly complete $r$ magnitude limited survey whereas tNCCSz is an incomplete $V$ magnitude limited survey.
    \item The star-galaxy separation is different.  SDSS uses a multi-layered approach using magnitudes, light profiles, model fits in 5 bands. NCCS has a simpler morphological cut in two bands.
    \item The SDSS survey covers only parts of the sky with low extinction (absorption by Galactic dust) whereas extinction is large throughout the small tNCCSz survey area.  No extinction correction was applied to the $V$ magnitudes to select spectroscopic targets for tNCCSz.
\end{itemize}
We next describe how we have accounted for these difference in order to make ``realistic'' mock tNCCSz catalogs from the SDSS catalog.

SDSS mocks are created from the SDSS spectroscopic sample \citep{2017AJ....154...28B} by selecting a set of a 3.5$^\circ$ radius disks on the sky centered on different directions, $(\alpha_0, \delta_0)$, on the sky. All objects within the chosen region, identified as galaxies (${\rm type}=3$), with a spectroscopic redshift $z < 0.5$, and with an SDSS $g$ brighter than $g < 19.5 $, are rotated toward the NCP. A rotational transformation is determined for each disk which rotates the SDSS region to overlap the tNCCSz disk centered on the NCP, i.e. a transformation which rotates $(\alpha_0, \delta_0)$ to $\delta=90^\circ$.  Each source is moved to the NCP region by this transformation.  The $V$ band extinction  ($A_\mathrm{V}$) at the rotated position is determined using the Planck extinction map.  An un-extincted $V$ magnitude is computed from the SDSS $g$ magnitude using $V = g - 0.3$. This $-0.3$ magnitude shift has been estimated using the color transformations of the SDSS photometric bands \footnote{See \href{http://www.sdss3.org/dr8/algorithms/sdssUBVRITransform.php}{http://www.sdss3.org/dr8/algorithms/sdssUBVRITransform.php} for SDSS to other photometric system conversion.} 
\citep{2006A&A...460..339J}. The un-extincted $V$ is corrected for extinction by subtracting $A_\mathrm{v}$. A fraction of the sources are dropped, taking into account the NCCS selection efficiency (eq. \ref{eq:effnccs}) and the fiber assignment efficiency (eq. \ref{eq:efffibass}).  This catalog of rotated $(\alpha, \delta)$'s and extincted $V$'s provides a mock catalog of tNCCSz targets. A subset of these targets are selected for the spectroscopic sample using the redshift determination efficiency function of eq. \ref{eq:effzwiyn} to provide an tNCCSz mock catalog or ``mock''.  This procedure was repeated for different values of $(\alpha_0, \delta_0)$ to provide 16 tNCCSz mocks.

\subsection{Visual Comparison}

\begin{figure}
    \centering
   \includegraphics[width=0.75\textwidth]{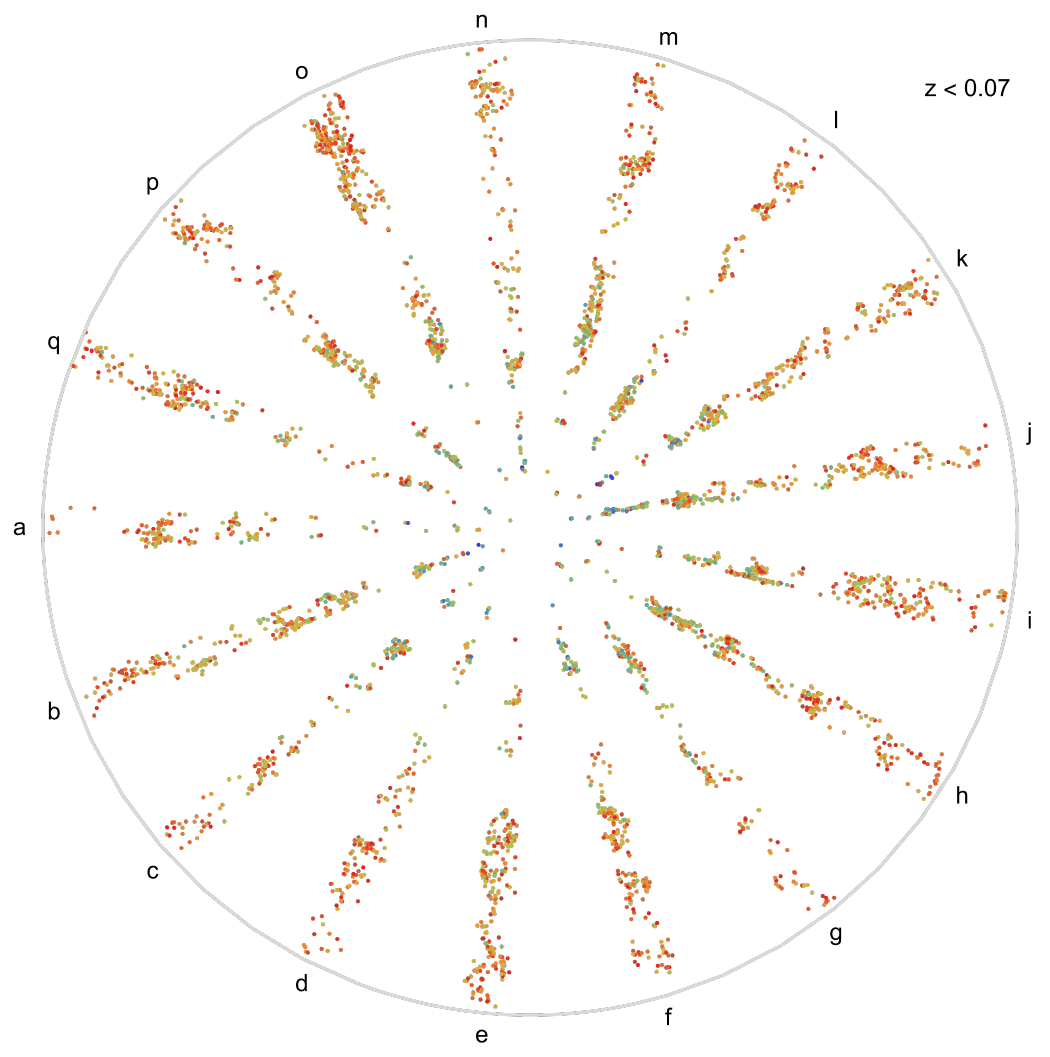}
\caption{Shown are 17 redshift space projections of pencil beam redshift surveys with geometry of the tNCCSz survey.  16 of these projections are of the SDSS based mocks described in sec.~\ref{sec:mocks} and one is of the actual tNCCSz survey.  The distance from the center gives the redshift which extends linearly from $z=0$ to $z=0.07$ which is the redshift range of the future 21m survey.  The pencil beams are distributed uniformly in angle around the center with no particular order and labeled by letters a-q.  Each point corresponds to one galaxy and the color of the points gives the V magnitude which increases from blue to red with red corresponding to the survey limit of $V=19$.  One can find the label corresponding to the tNCCSz survey somewhere below.
\label{fig:mockCarousel}}
\end{figure}

A visual comparison of the SDSS based mock catalogs described above with the tNCCSz is an efficient way to determine whether there are noticeable differences between SDSS and tNCCSz. We leave it to the reader to make this determination.  In Figure~\ref{fig:mockCarousel} we show a projection in redshift space of 13 mock catalogs and the tNCCSz catalog in a random order.  We would argue that no one projection stands out as being different and therefore our tNCCSz redshift catalog lies within the sample variance of similarly selected samples of the SDSS redshift catalog.  

Figure~\ref{fig:mockCarousel} also shows that within the small volume of the tNCCSz survey there is large variation in the numbers of galaxies, redshift distribution, and large scale structures.  As with all pencil beam redshift surveys the dominant visual feature are dense clumps of galaxies along the line of sight separated by underdense regions \cite{1991ApJ...379..482K}.  These are the intersection of the pencil beams with walls filaments and voids in the cosmic web.

\subsection{2-Point correlation function}
\label{sec:2point}

Another comparison we make with the SDSS mock redshift surveys is the 2-point correlation function as a function of galaxy separation, $\xi(r)$.  We use standard methodology to compute estimators of $\xi(r)$ which are described in the following subsections

\subsubsection{Random Catalogs}
\label{sec-rand-cat}
Estimators of $n$-point correlation functions (usually) make use of a ``random catalog''.  These are points given by a Poisson sampling of the survey volume after taking into account all selection functions and efficiencies.  These points are meant to represent a distribution with no clustering, i.e. zero $n$-point correlation functions.  We have generated such random tNCCSz galaxy catalogs using the efficiency functions defined in \S\ref{sec:selfuncs}. We start from the galaxy luminosity distribution defined as a Schechter function depending on the galaxy absolute $V$-band magnitude $M_V$ with the following parameters (see e.g \cite{2001AJ....122..714B}; \cite{2009A&A...508.1217Z}). 
\begin{eqnarray}
 \Phi(L) d L &= &\Phi^*  \left( \frac{L}{L^*}\right)^\alpha \exp \left( - \frac{L}{L^*} \right) \, d L
\hspace{10mm} \frac{L}{L^*}  =   10^{0.4 (M^* - M_V)}   \\
\Phi(M_V) d M_V & = & A \, \Phi^*  \left( \frac{L}{L^*}\right)^{\alpha+1} \exp \left( - \frac{L}{L^*} \right) \, \, d M_V  \hspace{10mm} A = 0.4 \ln(10) \label{eq-lum-func}\\  & &  \Phi^* = 0.0063 \, \mathrm{Mpc^{-3}}  \hspace{2mm} M^*=-20.5  \hspace{2mm} \alpha=-1.07 \label{eq:schechgallum}
\end{eqnarray}
Galaxies in the random catalog are initially uniformly distributed in the 3D space, with the galaxy number count proportional to the cosmological volume element, computed according to the standard $\Lambda CDM$ cosmology with Planck  \citep{2021A&A...652C...4P} parameters.  The apparent NCCS magnitude is then computed, applying the distance modulus, and the Galactic extinction $Av$, without any $k$-correction.  We then apply the NCCS survey selection efficiency (eq. \ref{eq:effnccs}), dropping a fraction of galaxies. 
We continue by applying the tNCCSz target selection cuts  ($13 < V_\mathrm{mag} < 19$), followed by the declination dependent fiber assignment efficiency (equation \ref{eq:efffibass}). Finally, the galaxies for which we have a redshift estimates are selected applying the $\epsilon_z(V_\mathrm{mag}) $ defined in equation \ref{eq:effzwiyn}. 

As shown in Figure~\ref{fig:magdistallsim}, the magnitude distribution of the galaxies in the NCCS catalog,
selected requiring ${\rm PESS} \geq 2$ and $ \delta > 81^\circ$, and the one corresponding to the Tianlai-WIYN sample with an estimated redshift agrees quite well with the magnitude distribution of the random catalog. The total number count for real data and random catalogs agree within a few percent, after a slight adjustment  of the $\Phi^*$ parameter (less than 5\%).

\subsubsection{Landy-Szalay Estimators}

We compute the correlation function  \citep{2013A&A...554A.131V} of the tNCCSz catalog using the Landy-Szalay estimator $\hat{\xi}_\mathrm{LS}(d_\mathrm{sep})$ \citep{1993ApJ...412...64L}, using the random catalogs generated as described above in section \ref{sec:mocks} to correct for the different selection effects, including the Galactic absorption. We have also computed the correlation function $\hat{\xi}_\mathrm{LS}(d_\mathrm{sep})$ for the SDSS mocks. Angular position and redshifts are converted into 3D positions using the standard $\Lambda$CDM cosmology. The auto-correlation function is computed as a normalized galaxy pairs count histogram, as a function of the pair separation distance in Mpc, for the real data $h_{DD}(d_\mathrm{sep})$, 
and for the corresponding random catalog $h_{RR}(d_\mathrm{sep})$. A cross correlation normalised pair count histogram is also computed for all (data $\times$ random) pairs. The  Landy-Szalay auto-correlation function estimator $\hat{\xi}_{LS}(d_\mathrm{sep})$ is then computed as:
\begin{eqnarray}
    \hat{\xi}_\mathrm{LS}(d_\mathrm{sep})& = &
    \frac{h_{DD} - 2 h_{DR} + h_{RR} }{h_{RR}} \left( d_\mathrm{sep} \right)
\end{eqnarray}
which is expected to be zero if the observed structuring is only due to observational biases or selection effects.

\begin{figure}
    \centering
   \includegraphics[width=0.95\textwidth]{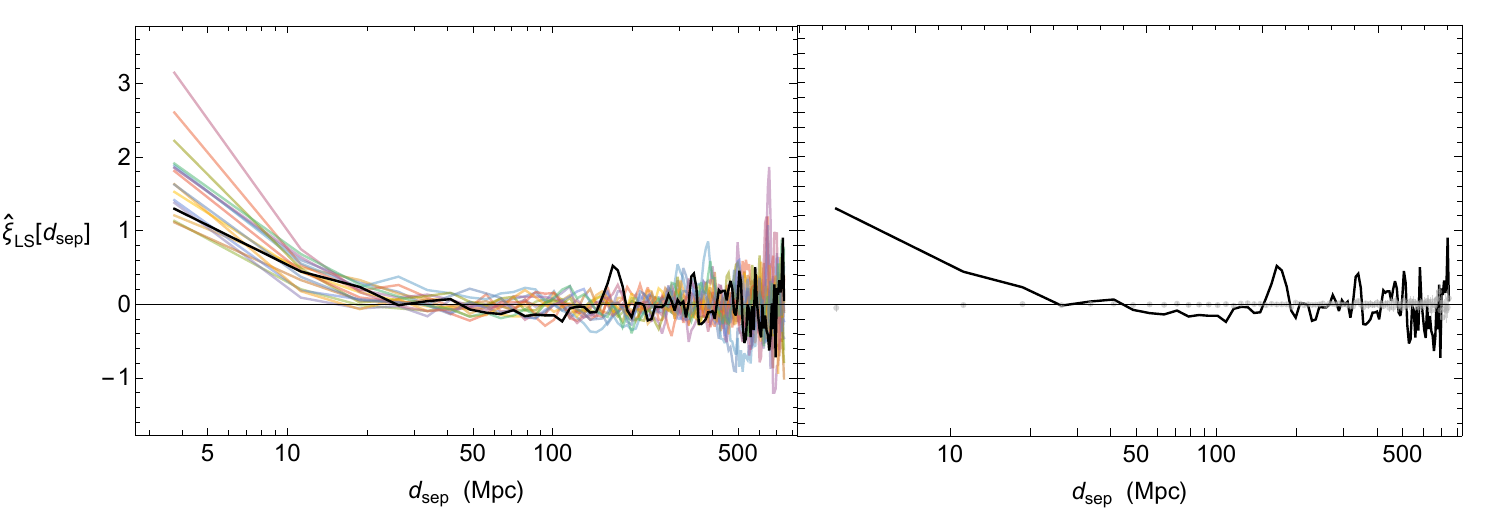}
   \caption{Left panel: The curves give Landy-Szalay estimators of the auto-correlation functions, $\hat{\xi}_\mathrm{LS}(d_\mathrm{sep})$, as a function of galaxy pair separation $d_\mathrm{sep}$ in $7.5\,\mathrm{Mpc}$ bins, computed from redshift catalogs as described in \S~\ref{sec:2point}.  The black curve is $\hat{\xi}(r)$ for the tNCCSz redshift survey, the colored curves are for 15 SDSS extracted mocks described in sec.~\ref{sec:mocks}. Right panel: The tNCCSz $\hat{\xi}(r)$ is plotted in black curve overlayed with the mean and standard deviation of the $\hat{\xi}(d_\mathrm{sep})$'s from the random catalog as gray points with error bars.  The latter, which should have no correlation, gives an estimate of the statistical uncertainty in the $\xi(r)$ estimators from the finite number of galaxies.  A much larger uncertainty comes from sample variance in the large scales structures in these relatively small cosmological volumes.  Sample variance leads to the large spread in $\hat{\xi}(r)$ at $r\lesssim10\,\mathrm{Mpc}$ as well as the large excursions from zero at $r\gtrsim20\,\mathrm{Mpc}$.  These excursions should and do differ between pencil beam surveys.
   \label{fig:xiofd}}
\end{figure}

The $\hat{\xi}_\mathrm{LS}$ estimators from the tNCCSz catalog are plotted amidst the  $\hat{\xi}_\mathrm{LS}$ estimators from the mock catalogs in Figure~\ref{fig:xiofd}.  The tNCCSz estimators lie within the distribution of estimators from the mock catalogs.  This visual comparison suggests no anomalous behavior of the tNCCSz catalog when compared with other catalogs.

Also plotted in Figure~\ref{fig:xiofd} is the statistical estimator mean and variance from the random catalog should be and is consistent with zero but only includes the statistical uncertainty from the finite number of galaxies.  A much larger contribution to the variation in these estimators comes from sample variance of large scale structures in this relatively small and elongated cosmological volume.  Large excursions from zero in the tNCCSz $\hat{\xi}_\mathrm{LS}$ are present at
$d_\mathrm{sep}\sim160,\, 330\,\mathrm{Mpc}$.   Such excursions are expected for pencil beam survey (see \cite{1991ApJ...379..482K}) because there are a small number of structures (voids, filaments etc.) of size comparable to the width of the survey along the line-of-sight.  Different pencil beams have different structures leading to large excursions at different separations, as illustrated in this figure.

\subsection{3D distribution}

\begin{figure}
    \centering
   \includegraphics[width=0.80\textwidth]{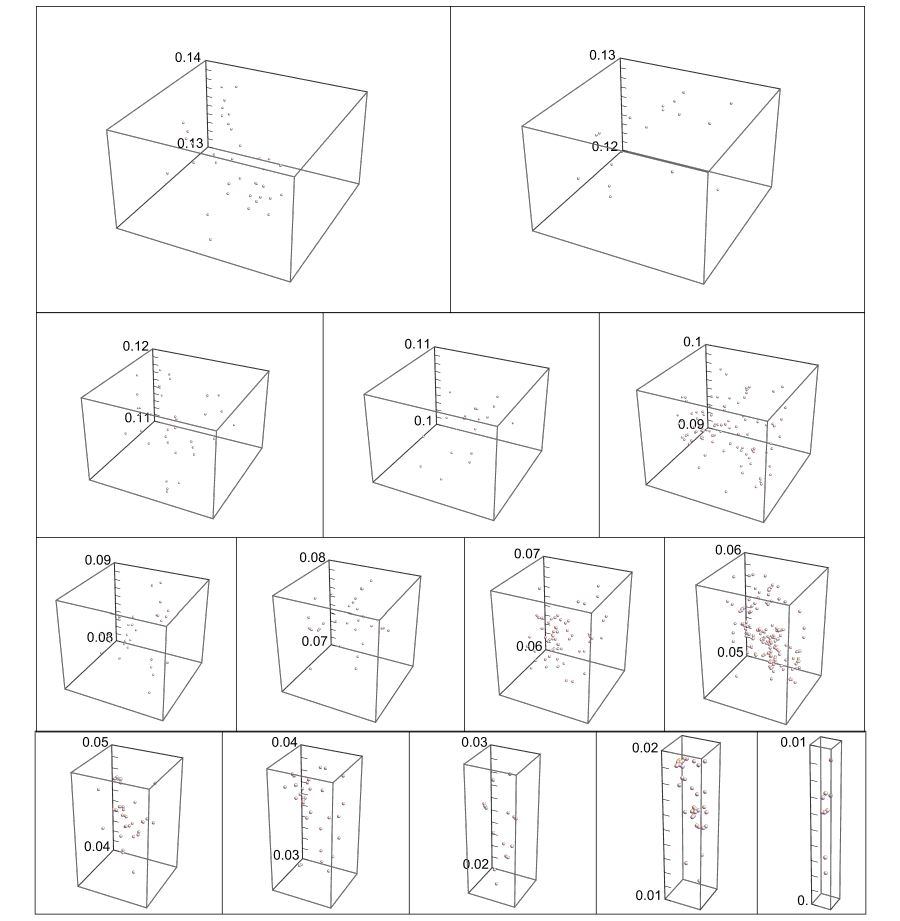}
   \caption{Shown are 14 redshift space boxes with height $\Delta z=0.01$ and width large enough to include all target galaxies in the survey.  Within each box all of the tNCCSz galaxies with measured redshifts ($Q\ge1$) are represented as by a sphere of diameter $1\,\mathrm{Mpc}$. The redshifts are labeled on the vertical axis.  The distance scale, say the distance between two galaxies in Mpc, can be gauged visually by the number of spheres that fit between the two galaxy spheres.  The box width and depth increases proportional to the redshift while the box height is the same for all boxes.  Even though the box volumes increases rapidly with redshift the number of galaxies in the boxes decreases because at large distances fewer galaxies are bright enough to make it into our survey.
   }
   \label{fig:tNCCSzRedshiftSpace}
\end{figure}

Our redshift catalog is small enough to visualize the redshift distribution of all the galaxies in a few figures.  In Figure~\ref{fig:tNCCSzRedshiftSpace} we show the redshift space distribution of galaxies in 14 slices of redshift thickness $\Delta z=0.1$ spanning $z\in[0,0.14]$.  The galaxies are represented by spheres of radius $1\,\mathrm{Mpc}$.  The comoving number density of galaxies with redshift decreases rapidly due to the magnitude limit of the survey, but not monotonically.  There are localized peaks in the redshift distribution which can also be seen in Figure~\ref{fig:mockCarousel}. Pencil beam survey labeled ``g'' is the tNCCSz survey.  The separation of these overdense regions give the peaks in $\xi_\mathrm{LS}(d_\mathrm{sep})$ of Figure~\ref{fig:xiofd}.  Evident structure can be seen down to the few Mpc scale.  Since the clustering of tNCCSz galaxies lies visually and quantitatively within the range of the clustering of SDSS we believe these structures are not artifacts of incompleteness. 

\subsection{Clusters}

Despite the sparse sampling of our redshift catalog we have attempted to detect galaxy clusters using a simple Friends of Friends (FoF) cluster finder algorithm (\cite{1982ApJ...257..423H}; \cite{2016A&A...588A..14T}). We allow a maximum redshift difference $\delta z=0.005$ ($=1500\,\mathrm{km/s}$) and a transverse distance of $d_\perp=0.5\,\mathrm{Mpc}$ for cluster members. We require a minimum number of cluster members of $N^\mathrm{gal}_\mathrm{min}=3$ for cluster detection; this threshold increases to $N^\mathrm{gal}_\mathrm{min}=4$ for $0.03<z<0.06$ and $N^\mathrm{gal}_\mathrm{min}=5$ for $z<0.03$.

\begin{figure}
    \centering
   \includegraphics[width=0.80\textwidth]{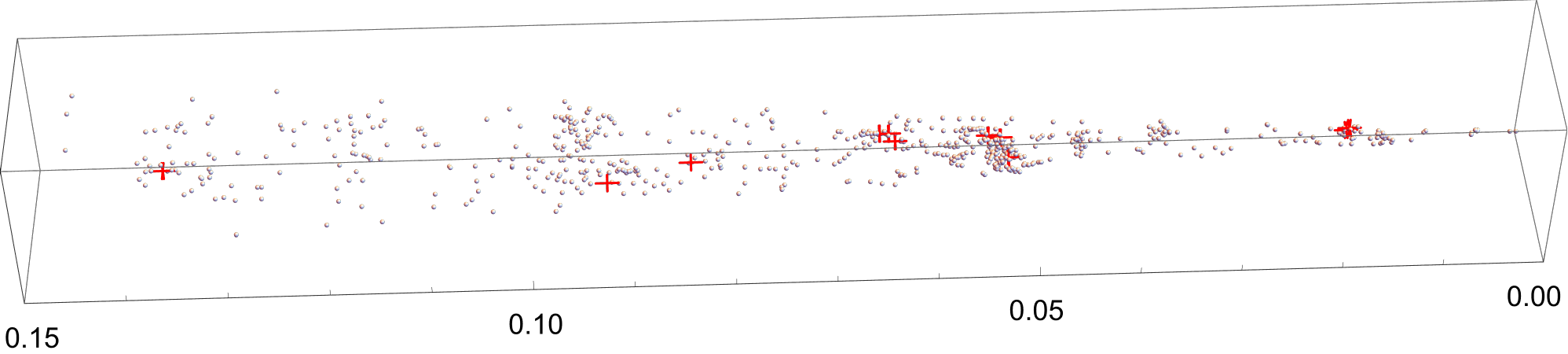}
   \caption{Shown is the full galaxy redshift survey with the positions of 11 galaxy clusters identified in Table~\ref{tab:clusters}. The galaxy positions are represented as $2\,\mathrm{Mpc}$ diameter spheres and the cluster positions as $10\,\mathrm{Mpc}$ 3D crosses in red.  These markers are of course much larger than the objects they represent.  Only the redshift direction is labeled. 
   }
   \label{fig:tNCCSzRedshiftSpaceClusters}
\end{figure}

We identified 11 clusters in the tNCCSz sample with the corresponding parameters listed in the Table~\ref{tab:clusters}. Two of these 11 likely galaxy clusters have a more compact core, and are also detected with $d_\perp = 0.35\,\mathrm{Mpc}$. The cluster position and redshift are computed as the average of the corresponding quantities for member galaxies.  The positions of these galaxy clusters amidst the galaxies are shown in Figure~\ref{fig:tNCCSzRedshiftSpaceClusters}. The only parameter characterizing the cluster is $N_{\rm gal}$, which corresponds to the number of galaxies grouped together. We have estimated sizes for the clusters using the tNCCSz redshift catalog, which corresponds to the average 3D comoving distance of member galaxies with respect to the estimated cluster position. A too large value of the cluster size included in the Table~\ref{tab:clusters} is a hint of an unreliable cluster. Note, however, that the galaxy 3D positions have an error of a few Mpc, due to the redshift uncertainties ($\delta z \sim 5\times10^{-4} \rightarrow 2\,\mathrm{Mpc}$). 

To check the validity of our FoF cluster finder and quantify roughly our cluster sample purity, we detected clusters on 14 SDSS extracted mocks, half near $\delta = 50^\circ$,
and the other half centered at $\delta = 25^\circ$. 
We have found an average of $\langle N_\mathrm{clus} \rangle \simeq 14$ per extracted mock, with a rather large dispersion as expected for pencil beam surveys. This number decreases to about $\langle N_\mathrm{clus} \rangle \simeq 8$ for $d_\perp = 0.35 \mathrm{Mpc}$. About 75\% of all detected clusters have a redshift $z \lesssim 0.073$, corresponding to the volume limited published cluster catalog \citep{2014yCat..35660001T}, used as a reference or truth catalog. We cross matched each of our cluster catalog detected on SDSS mocks with the above mentioned reference catalog, and we have a matching fraction $\sim 85 \%$ among the clusters in the redshift range covered by the reference catalog. The matching fraction drops to less than 10\% if the extracted reference cluster catalogs are shuffled. As expected, the \cite{2014yCat..35660001T} catalog is richer than the list of clusters detected on SDSS extracted mocks and includes groups with only two galaxy members. Requiring minimal number of associated galaxies $(N_{\rm gal})$ as for the tNCCSz clusters, the average number of clusters found in the \cite{2014yCat..35660001T} catalog for each mock $ \langle N_\mathrm{clus}^\mathrm{Tempel} \rangle \simeq 27 $ is about twice the number of clusters identified by us in SDSS mocks.

\begin{table}[htbp]
\begin{tabular}{|l|c|c|c|c|c|c|}
\hline
ClusId & RA (deg) & Dec (deg) & Redshift & $N_{\rm gal}$ & Size (Mpc) & Core-NGal \\
\hline
CL-NCCS-W-1 & 313.409 & 87.6539 & 0.01894  & 5  & 1.2 & - \\
CL-NCCS-W-2 & 274.171 & 87.7034 & 0.01922 & 5  & 1.15 & - \\
CL-NCCS-W-3 & 174.534 & 88.255 & 0.05278 & 5  & 3.9 & - \\
CL-NCCS-W-4 & 54.0181 & 88.9466  & 0.05354 & 5  & 2.8 & - \\
CL-NCCS-W-5 & 250.655 & 87.6105 & 0.054675 & 4  & 1.1 & - \\ 
CL-NCCS-W-6 & 251.414 & 88.6894 & 0.06395 &  4 & 4.2 & - \\ 
CL-NCCS-W-7 & 331.568 & 88.6983 & 0.06466 &  3 & 1.9 & - \\ 
CL-NCCS-W-8 & 263.056 & 87.9489 & 0.06548 & 5  & 0.85 & 4 \\
CL-NCCS-W-9 & 212.681 & 87.0845 & 0.0843 & 3  & 0.65 & 3 \\ 
CL-NCCS-W-10 & 83.1438 & 87.2961 & 0.09256 & 3  & 4.6 & - \\
CL-NCCS-W-11 & 222.246 & 86.7443 & 0.13726 & 3  & 4.8 & - \\
\hline
\end{tabular}
\caption{List of clusters detected in the tNCCSz redshift catalog. The cluster angular position $(\alpha, \delta)$ in degrees is given in the two columns labeled RA and Dec, followed by the redshift and the number of galaxies associated to the cluster candidate in the next two columns. The column labeled {\it Size} correspond to the average of the member galaxy distances to their mean, taken as the cluster center. The last column gives the number of galaxies contained in the cluster core, for compact clusters, detected with a transverse separation distance $d_\perp = 0.35 \mathrm{Mpc}$. Large size values hints toward unreliable cluster candidates (CL-NCCS-W-3 , 6, 10, 11).  \label{tab:clusters}}
\end{table}

\subsection{HI mass fraction}

The motivation for this survey was to localize the HI 
in a survey volume of $\sim3^\circ$ around the NCP out to $z\approx0.07$.  This volume will be surveyed for 21,cm emission by the Tianlai Dish Array described in \cite{2021MNRAS.506.3455W}.  We have accomplished this localization in 3D with this optical redshift survey of galaxies, all of which will contain some HI. We have only determined the redshifts of a fraction of bright galaxies, those with $V<19$. Due to the high extinction near the NCP this corresponds to an extincted magnitude limit of $V\lesssim18$. Such bright galaxies contain only a fraction of the HI.  Thus, this survey only directly localizes a tracer of the total HI content of the survey volume. However, since the spatial distribution of this tracer will be strongly correlated that of the remainder of the HI containing galaxies we have, in a statistical sense, also indirectly localized all of the HI in the survey volume.

\begin{figure}
    \centering
   \includegraphics[width=0.80\textwidth]{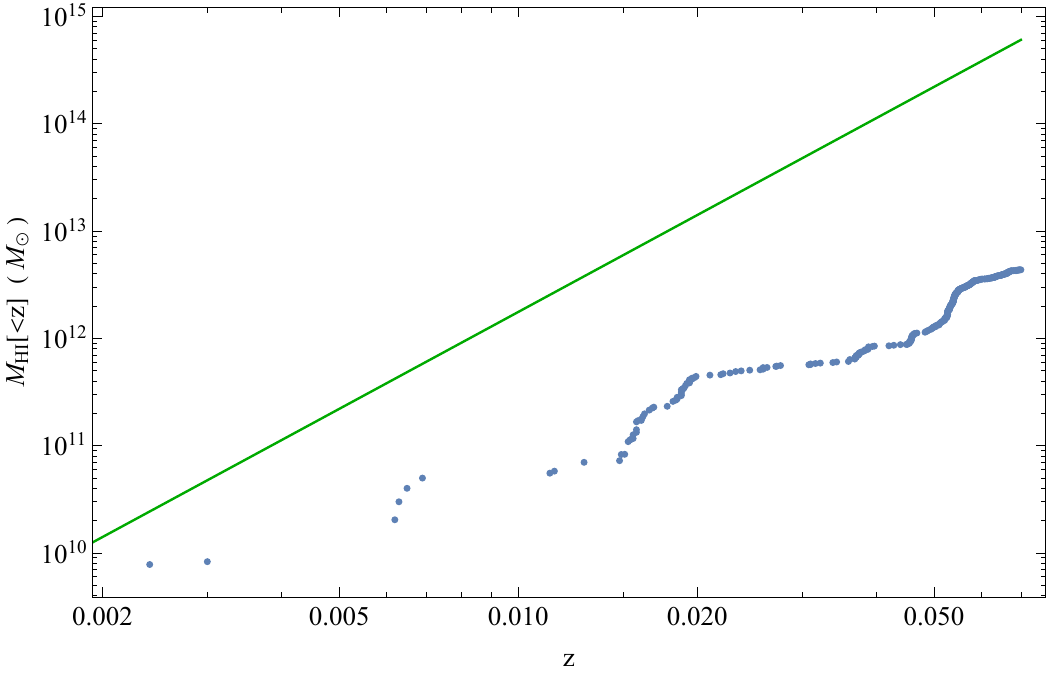}
   \caption{In this figure we plot the cumulative neutral hydrogen (HI) mass with redshift, i.e. the mass of HI with redshift less than $z$ in the survey area on the sky, $\delta\le87^\circ$.  The green curve gives this cumulative distribution if all of the HI were spread out uniformly (see \cite{10.1093/mnras/stz2038}).  The blue dots are \emph{rough} estimates for the HI in the galaxies of the tNCCSz survey, with each dot representing the addition of a tNCCSz galaxy.  The raggedness of this curve is real, representing the large inhomogeneities as a function of $z$ expected for pencil beam surveys (see \cite{1991ApJ...379..482K}).  The same inhomogeneities should be present in the full distribution of HI including HI in galaxies not included in tNCCSz.
   }
   \label{fig:HIdensity}
\end{figure}

We can estimate the fraction of HI contained in the sample of galaxies for which we have obtained redshifts (tNCCSz) by estimating the HI mass in each of these galaxies.  Although we have no direct measurement of the HI content in any of our galaxies, we can use the ALFALFA survey \cite{Haynes:2011hi} to find similar galaxies for which the HI mass has been measured.  Specifically, for each of the tNCCSz galaxies ($Q_{\rm auto}\ge2$) we correct the apparent VRI magnitudes for extinction and use the measured redshift ($z_{\rm auto}$) to determine their absolute VRI magnitudes.  We then find the ALFALFA survey galaxy which is closest in absolute VRI space and assign the mass of that ALFALFA galaxy to the tNCCSz galaxy.  This procedure gives a rough estimate of the HI mass that might be expected in the tNCCSz galaxy.  We compare the estimated HI masses of the galaxies for which we have determined redshifts with the mean HI density of the universe at low redshifts.  This is estimated by \cite{10.1093/mnras/stz2038} to be $\Omega_\mathrm{HI}=0.0004$ or $\bar{\rho}_\mathrm{HI}=4.5\times10^7\,M_\odot/\mathrm{Mpc}^3$.  In Figure~\ref{fig:HIdensity} we plot the cumulative HI mass in our survey volume as a function of redshift using 1) the estimated masses of the tNCCSz galaxies with measured redshifts, and 2) a uniform $\bar{\rho}_\mathrm{HI}$ density.  The ratio of the two is a rough estimate of the fraction of HI contained in these galaxies. This ratio varies from $\sim0.5$ at the lowest $z$ to $\sim0.01$ at $z\sim0.07$. This HI incompleteness is a function of both the galaxy redshift incompleteness at the magnitude limit and the magnitude limit itself - a deeper, more complete survey is better. Nevertheless, the HI which we have localized provides a good template for all the HI in the survey volume.  This will be used to compare with future measurements by the Tianlai Dish Array survey described in \cite{2022MNRAS.517.4637P}.

%% file: sec_6_conclusions.tex
\section{Summary}
\label{sec:conc}

In this paper we have presented the results of a redshift survey of galaxies with $m_V<19$ (including extinction) within $\sim3^\circ$ of the North Celestial Pole (NCP) based on a published photometric catalog \citep{Gorbikov_2014}.  Because of the proximity to the Galactic Plane, high extinction and difficulties associated with observations very near to the celestial poles no redshift survey of this region had been attempted until now.  The motivation for this survey is to compare a future 21\,cm (radio) redshift survey by the Tianlai project 
with an optical redshift survey.  The non-tracking transit radio telescope can obtain the deepest exposure in a given amount of observation time at the NCP.  Hence the NCP has advantages for radio observations even though it is not ideal for extra-Galactic optical astronomy.  

Out of 2102 potential targets we have only obtained 787 redshifts for a variety of reasons.   We have shown that the clustering properties of our survey is compatible with that found in SDSS after adjusting for different incompleteness.

Our overall goal is to develop the technique of hydrogen intensity mapping (HIM).  It is expected that HIM will be able map cosmological large scale structure traced by galaxies back to the epoch of reionization and the large scale structure traced by gas even beyond the epoch of reionization, into the dark ages.  

Several surveys observe the NCP region for the same reason the Tianlai project does -- to obtain long, continuous observations of a limited region of the sky.  Other HIM programs observe the NCP region, but to our knowledge only the Tianlai arrays can currently  be tuned to overlap the redshift range of the galaxy redshift survey described here. However, this survey may also be useful for finding optical counterparts for transient radio phenomena, such as Fast Radio Bursts (FRBs).   Several such radio transient surveys observe the NCP: the LOFAR Multifrequency Snapshot Sky Survey (MSSS) \citep{2016MNRAS.456.2321S}, the CHIME/FRB project \citep{2021ApJS..257...59C}.

%% file: appendix_additional_spectra.tex
\section{Additional spectra measurements}
\label{sec:appendix}
During the  first WIYN HYDRA observing night, on Feb. 1 2020, 
we did not apply the  V magnitude selection $V<18.995$ that was subsequently used to build the tNCCSz sample. As a result, 208 spectra were recorded for NCCS objects with $V \geq 18.995$. We subjected these spectra to the semi-automatic redshift reconstruction procedure described in Section \ref{subsec:semiauto}. 

\begin{figure}
    \centering
\begin{tabular}{cc}
    \includegraphics[width=0.5\linewidth]{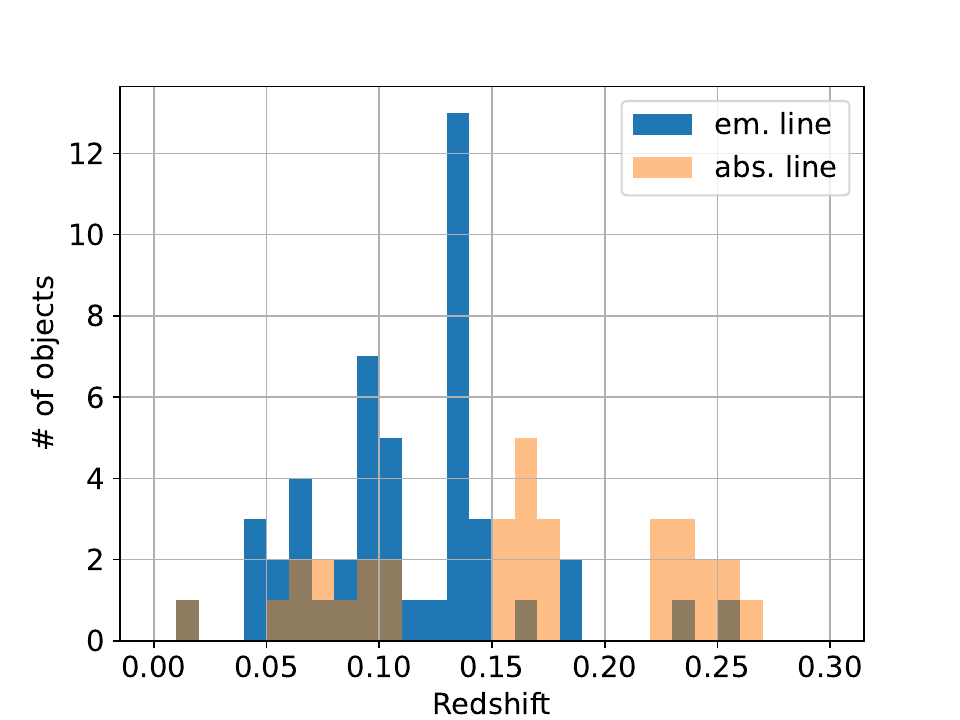}
&  
    \includegraphics[width=0.5\linewidth]{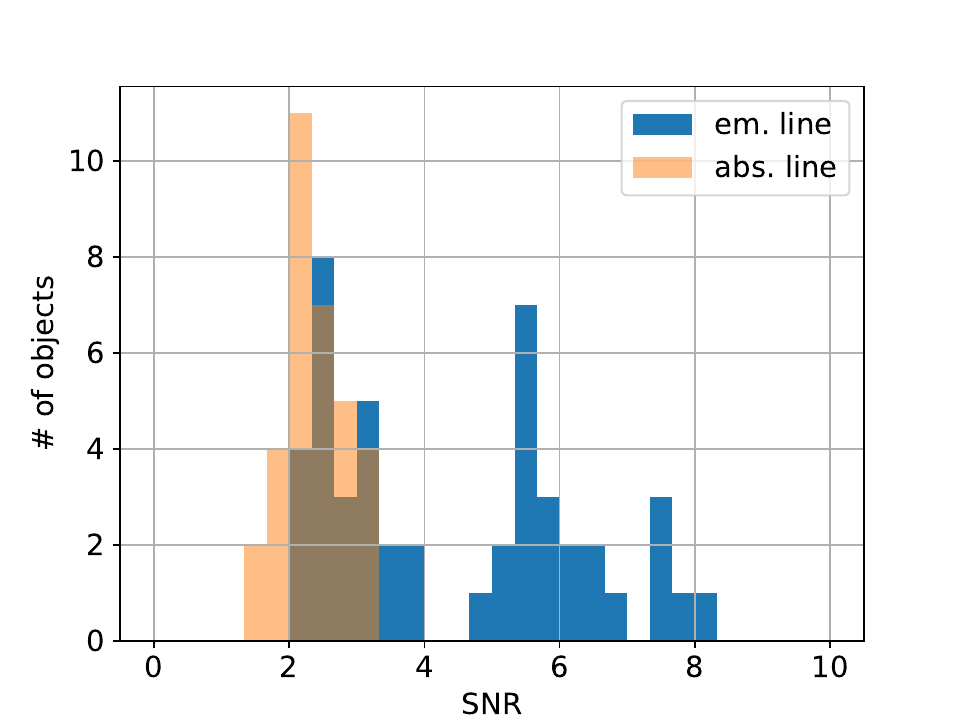}
\end{tabular}

    \caption{Redshift (left) and SNR (right) distribution from the Feb. 1 2020 additional dataset. Distributions for objects with a redshift based on an emission line template are represented in blue, and for objects with a redshift estimation based on an absorption line template are in orange.}
    \label{fig:annex}
\end{figure}

For completeness we include these spectra and results in this data release. From these 206 spectra, we reconstructed 48 emission line based redshifts and 33 absorption line based ones.   We show in Figure \ref{fig:annex} the redhift distributions and SNR for these two samples. As expected from the lower photometric magnitudes of these objects, the redshifts we reconstruct are higher than those in  the tNCCSz catalog. 
While some high  SNR spectra have been found, in general the SNR we reconstruct are lower than those observed in the  tNCCSz catalog for the same reason.

%% file: tnccsz_wiyn_ss.bib
@article{Baldry_2014,
   title={Galaxy And Mass Assembly (GAMA): AUTOZ spectral redshift measurements, confidence and errors},
   volume={441},
   ISSN={1365-2966},
   url={http://dx.doi.org/10.1093/mnras/stu727},
   DOI={10.1093/mnras/stu727},
   number={3},
   journal={Monthly Notices of the Royal Astronomical Society},
   publisher={Oxford University Press (OUP)},
   author={Baldry, I. K. and Alpaslan, M. and Bauer, A. E. and Bland-Hawthorn, J. and Brough, S. and Cluver, M. E. and Croom, S. M. and Davies, L. J. M. and Driver, S. P. and Gunawardhana, M. L. P. and Holwerda, B. W. and Hopkins, A. M. and Kelvin, L. S. and Liske, J. and Lopez-Sanchez, A. R. and Loveday, J. and Norberg, P. and Peacock, J. and Robotham, A. S. G. and Taylor, E. N.},
   year={2014},
   month=may, pages={2440–2451} }

@article{Brown_2014,
   title="{AN ATLAS OF GALAXY SPECTRAL ENERGY DISTRIBUTIONS FROM THE ULTRAVIOLET TO THE MID-INFRARED}",
   volume={212},
   ISSN={1538-4365},
   url={http://dx.doi.org/10.1088/0067-0049/212/2/18},
   DOI={10.1088/0067-0049/212/2/18},
   number={2},
   journal={The Astrophysical Journal Supplement Series},
   publisher={American Astronomical Society},
   author={Brown, Michael J. I. and Moustakas, John and Smith, J.-D. T. and da Cunha, Elisabete and Jarrett, T. H. and Imanishi, Masatoshi and Armus, Lee and Brandl, Bernhard R. and Peek, J. E. G.},
   year={2014},
   month=may, pages={18} }

@ARTICLE{Paul2023,
       author = {{Paul}, Sourabh and {Santos}, Mario G. and {Chen}, Zhaoting and {Wolz}, Laura},
        title = "{A first detection of neutral hydrogen intensity mapping on Mpc scales at $z\approx 0.32$ and $z\approx 0.44$}",
      journal = {arXiv e-prints},
         year = 2023,
        month = jan,
          eid = {arXiv:2301.11943},
        pages = {arXiv:2301.11943},
          doi = {10.48550/arXiv.2301.11943},
archivePrefix = {arXiv},
       eprint = {2301.11943},
 primaryClass = {astro-ph.CO},
       adsurl = {https://ui.adsabs.harvard.edu/abs/2023arXiv230111943P}
}

@misc{CHIME2025,
	title = {Detection of the {Cosmological} 21 cm {Signal} in {Auto}-correlation at z {\textasciitilde} 1 with the {Canadian} {Hydrogen} {Intensity} {Mapping} {Experiment}},
	doi = {10.48550/arXiv.2511.19620},
	publisher = {arXiv},
	author = {Amiri, Mandana and Bandura, Kevin and Chakraborty, Arnab and Cliche, Jean-François and Dobbs, Matt and Foreman, Simon and Gray, Liam and Halpern, Mark and Hill, Alex S. and Hinshaw, Gary and Höfer, Carolin and Joseph, Albin and Kruger, Nolan and Landecker, T. L. and Lieshout, Rik van and MacEachern, Joshua and Masui, Kiyoshi W. and Mena-Parra, Juan and Miller, Kyle and Milutinovic, Nikola and Mirhosseini, Arash and Newburgh, Laura and Ordog, Anna and Pen, Ue-Li and Pinsonneault-Marotte, Tristan and Reda, Alex and Renard, Andre and Sakaguri, Kana and Shaw, J. Richard and Shaikh, Shabbir and Siegel, Seth R. and Singh, Saurabh and Spear, David and Uchibori, Yukari and Vanderlinde, Keith and Wang, Haochen and Wiebe, Donald V. and Wulf, Dallas},
	month = nov,
	year = {2025},
	note = {arXiv:2511.19620 [astro-ph]},
}

@ARTICLE{Kern2019,
       author = {{Kern}, Nicholas S. and {Parsons}, Aaron R. and {Dillon}, Joshua S. and
         {Lanman}, Adam E. and {Fagnoni}, Nicolas and {de Lera Acedo}, Eloy},
        title = "{Mitigating Internal Instrument Coupling for 21 cm Cosmology. I. Temporal and Spectral Modeling in Simulations}",
      journal = {\apj},
         year = 2019,
        month = oct,
       volume = {884},
       number = {2},
          eid = {105},
        pages = {105},
          doi = {10.3847/1538-4357/ab3e73},
       adsurl = {https://ui.adsabs.harvard.edu/abs/2019ApJ...884..105K}
}

@ARTICLE{Kern2020,
       author = {{Kern}, Nicholas S. and {Parsons}, Aaron R. and {Dillon}, Joshua S. and
         {Lanman}, Adam E. and {Liu}, Adrian and {Bull}, Philip and
         {Ewall-Wice}, Aaron and {Abdurashidova}, Zara and {Aguirre}, James E. and
         {Alexander}, Paul and {Ali}, Zaki S. and {Balfour}, Yanga and
         {Beardsley}, Adam P. and {Bernardi}, Gianni and {Bowman}, Judd D. and
         {Bradley}, Richard F. and {Burba}, Jacob and {Carilli}, Chris L. and
         {Cheng}, Carina and {DeBoer}, David R. and {Dexter}, Matt and
         {de Lera Acedo}, Eloy and {Fagnoni}, Nicolas and {Fritz}, Randall and
         {Furlanetto}, Steve R. and {Glendenning}, Brian and {Gorthi}, Deepthi and
         {Greig}, Bradley and {Grobbelaar}, Jasper and {Halday}, Ziyaad and
         {Hazelton}, Bryna J. and {Hewitt}, Jacqueline N. and {Hickish}, Jack and
         {Jacobs}, Daniel C. and {Julius}, Austin and {Kerrigan}, Joshua and
         {Kittiwisit}, Piyanat and {Kohn}, Saul A. and {Kolopanis}, Matthew and
         {La Plante}, Paul and {Lekalake}, Telalo and {MacMahon}, David and
         {Malan}, Lourence and {Malgas}, Cresshim and {Maree}, Matthys and
         {Martinot}, Zachary E. and {Matsetela}, Eunice and {Mesinger}, Andrei and
         {Molewa}, Mathakane and {Morales}, Miguel F. and
         {Mosiane}, Tshegofalang and {Murray}, Steven G. and
         {Neben}, Abraham R. and {Parsons}, Aaron R. and {Patra}, Nipanjana and
         {Pieterse}, Samantha and {Pober}, Jonathan C. and {Razavi-Ghods}, Nima and
         {Ringuette}, Jon and {Robnett}, James and {Rosie}, Kathryn and
         {Sims}, Peter and {Smith}, Craig and {Syce}, Angelo and
         {Thyagarajan}, Nithyanandan and {Williams}, Peter K.~G. and
         {Zheng}, Haoxuan},
        title = "{Mitigating Internal Instrument Coupling for 21 cm Cosmology. II. A Method Demonstration with the Hydrogen Epoch of Reionization Array}",
      journal = {\apj},
         year = 2020,
        month = jan,
       volume = {888},
       number = {2},
          eid = {70},
        pages = {70},
          doi = {10.3847/1538-4357/ab5e8a},
archivePrefix = {arXiv},
       eprint = {1909.11733},
 primaryClass = {astro-ph.IM},
       adsurl = {https://ui.adsabs.harvard.edu/abs/2020ApJ...888...70K}
}

@ARTICLE{2022MNRAS.517.4637P,
       author = {{Perdereau}, Olivier and {Ansari}, R{\'e}za and {Stebbins}, Albert and {Timbie}, Peter T. and {Chen}, Xuelei and {Wu}, Fengquan and {Li}, Jixia and {Marriner}, John P. and {Tucker}, Gregory S. and {Cong}, Yanping and {Das}, Santanu and {Li}, Yichao and {Liu}, Yingfeng and {Magneville}, Christophe and {Peterson}, Jeffrey B. and {Phan}, Anh and {Robinthal}, Lily and {Sun}, Shijie and {Wang}, Yougang and {Wu}, Yanlin and {Xu}, Yidong and {Yu}, Kaifeng and {Yu}, Zijie and {Zhang}, Jiao and {Zhang}, Juyong and {Zuo}, Shifan},
        title = "{The Tianlai dish array low-z surveys forecasts}",
      journal = {\mnras},
         year = 2022,
        month = dec,
       volume = {517},
       number = {3},
        pages = {4637-4655},
          doi = {10.1093/mnras/stac2832},
archivePrefix = {arXiv},
       eprint = {2205.06086},
 primaryClass = {astro-ph.CO},
       adsurl = {https://ui.adsabs.harvard.edu/abs/2022MNRAS.517.4637P}
}

@ARTICLE{2021MNRAS.506.3455W,
       author = {{Wu}, Fengquan and {Li}, Jixia and {Zuo}, Shifan and {Chen}, Xuelei and {Das}, Santanu and {Marriner}, John P. and {Oxholm}, Trevor M. and {Phan}, Anh and {Stebbins}, Albert and {Timbie}, Peter T. and {Ansari}, Reza and {Campagne}, Jean-Eric and {Chen}, Zhiping and {Cong}, Yanping and {Huang}, Qizhi and {Kwak}, Juhun and {Li}, Yichao and {Liu}, Tao and {Liu}, Yingfeng and {Niu}, Chenhui and {Osinga}, Calvin and {Perdereau}, Olivier and {Peterson}, Jeffrey B. and {Podczerwinski}, John and {Shi}, Huli and {Siebert}, Gage and {Sun}, Shijie and {Tian}, Haijun and {Tucker}, Gregory S. and {Wang}, Qunxiong and {Wang}, Rongli and {Wang}, Yougang and {Wu}, Yanlin and {Xu}, Yidong and {Yu}, Kaifeng and {Yu}, Zijie and {Zhang}, Jiao and {Zhang}, Juyong and {Zhu}, Jialu},
        title = "{The Tianlai dish pathfinder array: design, operation, and performance of a prototype transit radio interferometer}",
      journal = {\mnras},
         year = 2021,
        month = sep,
       volume = {506},
       number = {3},
        pages = {3455-3482},
          doi = {10.1093/mnras/stab1802},
archivePrefix = {arXiv},
       eprint = {2011.05946},
 primaryClass = {astro-ph.IM},
       adsurl = {https://ui.adsabs.harvard.edu/abs/2021MNRAS.506.3455W}
}

@ARTICLE{2020SCPMA..6329862L,
       author = {{Li}, JiXia and {Zuo}, ShiFan and {Wu}, FengQuan and {Wang}, YouGang and {Zhang}, JuYong and {Sun}, ShiJie and {Xu}, YiDong and {Yu}, ZiJie and {Ansari}, Reza and {Li}, YiChao and {Stebbins}, Albert and {Timbie}, Peter and {Cong}, YanPing and {Geng}, JingChao and {Hao}, Jie and {Huang}, QiZhi and {Li}, JianBin and {Li}, Rui and {Liu}, DongHao and {Liu}, YingFeng and {Liu}, Tao and {Marriner}, John P. and {Niu}, ChenHui and {Pen}, Ue-Li and {Peterson}, Jeffery B. and {Shi}, HuLi and {Shu}, Lin and {Song}, YaFang and {Tian}, HaiJun and {Wang}, GuiSong and {Wang}, QunXiong and {Wang}, RongLi and {Wang}, WeiXia and {Wang}, Xin and {Yu}, KaiFeng and {Zhang}, Jiao and {Zhu}, BoQin and {Zhu}, JiaLu and {Chen}, XueLei},
        title = "{The Tianlai Cylinder Pathfinder array: System functions and basic performance analysis}",
      journal = {Science China Physics, Mechanics, and Astronomy},
         year = 2020,
        month = sep,
       volume = {63},
       number = {12},
          eid = {129862},
        pages = {129862},
          doi = {10.1007/s11433-020-1594-8},
archivePrefix = {arXiv},
       eprint = {2006.05605},
 primaryClass = {astro-ph.IM},
       adsurl = {https://ui-adsabs-harvard-edu.ezproxy.library.wisc.edu/abs/2020SCPMA..6329862L}
}

@ARTICLE{2008PhRvL.100i1303C,
       author = {{Chang}, Tzu-Ching and {Pen}, Ue-Li and {Peterson}, Jeffrey B. and {McDonald}, Patrick},
        title = "{Baryon Acoustic Oscillation Intensity Mapping of Dark Energy}",
      journal = {\prl},
         year = 2008,
        month = mar,
       volume = {100},
       number = {9},
          eid = {091303},
        pages = {091303},
          doi = {10.1103/PhysRevLett.100.091303},
archivePrefix = {arXiv},
       eprint = {0709.3672},
 primaryClass = {astro-ph},
       adsurl = {https://ui-adsabs-harvard-edu.ezproxy.library.wisc.edu/abs/2008PhRvL.100i1303C}
}

@ARTICLE{2008PhRvL.100p1301L,
       author = {{Loeb}, Abraham and {Wyithe}, J. Stuart B.},
        title = "{Possibility of Precise Measurement of the Cosmological Power Spectrum with a Dedicated Survey of 21cm Emission after Reionization}",
      journal = {\prl},
         year = 2008,
        month = apr,
       volume = {100},
       number = {16},
          eid = {161301},
        pages = {161301},
          doi = {10.1103/PhysRevLett.100.161301},
archivePrefix = {arXiv},
       eprint = {0801.1677},
 primaryClass = {astro-ph},
       adsurl = {https://ui-adsabs-harvard-edu.ezproxy.library.wisc.edu/abs/2008PhRvL.100p1301L}
}

@ARTICLE{2020PASP..132f2001L,
       author = {{Liu}, Adrian and {Shaw}, J. Richard},
        title = "{Data Analysis for Precision 21 cm Cosmology}",
      journal = {\pasp},
         year = 2020,
        month = jun,
       volume = {132},
       number = {1012},
          eid = {062001},
        pages = {062001},
          doi = {10.1088/1538-3873/ab5bfd},
archivePrefix = {arXiv},
       eprint = {1907.08211},
 primaryClass = {astro-ph.IM},
       adsurl = {https://ui-adsabs-harvard-edu.ezproxy.library.wisc.edu/abs/2020PASP..132f2001L}
}

@ARTICLE{2019MNRAS.486.5124O,
       author = {{Obuljen}, Andrej and {Alonso}, David and {Villaescusa-Navarro}, Francisco and {Yoon}, Ilsang and {Jones}, Michael},
        title = "{The H I content of dark matter haloes at z {\ensuremath{\approx}} 0 from ALFALFA}",
      journal = {\mnras},
         year = 2019,
        month = jul,
       volume = {486},
       number = {4},
        pages = {5124-5138},
          doi = {10.1093/mnras/stz1118},
archivePrefix = {arXiv},
       eprint = {1805.00934},
 primaryClass = {astro-ph.CO},
       adsurl = {https://ui-adsabs-harvard-edu.ezproxy.library.wisc.edu/abs/2019MNRAS.486.5124O}
}

@ARTICLE{2024ApJ...963...23A,
       author = {{Amiri}, Mandana and {Bandura}, Kevin and {Chakraborty}, Arnab and {Dobbs}, Matt and {Fandino}, Mateus and {Foreman}, Simon and {Gan}, Hyoyin and {Halpern}, Mark and {Hill}, Alex S. and {Hinshaw}, Gary and {H{\"o}fer}, Carolin and {Landecker}, T.~L. and {Li}, Zack and {MacEachern}, Joshua and {Masui}, Kiyoshi and {Mena-Parra}, Juan and {Milutinovic}, Nikola and {Mirhosseini}, Arash and {Newburgh}, Laura and {Ordog}, Anna and {Paul}, Sourabh and {Pen}, Ue-Li and {Pinsonneault-Marotte}, Tristan and {Reda}, Alex and {Shaw}, J. Richard and {Siegel}, Seth R. and {Vanderlinde}, Keith and {Wang}, Haochen and {Wiebe}, D.~V. and {Wulf}, Dallas and {The Chime Collaboration}},
        title = "{A Detection of Cosmological 21 cm Emission from CHIME in Cross-correlation with eBOSS Measurements of the Ly{\ensuremath{\alpha}} Forest}",
      journal = {\apj},
         year = 2024,
        month = mar,
       volume = {963},
       number = {1},
          eid = {23},
        pages = {23},
          doi = {10.3847/1538-4357/ad0f1d},
archivePrefix = {arXiv},
       eprint = {2309.04404},
 primaryClass = {astro-ph.CO},
       adsurl = {https://ui-adsabs-harvard-edu.ezproxy.library.wisc.edu/abs/2024ApJ...963...23A}
}

@ARTICLE{2023ApJ...947...16A,
       author = {{Amiri}, Mandana and {Bandura}, Kevin and {Chen}, Tianyue and {Deng}, Meiling and {Dobbs}, Matt and {Fandino}, Mateus and {Foreman}, Simon and {Halpern}, Mark and {Hill}, Alex S. and {Hinshaw}, Gary and {H{\"o}fer}, Carolin and {Kania}, Joseph and {Landecker}, T.~L. and {MacEachern}, Joshua and {Masui}, Kiyoshi and {Mena-Parra}, Juan and {Milutinovic}, Nikola and {Mirhosseini}, Arash and {Newburgh}, Laura and {Ordog}, Anna and {Pen}, Ue-Li and {Pinsonneault-Marotte}, Tristan and {Polzin}, Ava and {Reda}, Alex and {Renard}, Andre and {Shaw}, J. Richard and {Siegel}, Seth R. and {Singh}, Saurabh and {Vanderlinde}, Keith and {Wang}, Haochen and {Wiebe}, Donald V. and {Wulf}, Dallas and {CHIME Collaboration}},
        title = "{Detection of Cosmological 21 cm Emission with the Canadian Hydrogen Intensity Mapping Experiment}",
      journal = {\apj},
         year = 2023,
        month = apr,
       volume = {947},
       number = {1},
          eid = {16},
        pages = {16},
          doi = {10.3847/1538-4357/acb13f},
archivePrefix = {arXiv},
       eprint = {2202.01242},
 primaryClass = {astro-ph.CO},
       adsurl = {https://ui.adsabs.harvard.edu/abs/2023ApJ...947...16A}
}

@ARTICLE{2023MNRAS.518.6262C,
       author = {{Cunnington}, Steven and {Li}, Yichao and {Santos}, Mario G. and {Wang}, Jingying and {Carucci}, Isabella P. and {Irfan}, Melis O. and {Pourtsidou}, Alkistis and {Spinelli}, Marta and {Wolz}, Laura and {Soares}, Paula S. and {Blake}, Chris and {Bull}, Philip and {Engelbrecht}, Brandon and {Fonseca}, Jos{\'e} and {Grainge}, Keith and {Ma}, Yin-Zhe},
        title = "{H I intensity mapping with MeerKAT: power spectrum detection in cross-correlation with WiggleZ galaxies}",
      journal = {\mnras},
         year = 2023,
        month = feb,
       volume = {518},
       number = {4},
        pages = {6262-6272},
          doi = {10.1093/mnras/stac3060},
archivePrefix = {arXiv},
       eprint = {2206.01579},
 primaryClass = {astro-ph.CO},
       adsurl = {https://ui-adsabs-harvard-edu.ezproxy.library.wisc.edu/abs/2023MNRAS.518.6262C}
}

@article{Gorbikov_2014,
   title={An optical-UV-IR survey of the North Celestial Cap – I. The catalogue},
   volume={443},
   ISSN={0035-8711},
   url={http://dx.doi.org/10.1093/mnras/stu1183},
   DOI={10.1093/mnras/stu1183},
   number={1},
   journal={Monthly Notices of the Royal Astronomical Society},
   publisher={Oxford University Press (OUP)},
   author={Gorbikov, Evgeny and Brosch, Noah},
   year={2014},
   month=jul, pages={725–737} }

@article{Haynes:2011hi,
    author = "Haynes, Martha P. and others",
    title = "{The Arecibo Legacy Fast ALFA Survey: The alpha.40 HI Source Catalog, its Characteristics and their Impact on the Derivation of the HI Mass Function}",
    eprint = "1109.0027",
    archivePrefix = "arXiv",
    primaryClass = "astro-ph.CO",
    doi = "10.1088/0004-6256/142/5/170",
    journal = "Astron. J.",
    volume = "142",
    pages = "170",
    year = "2011"
}

@article{10.1093/mnras/stz2038,
    author = {Hu, Wenkai and Hoppmann, Laura and Staveley-Smith, Lister and Geréb, Katinka and Oosterloo, Tom and Morganti, Raffaella and Catinella, Barbara and Cortese, Luca and Lagos, Claudia del P and Meyer, Martin},
    title = {An accurate low-redshift measurement of the cosmic neutral hydrogen density},
    journal = {Monthly Notices of the Royal Astronomical Society},
    volume = {489},
    number = {2},
    pages = {1619-1632},
    year = {2019},
    month = {07},
    issn = {0035-8711},
    doi = {10.1093/mnras/stz2038},
    url = {"https://doi.org/10.1093/mnras/stz2038"},
    eprint = {"https://academic.oup.com/mnras/article-pdf/489/2/1619/29399950/stz2038.pdf"}
}

@ARTICLE{2016A&A...586A.132P,
       author = {{Planck Collaboration} and {Ade}, P.~A.~R. and {Aghanim}, N. and {Alves}, M.~I.~R. and {Aniano}, G. and {Arnaud}, M. and {Ashdown}, M. and {Aumont}, J. and {Baccigalupi}, C. and {Banday}, A.~J. and {Barreiro}, R.~B. and {Bartolo}, N. and {Battaner}, E. and {Benabed}, K. and {Benoit-L{\'e}vy}, A. and {Bernard}, J. -P. and {Bersanelli}, M. and {Bielewicz}, P. and {Bonaldi}, A. and {Bonavera}, L. and {Bond}, J.~R. and {Borrill}, J. and {Bouchet}, F.~R. and {Boulanger}, F. and {Burigana}, C. and {Butler}, R.~C. and {Calabrese}, E. and {Cardoso}, J. -F. and {Catalano}, A. and {Chamballu}, A. and {Chiang}, H.~C. and {Christensen}, P.~R. and {Clements}, D.~L. and {Colombi}, S. and {Colombo}, L.~P.~L. and {Couchot}, F. and {Crill}, B.~P. and {Curto}, A. and {Cuttaia}, F. and {Danese}, L. and {Davies}, R.~D. and {Davis}, R.~J. and {de Bernardis}, P. and {de Rosa}, A. and {de Zotti}, G. and {Delabrouille}, J. and {Dickinson}, C. and {Diego}, J.~M. and {Dole}, H. and {Donzelli}, S. and {Dor{\'e}}, O. and {Douspis}, M. and {Draine}, B.~T. and {Ducout}, A. and {Dupac}, X. and {Efstathiou}, G. and {Elsner}, F. and {En{\ss}lin}, T.~A. and {Eriksen}, H.~K. and {Falgarone}, E. and {Finelli}, F. and {Forni}, O. and {Frailis}, M. and {Fraisse}, A.~A. and {Franceschi}, E. and {Frejsel}, A. and {Galeotta}, S. and {Galli}, S. and {Ganga}, K. and {Ghosh}, T. and {Giard}, M. and {Gjerl{\o}w}, E. and {Gonz{\'a}lez-Nuevo}, J. and {G{\'o}rski}, K.~M. and {Gregorio}, A. and {Gruppuso}, A. and {Guillet}, V. and {Hansen}, F.~K. and {Hanson}, D. and {Harrison}, D.~L. and {Henrot-Versill{\'e}}, S. and {Hern{\'a}ndez-Monteagudo}, C. and {Herranz}, D. and {Hildebrandt}, S.~R. and {Hivon}, E. and {Holmes}, W.~A. and {Hovest}, W. and {Huffenberger}, K.~M. and {Hurier}, G. and {Jaffe}, A.~H. and {Jaffe}, T.~R. and {Jones}, W.~C. and {Keih{\"a}nen}, E. and {Keskitalo}, R. and {Kisner}, T.~S. and {Kneissl}, R. and {Knoche}, J. and {Kunz}, M. and {Kurki-Suonio}, H. and {Lagache}, G. and {Lamarre}, J. -M. and {Lasenby}, A. and {Lattanzi}, M. and {Lawrence}, C.~R. and {Leonardi}, R. and {Levrier}, F. and {Liguori}, M. and {Lilje}, P.~B. and {Linden-V{\o}rnle}, M. and {L{\'o}pez-Caniego}, M. and {Lubin}, P.~M. and {Mac{\'\i}as-P{\'e}rez}, J.~F. and {Maffei}, B. and {Maino}, D. and {Mandolesi}, N. and {Maris}, M. and {Marshall}, D.~J. and {Martin}, P.~G. and {Mart{\'\i}nez-Gonz{\'a}lez}, E. and {Masi}, S. and {Matarrese}, S. and {Mazzotta}, P. and {Melchiorri}, A. and {Mendes}, L. and {Mennella}, A. and {Migliaccio}, M. and {Miville-Desch{\^e}nes}, M. -A. and {Moneti}, A. and {Montier}, L. and {Morgante}, G. and {Mortlock}, D. and {Munshi}, D. and {Murphy}, J.~A. and {Naselsky}, P. and {Natoli}, P. and {N{\o}rgaard-Nielsen}, H.~U. and {Novikov}, D. and {Novikov}, I. and {Oxborrow}, C.~A. and {Pagano}, L. and {Pajot}, F. and {Paladini}, R. and {Paoletti}, D. and {Pasian}, F. and {Perdereau}, O. and {Perotto}, L. and {Perrotta}, F. and {Pettorino}, V. and {Piacentini}, F. and {Piat}, M. and {Plaszczynski}, S. and {Pointecouteau}, E. and {Polenta}, G. and {Ponthieu}, N. and {Popa}, L. and {Pratt}, G.~W. and {Prunet}, S. and {Puget}, J. -L. and {Rachen}, J.~P. and {Reach}, W.~T. and {Rebolo}, R. and {Reinecke}, M. and {Remazeilles}, M. and {Renault}, C. and {Ristorcelli}, I. and {Rocha}, G. and {Roudier}, G. and {Rubi{\~n}o-Mart{\'\i}n}, J.~A. and {Rusholme}, B. and {Sandri}, M. and {Santos}, D. and {Scott}, D. and {Spencer}, L.~D. and {Stolyarov}, V. and {Sudiwala}, R. and {Sunyaev}, R. and {Sutton}, D. and {Suur-Uski}, A. -S. and {Sygnet}, J. -F. and {Tauber}, J.~A. and {Terenzi}, L. and {Toffolatti}, L. and {Tomasi}, M. and {Tristram}, M. and {Tucci}, M. and {Umana}, G. and {Valenziano}, L. and {Valiviita}, J. and {Van Tent}, B. and {Vielva}, P. and {Villa}, F. and {Wade}, L.~A. and {Wandelt}, B.~D. and {Wehus}, I.~K. and {Ysard}, N. and {Yvon}, D. and {Zacchei}, A. and {Zonca}, A.},
        title = "{Planck intermediate results. XXIX. All-sky dust modelling with Planck, IRAS, and WISE observations}",
      journal = {\aap},
         year = 2016,
        month = feb,
       volume = {586},
          eid = {A132},
        pages = {A132},
          doi = {10.1051/0004-6361/201424945},
archivePrefix = {arXiv},
       eprint = {1409.2495},
 primaryClass = {astro-ph.GA},
       adsurl = {https://ui.adsabs.harvard.edu/abs/2016A&A...586A.132P}
}

@ARTICLE{2021ApJS..257...59C,
       author = {{CHIME/FRB Collaboration} and {Amiri}, Mandana and {Andersen}, Bridget C. and {Bandura}, Kevin and {Berger}, Sabrina and {Bhardwaj}, Mohit and {Boyce}, Michelle M. and {Boyle}, P.~J. and {Brar}, Charanjot and {Breitman}, Daniela and {Cassanelli}, Tomas and {Chawla}, Pragya and {Chen}, Tianyue and {Cliche}, J. -F. and {Cook}, Amanda and {Cubranic}, Davor and {Curtin}, Alice P. and {Deng}, Meiling and {Dobbs}, Matt and {Dong}, Fengqiu Adam and {Eadie}, Gwendolyn and {Fandino}, Mateus and {Fonseca}, Emmanuel and {Gaensler}, B.~M. and {Giri}, Utkarsh and {Good}, Deborah C. and {Halpern}, Mark and {Hill}, Alex S. and {Hinshaw}, Gary and {Josephy}, Alexander and {Kaczmarek}, Jane F. and {Kader}, Zarif and {Kania}, Joseph W. and {Kaspi}, Victoria M. and {Landecker}, T.~L. and {Lang}, Dustin and {Leung}, Calvin and {Li}, Dongzi and {Lin}, Hsiu-Hsien and {Masui}, Kiyoshi W. and {McKinven}, Ryan and {Mena-Parra}, Juan and {Merryfield}, Marcus and {Meyers}, Bradley W. and {Michilli}, Daniele and {Milutinovic}, Nikola and {Mirhosseini}, Arash and {M{\"u}nchmeyer}, Moritz and {Naidu}, Arun and {Newburgh}, Laura and {Ng}, Cherry and {Patel}, Chitrang and {Pen}, Ue-Li and {Petroff}, Emily and {Pinsonneault-Marotte}, Tristan and {Pleunis}, Ziggy and {Rafiei-Ravandi}, Masoud and {Rahman}, Mubdi and {Ransom}, Scott M. and {Renard}, Andre and {Sanghavi}, Pranav and {Scholz}, Paul and {Shaw}, J. Richard and {Shin}, Kaitlyn and {Siegel}, Seth R. and {Sikora}, Andrew E. and {Singh}, Saurabh and {Smith}, Kendrick M. and {Stairs}, Ingrid and {Tan}, Chia Min and {Tendulkar}, S.~P. and {Vanderlinde}, Keith and {Wang}, Haochen and {Wulf}, Dallas and {Zwaniga}, A.~V.},
        title = "{The First CHIME/FRB Fast Radio Burst Catalog}",
      journal = {\apjs},
         year = 2021,
        month = dec,
       volume = {257},
       number = {2},
          eid = {59},
        pages = {59},
          doi = {10.3847/1538-4365/ac33ab},
archivePrefix = {arXiv},
       eprint = {2106.04352},
 primaryClass = {astro-ph.HE},
       adsurl = {https://ui-adsabs-harvard-edu.ezproxy.library.wisc.edu/abs/2021ApJS..257...59C}
}

@ARTICLE{2016MNRAS.456.2321S,
       author = {{Stewart}, A.~J. and {Fender}, R.~P. and {Broderick}, J.~W. and {Hassall}, T.~E. and {Mu{\~n}oz-Darias}, T. and {Rowlinson}, A. and {Swinbank}, J.~D. and {Staley}, T.~D. and {Molenaar}, G.~J. and {Scheers}, B. and {Grobler}, T.~L. and {Pietka}, M. and {Heald}, G. and {McKean}, J.~P. and {Bell}, M.~E. and {Bonafede}, A. and {Breton}, R.~P. and {Carbone}, D. and {Cendes}, Y. and {Clarke}, A.~O. and {Corbel}, S. and {de Gasperin}, F. and {Eisl{\"o}ffel}, J. and {Falcke}, H. and {Ferrari}, C. and {Grie{\ss}meier}, J. -M. and {Hardcastle}, M.~J. and {Heesen}, V. and {Hessels}, J.~W.~T. and {Horneffer}, A. and {Iacobelli}, M. and {Jonker}, P. and {Karastergiou}, A. and {Kokotanekov}, G. and {Kondratiev}, V.~I. and {Kuniyoshi}, M. and {Law}, C.~J. and {van Leeuwen}, J. and {Markoff}, S. and {Miller-Jones}, J.~C.~A. and {Mulcahy}, D. and {Orru}, E. and {Pandey-Pommier}, M. and {Pratley}, L. and {Rol}, E. and {R{\"o}ttgering}, H.~J.~A. and {Scaife}, A.~M.~M. and {Shulevski}, A. and {Sobey}, C.~A. and {Stappers}, B.~W. and {Tasse}, C. and {van der Horst}, A.~J. and {van Velzen}, S. and {van Weeren}, R.~J. and {Wijers}, R.~A.~M.~J. and {Wijnands}, R. and {Wise}, M. and {Zarka}, P. and {Alexov}, A. and {Anderson}, J. and {Asgekar}, A. and {Avruch}, I.~M. and {Bentum}, M.~J. and {Bernardi}, G. and {Best}, P. and {Breitling}, F. and {Br{\"u}ggen}, M. and {Butcher}, H.~R. and {Ciardi}, B. and {Conway}, J.~E. and {Corstanje}, A. and {de Geus}, E. and {Deller}, A. and {Duscha}, S. and {Frieswijk}, W. and {Garrett}, M.~A. and {Gunst}, A.~W. and {van Haarlem}, M.~P. and {Hoeft}, M. and {H{\"o}randel}, J. and {Juette}, E. and {Kuper}, G. and {Loose}, M. and {Maat}, P. and {McFadden}, R. and {McKay-Bukowski}, D. and {Moldon}, J. and {Munk}, H. and {Norden}, M.~J. and {Paas}, H. and {Polatidis}, A.~G. and {Schwarz}, D. and {Sluman}, J. and {Smirnov}, O. and {Steinmetz}, M. and {Thoudam}, S. and {Toribio}, M.~C. and {Vermeulen}, R. and {Vocks}, C. and {Wijnholds}, S.~J. and {Wucknitz}, O. and {Yatawatta}, S.},
        title = "{LOFAR MSSS: detection of a low-frequency radio transient in 400 h of monitoring of the North Celestial Pole}",
      journal = {\mnras},
         year = 2016,
        month = mar,
       volume = {456},
       number = {3},
        pages = {2321-2342},
          doi = {10.1093/mnras/stv2797},
archivePrefix = {arXiv},
       eprint = {1512.00014},
 primaryClass = {astro-ph.HE},
       adsurl = {https://ui-adsabs-harvard-edu.ezproxy.library.wisc.edu/abs/2016MNRAS.456.2321S}
}

@article{gaia_collaboration_gaia_2023,
	title = {Gaia {Data} {Release} 3 - {Summary} of the content and survey properties},
	volume = {674},
	url = {https://doi.org/10.1051/0004-6361/202243940},
	doi = {10.1051/0004-6361/202243940},
	journal = {A\&A},
	author = {{Gaia Collaboration} and {Vallenari, A.} and {Brown, A. G. A.} and {Prusti, T.} and {de Bruijne, J. H. J.} and {Arenou, F.} and {Babusiaux, C.} and {Biermann, M.} and {Creevey, O. L.} and {Ducourant, C.} and {Evans, D. W.} and {Eyer, L.} and {Guerra, R.} and {Hutton, A.} and {Jordi, C.} and {Klioner, S. A.} and {Lammers, U. L.} and {Lindegren, L.} and {Luri, X.} and {Mignard, F.} and {Panem, C.} and {Pourbaix, D.} and {Randich, S.} and {Sartoretti, P.} and {Soubiran, C.} and {Tanga, P.} and {Walton, N. A.} and {Bailer-Jones, C. A. L.} and {Bastian, U.} and {Drimmel, R.} and {Jansen, F.} and {Katz, D.} and {Lattanzi, M. G.} and {van Leeuwen, F.} and {Bakker, J.} and {Cacciari, C.} and {Castañeda, J.} and {De Angeli, F.} and {Fabricius, C.} and {Fouesneau, M.} and {Frémat, Y.} and {Galluccio, L.} and {Guerrier, A.} and {Heiter, U.} and {Masana, E.} and {Messineo, R.} and {Mowlavi, N.} and {Nicolas, C.} and {Nienartowicz, K.} and {Pailler, F.} and {Panuzzo, P.} and {Riclet, F.} and {Roux, W.} and {Seabroke, G. M.} and {Sordo, R.} and {Thévenin, F.} and {Gracia-Abril, G.} and {Portell, J.} and {Teyssier, D.} and {Altmann, M.} and {Andrae, R.} and {Audard, M.} and {Bellas-Velidis, I.} and {Benson, K.} and {Berthier, J.} and {Blomme, R.} and {Burgess, P. W.} and {Busonero, D.} and {Busso, G.} and {Cánovas, H.} and {Carry, B.} and {Cellino, A.} and {Cheek, N.} and {Clementini, G.} and {Damerdji, Y.} and {Davidson, M.} and {de Teodoro, P.} and {Nuñez Campos, M.} and {Delchambre, L.} and {Dell’Oro, A.} and {Esquej, P.} and {Fernández-Hernández, J.} and {Fraile, E.} and {Garabato, D.} and {García-Lario, P.} and {Gosset, E.} and {Haigron, R.} and {Halbwachs, J.-L.} and {Hambly, N. C.} and {Harrison, D. L.} and {Hernández, J.} and {Hestroffer, D.} and {Hodgkin, S. T.} and {Holl, B.} and {Janßen, K.} and {Jevardat de Fombelle, G.} and {Jordan, S.} and {Krone-Martins, A.} and {Lanzafame, A. C.} and {Löffler, W.} and {Marchal, O.} and {Marrese, P. M.} and {Moitinho, A.} and {Muinonen, K.} and {Osborne, P.} and {Pancino, E.} and {Pauwels, T.} and {Recio-Blanco, A.} and {Reylé, C.} and {Riello, M.} and {Rimoldini, L.} and {Roegiers, T.} and {Rybizki, J.} and {Sarro, L. M.} and {Siopis, C.} and {Smith, M.} and {Sozzetti, A.} and {Utrilla, E.} and {van Leeuwen, M.} and {Abbas, U.} and {Ábrahám, P.} and {Abreu Aramburu, A.} and {Aerts, C.} and {Aguado, J. J.} and {Ajaj, M.} and {Aldea-Montero, F.} and {Altavilla, G.} and {Álvarez, M. A.} and {Alves, J.} and {Anders, F.} and {Anderson, R. I.} and {Anglada Varela, E.} and {Antoja, T.} and {Baines, D.} and {Baker, S. G.} and {Balaguer-Núñez, L.} and {Balbinot, E.} and {Balog, Z.} and {Barache, C.} and {Barbato, D.} and {Barros, M.} and {Barstow, M. A.} and {Bartolomé, S.} and {Bassilana, J.-L.} and {Bauchet, N.} and {Becciani, U.} and {Bellazzini, M.} and {Berihuete, A.} and {Bernet, M.} and {Bertone, S.} and {Bianchi, L.} and {Binnenfeld, A.} and {Blanco-Cuaresma, S.} and {Blazere, A.} and {Boch, T.} and {Bombrun, A.} and {Bossini, D.} and {Bouquillon, S.} and {Bragaglia, A.} and {Bramante, L.} and {Breedt, E.} and {Bressan, A.} and {Brouillet, N.} and {Brugaletta, E.} and {Bucciarelli, B.} and {Burlacu, A.} and {Butkevich, A. G.} and {Buzzi, R.} and {Caffau, E.} and {Cancelliere, R.} and {Cantat-Gaudin, T.} and {Carballo, R.} and {Carlucci, T.} and {Carnerero, M. I.} and {Carrasco, J. M.} and {Casamiquela, L.} and {Castellani, M.} and {Castro-Ginard, A.} and {Chaoul, L.} and {Charlot, P.} and {Chemin, L.} and {Chiaramida, V.} and {Chiavassa, A.} and {Chornay, N.} and {Comoretto, G.} and {Contursi, G.} and {Cooper, W. J.} and {Cornez, T.} and {Cowell, S.} and {Crifo, F.} and {Cropper, M.} and {Crosta, M.} and {Crowley, C.} and {Dafonte, C.} and {Dapergolas, A.} and {David, M.} and {David, P.} and {de Laverny, P.} and {De Luise, F.} and {De March, R.} and {De Ridder, J.} and {de Souza, R.} and {de Torres, A.} and {del Peloso, E. F.} and {del Pozo, E.} and {Delbo, M.} and {Delgado, A.} and {Delisle, J.-B.} and {Demouchy, C.} and {Dharmawardena, T. E.} and {Di Matteo, P.} and {Diakite, S.} and {Diener, C.} and {Distefano, E.} and {Dolding, C.} and {Edvardsson, B.} and {Enke, H.} and {Fabre, C.} and {Fabrizio, M.} and {Faigler, S.} and {Fedorets, G.} and {Fernique, P.} and {Fienga, A.} and {Figueras, F.} and {Fournier, Y.} and {Fouron, C.} and {Fragkoudi, F.} and {Gai, M.} and {Garcia-Gutierrez, A.} and {Garcia-Reinaldos, M.} and {García-Torres, M.} and {Garofalo, A.} and {Gavel, A.} and {Gavras, P.} and {Gerlach, E.} and {Geyer, R.} and {Giacobbe, P.} and {Gilmore, G.} and {Girona, S.} and {Giuffrida, G.} and {Gomel, R.} and {Gomez, A.} and {González-Núñez, J.} and {González-Santamaría, I.} and {González-Vidal, J. J.} and {Granvik, M.} and {Guillout, P.} and {Guiraud, J.} and {Gutiérrez-Sánchez, R.} and {Guy, L. P.} and {Hatzidimitriou, D.} and {Hauser, M.} and {Haywood, M.} and {Helmer, A.} and {Helmi, A.} and {Sarmiento, M. H.} and {Hidalgo, S. L.} and {Hilger, T.} and {Hładczuk, N.} and {Hobbs, D.} and {Holland, G.} and {Huckle, H. E.} and {Jardine, K.} and {Jasniewicz, G.} and {Jean-Antoine Piccolo, A.} and {Jiménez-Arranz, Ó.} and {Jorissen, A.} and {Juaristi Campillo, J.} and {Julbe, F.} and {Karbevska, L.} and {Kervella, P.} and {Khanna, S.} and {Kontizas, M.} and {Kordopatis, G.} and {Korn, A. J.} and {Kóspál, Á} and {Kostrzewa-Rutkowska, Z.} and {Kruszyńska, K.} and {Kun, M.} and {Laizeau, P.} and {Lambert, S.} and {Lanza, A. F.} and {Lasne, Y.} and {Le Campion, J.-F.} and {Lebreton, Y.} and {Lebzelter, T.} and {Leccia, S.} and {Leclerc, N.} and {Lecoeur-Taibi, I.} and {Liao, S.} and {Licata, E. L.} and {Lindstrøm, H. E. P.} and {Lister, T. A.} and {Livanou, E.} and {Lobel, A.} and {Lorca, A.} and {Loup, C.} and {Madrero Pardo, P.} and {Magdaleno Romeo, A.} and {Managau, S.} and {Mann, R. G.} and {Manteiga, M.} and {Marchant, J. M.} and {Marconi, M.} and {Marcos, J.} and {Marcos Santos, M. M. S.} and {Marín Pina, D.} and {Marinoni, S.} and {Marocco, F.} and {Marshall, D. J.} and {Martin Polo, L.} and {Martín-Fleitas, J. M.} and {Marton, G.} and {Mary, N.} and {Masip, A.} and {Massari, D.} and {Mastrobuono-Battisti, A.} and {Mazeh, T.} and {McMillan, P. J.} and {Messina, S.} and {Michalik, D.} and {Millar, N. R.} and {Mints, A.} and {Molina, D.} and {Molinaro, R.} and {Molnár, L.} and {Monari, G.} and {Monguió, M.} and {Montegriffo, P.} and {Montero, A.} and {Mor, R.} and {Mora, A.} and {Morbidelli, R.} and {Morel, T.} and {Morris, D.} and {Muraveva, T.} and {Murphy, C. P.} and {Musella, I.} and {Nagy, Z.} and {Noval, L.} and {Ocaña, F.} and {Ogden, A.} and {Ordenovic, C.} and {Osinde, J. O.} and {Pagani, C.} and {Pagano, I.} and {Palaversa, L.} and {Palicio, P. A.} and {Pallas-Quintela, L.} and {Panahi, A.} and {Payne-Wardenaar, S.} and {Peñalosa Esteller, X.} and {Penttilä, A.} and {Pichon, B.} and {Piersimoni, A. M.} and {Pineau, F.-X.} and {Plachy, E.} and {Plum, G.} and {Poggio, E.} and {Prša, A.} and {Pulone, L.} and {Racero, E.} and {Ragaini, S.} and {Rainer, M.} and {Raiteri, C. M.} and {Rambaux, N.} and {Ramos, P.} and {Ramos-Lerate, M.} and {Re Fiorentin, P.} and {Regibo, S.} and {Richards, P. J.} and {Rios Diaz, C.} and {Ripepi, V.} and {Riva, A.} and {Rix, H.-W.} and {Rixon, G.} and {Robichon, N.} and {Robin, A. C.} and {Robin, C.} and {Roelens, M.} and {Rogues, H. R. O.} and {Rohrbasser, L.} and {Romero-Gómez, M.} and {Rowell, N.} and {Royer, F.} and {Ruz Mieres, D.} and {Rybicki, K. A.} and {Sadowski, G.} and {Sáez Núñez, A.} and {Sagristà Sellés, A.} and {Sahlmann, J.} and {Salguero, E.} and {Samaras, N.} and {Sanchez Gimenez, V.} and {Sanna, N.} and {Santoveña, R.} and {Sarasso, M.} and {Schultheis, M.} and {Sciacca, E.} and {Segol, M.} and {Segovia, J. C.} and {Ségransan, D.} and {Semeux, D.} and {Shahaf, S.} and {Siddiqui, H. I.} and {Siebert, A.} and {Siltala, L.} and {Silvelo, A.} and {Slezak, E.} and {Slezak, I.} and {Smart, R. L.} and {Snaith, O. N.} and {Solano, E.} and {Solitro, F.} and {Souami, D.} and {Souchay, J.} and {Spagna, A.} and {Spina, L.} and {Spoto, F.} and {Steele, I. A.} and {Steidelmüller, H.} and {Stephenson, C. A.} and {Süveges, M.} and {Surdej, J.} and {Szabados, L.} and {Szegedi-Elek, E.} and {Taris, F.} and {Taylor, M. B.} and {Teixeira, R.} and {Tolomei, L.} and {Tonello, N.} and {Torra, F.} and {Torra, J.} and {Torralba Elipe, G.} and {Trabucchi, M.} and {Tsounis, A. T.} and {Turon, C.} and {Ulla, A.} and {Unger, N.} and {Vaillant, M. V.} and {van Dillen, E.} and {van Reeven, W.} and {Vanel, O.} and {Vecchiato, A.} and {Viala, Y.} and {Vicente, D.} and {Voutsinas, S.} and {Weiler, M.} and {Wevers, T.} and {Wyrzykowski, Ł.} and {Yoldas, A.} and {Yvard, P.} and {Zhao, H.} and {Zorec, J.} and {Zucker, S.} and {Zwitter, T.}},
	year = {2023},
	pages = {A1},
}

@article{huchra_2mass_2012,
	title = {{THE} {2MASS} {REDSHIFT} {SURVEY}—{DESCRIPTION} {AND} {DATA} {RELEASE}},
	volume = {199},
	issn = {0067-0049},
	url = {https://dx.doi.org/10.1088/0067-0049/199/2/26},
	doi = {10.1088/0067-0049/199/2/26},
	language = {en},
	number = {2},
	urldate = {2025-05-05},
	journal = {ApJS},
	author = {Huchra, John P. and Macri, Lucas M. and Masters, Karen L. and Jarrett, Thomas H. and Berlind, Perry and Calkins, Michael and Crook, Aidan C. and Cutri, Roc and Erdoğdu, Pirin and Falco, Emilio and George, Teddy and Hutcheson, Conrad M. and Lahav, Ofer and Mader, Jeff and Mink, Jessica D. and Martimbeau, Nathalie and Schneider, Stephen and Skrutskie, Michael and Tokarz, Susan and Westover, Michael},
	month = mar,
	year = {2012},
	note = {Publisher: The American Astronomical Society},
	pages = {26},
}

@INPROCEEDINGS{1995SPIE.2476...56B,
       author = {{Barden}, Samuel C. and {Armandroff}, Taft},
        title = "{Performance of the WIYN fiber-fed MOS system: Hydra}",
    booktitle = {Fiber Optics in Astronomical Applications},
         year = 1995,
       editor = {{Barden}, Samuel C.},
       series = {Society of Photo-Optical Instrumentation Engineers (SPIE) Conference Series},
       volume = {2476},
        month = jun,
        pages = {56-67},
          doi = {10.1117/12.211839},
       adsurl = {https://ui.adsabs.harvard.edu/abs/1995SPIE.2476...56B}
}

@ARTICLE{2009A&A...508.1217Z,
       author = {{Zucca}, E. and {Bardelli}, S. and {Bolzonella}, M. and {Zamorani}, G. and {Ilbert}, O. and {Pozzetti}, L. and {Mignoli}, M. and {Kova{\v{c}}}, K. and {Lilly}, S. and {Tresse}, L. and {Tasca}, L. and {Cassata}, P. and {Halliday}, C. and {Vergani}, D. and {Caputi}, K. and {Carollo}, C.~M. and {Contini}, T. and {Kneib}, J. -P. and {Le F{\`e}vre}, O. and {Mainieri}, V. and {Renzini}, A. and {Scodeggio}, M. and {Bongiorno}, A. and {Coppa}, G. and {Cucciati}, O. and {de La Torre}, S. and {de Ravel}, L. and {Franzetti}, P. and {Garilli}, B. and {Iovino}, A. and {Kampczyk}, P. and {Knobel}, C. and {Lamareille}, F. and {Le Borgne}, J. -F. and {Le Brun}, V. and {Maier}, C. and {Pell{\`o}}, R. and {Peng}, Y. and {Perez-Montero}, E. and {Ricciardelli}, E. and {Silverman}, J.~D. and {Tanaka}, M. and {Abbas}, U. and {Bottini}, D. and {Cappi}, A. and {Cimatti}, A. and {Guzzo}, L. and {Koekemoer}, A.~M. and {Leauthaud}, A. and {Maccagni}, D. and {Marinoni}, C. and {McCracken}, H.~J. and {Memeo}, P. and {Meneux}, B. and {Moresco}, M. and {Oesch}, P. and {Porciani}, C. and {Scaramella}, R. and {Arnouts}, S. and {Aussel}, H. and {Capak}, P. and {Kartaltepe}, J. and {Salvato}, M. and {Sanders}, D. and {Scoville}, N. and {Taniguchi}, Y. and {Thompson}, D.},
        title = "{The zCOSMOS survey: the role of the environment in the evolution of the luminosity function of different galaxy types}",
      journal = {\aap},
         year = 2009,
        month = dec,
       volume = {508},
       number = {3},
        pages = {1217-1234},
          doi = {10.1051/0004-6361/200912665},
archivePrefix = {arXiv},
       eprint = {0909.4674},
 primaryClass = {astro-ph.CO},
       adsurl = {https://ui.adsabs.harvard.edu/abs/2009A&A...508.1217Z}
}

@ARTICLE{2001AJ....122..714B,
       author = {{Brown}, Warren R. and {Geller}, Margaret J. and {Fabricant}, Daniel G. and {Kurtz}, Michael J.},
        title = "{V- and R-band Galaxy Luminosity Functions and Low Surface Brightness Galaxies in the Century Survey}",
      journal = {\aj},
         year = 2001,
        month = aug,
       volume = {122},
       number = {2},
        pages = {714-728},
          doi = {10.1086/321176},
archivePrefix = {arXiv},
       eprint = {astro-ph/0105186},
 primaryClass = {astro-ph},
       adsurl = {https://ui.adsabs.harvard.edu/abs/2001AJ....122..714B}
}

@ARTICLE{2021A&A...652C...4P,
       author = {{Planck Collaboration} and {Aghanim}, N. and {Akrami}, Y. and {Ashdown}, M. and {Aumont}, J. and {Baccigalupi}, C. and {Ballardini}, M. and {Banday}, A.~J. and {Barreiro}, R.~B. and {Bartolo}, N. and {Basak}, S. and {Battye}, R. and {Benabed}, K. and {Bernard}, J. -P. and {Bersanelli}, M. and {Bielewicz}, P. and {Bock}, J.~J. and {Bond}, J.~R. and {Borrill}, J. and {Bouchet}, F.~R. and {Boulanger}, F. and {Bucher}, M. and {Burigana}, C. and {Butler}, R.~C. and {Calabrese}, E. and {Cardoso}, J. -F. and {Carron}, J. and {Challinor}, A. and {Chiang}, H.~C. and {Chluba}, J. and {Colombo}, L.~P.~L. and {Combet}, C. and {Contreras}, D. and {Crill}, B.~P. and {Cuttaia}, F. and {de Bernardis}, P. and {de Zotti}, G. and {Delabrouille}, J. and {Delouis}, J. -M. and {Di Valentino}, E. and {Diego}, J.~M. and {Dor{\'e}}, O. and {Douspis}, M. and {Ducout}, A. and {Dupac}, X. and {Dusini}, S. and {Efstathiou}, G. and {Elsner}, F. and {En{\ss}lin}, T.~A. and {Eriksen}, H.~K. and {Fantaye}, Y. and {Farhang}, M. and {Fergusson}, J. and {Fernandez-Cobos}, R. and {Finelli}, F. and {Forastieri}, F. and {Frailis}, M. and {Fraisse}, A.~A. and {Franceschi}, E. and {Frolov}, A. and {Galeotta}, S. and {Galli}, S. and {Ganga}, K. and {G{\'e}nova-Santos}, R.~T. and {Gerbino}, M. and {Ghosh}, T. and {Gonz{\'a}lez-Nuevo}, J. and {G{\'o}rski}, K.~M. and {Gratton}, S. and {Gruppuso}, A. and {Gudmundsson}, J.~E. and {Hamann}, J. and {Handley}, W. and {Hansen}, F.~K. and {Herranz}, D. and {Hildebrandt}, S.~R. and {Hivon}, E. and {Huang}, Z. and {Jaffe}, A.~H. and {Jones}, W.~C. and {Karakci}, A. and {Keih{\"a}nen}, E. and {Keskitalo}, R. and {Kiiveri}, K. and {Kim}, J. and {Kisner}, T.~S. and {Knox}, L. and {Krachmalnicoff}, N. and {Kunz}, M. and {Kurki-Suonio}, H. and {Lagache}, G. and {Lamarre}, J. -M. and {Lasenby}, A. and {Lattanzi}, M. and {Lawrence}, C.~R. and {Le Jeune}, M. and {Lemos}, P. and {Lesgourgues}, J. and {Levrier}, F. and {Lewis}, A. and {Liguori}, M. and {Lilje}, P.~B. and {Lilley}, M. and {Lindholm}, V. and {L{\'o}pez-Caniego}, M. and {Lubin}, P.~M. and {Ma}, Y. -Z. and {Mac{\'\i}as-P{\'e}rez}, J.~F. and {Maggio}, G. and {Maino}, D. and {Mandolesi}, N. and {Mangilli}, A. and {Marcos-Caballero}, A. and {Maris}, M. and {Martin}, P.~G. and {Martinelli}, M. and {Mart{\'\i}nez-Gonz{\'a}lez}, E. and {Matarrese}, S. and {Mauri}, N. and {McEwen}, J.~D. and {Meinhold}, P.~R. and {Melchiorri}, A. and {Mennella}, A. and {Migliaccio}, M. and {Millea}, M. and {Mitra}, S. and {Miville-Desch{\^e}nes}, M. -A. and {Molinari}, D. and {Montier}, L. and {Morgante}, G. and {Moss}, A. and {Natoli}, P. and {N{\o}rgaard-Nielsen}, H.~U. and {Pagano}, L. and {Paoletti}, D. and {Partridge}, B. and {Patanchon}, G. and {Peiris}, H.~V. and {Perrotta}, F. and {Pettorino}, V. and {Piacentini}, F. and {Polastri}, L. and {Polenta}, G. and {Puget}, J. -L. and {Rachen}, J.~P. and {Reinecke}, M. and {Remazeilles}, M. and {Renzi}, A. and {Rocha}, G. and {Rosset}, C. and {Roudier}, G. and {Rubi{\~n}o-Mart{\'\i}n}, J.~A. and {Ruiz-Granados}, B. and {Salvati}, L. and {Sandri}, M. and {Savelainen}, M. and {Scott}, D. and {Shellard}, E.~P.~S. and {Sirignano}, C. and {Sirri}, G. and {Spencer}, L.~D. and {Sunyaev}, R. and {Suur-Uski}, A. -S. and {Tauber}, J.~A. and {Tavagnacco}, D. and {Tenti}, M. and {Toffolatti}, L. and {Tomasi}, M. and {Trombetti}, T. and {Valenziano}, L. and {Valiviita}, J. and {Van Tent}, B. and {Vibert}, L. and {Vielva}, P. and {Villa}, F. and {Vittorio}, N. and {Wandelt}, B.~D. and {Wehus}, I.~K. and {White}, M. and {White}, S.~D.~M. and {Zacchei}, A. and {Zonca}, A.},
        title = "{Planck 2018 results. VI. Cosmological parameters (Corrigendum)}",
      journal = {\aap},
         year = 2021,
        month = aug,
       volume = {652},
          eid = {C4},
        pages = {C4},
          doi = {10.1051/0004-6361/201833910e},
       adsurl = {https://ui.adsabs.harvard.edu/abs/2021A&A...652C...4P}
}

@ARTICLE{2017AJ....154...28B,
       author = {{Blanton}, Michael R. and {Bershady}, Matthew A. and {Abolfathi}, Bela and {Albareti}, Franco D. and {Allende Prieto}, Carlos and {Almeida}, Andres and {Alonso-Garc{\'\i}a}, Javier and {Anders}, Friedrich and {Anderson}, Scott F. and {Andrews}, Brett and {Aquino-Ort{\'\i}z}, Erik and {Arag{\'o}n-Salamanca}, Alfonso and {Argudo-Fern{\'a}ndez}, Maria and {Armengaud}, Eric and {Aubourg}, Eric and {Avila-Reese}, Vladimir and {Badenes}, Carles and {Bailey}, Stephen and {Barger}, Kathleen A. and {Barrera-Ballesteros}, Jorge and {Bartosz}, Curtis and {Bates}, Dominic and {Baumgarten}, Falk and {Bautista}, Julian and {Beaton}, Rachael and {Beers}, Timothy C. and {Belfiore}, Francesco and {Bender}, Chad F. and {Berlind}, Andreas A. and {Bernardi}, Mariangela and {Beutler}, Florian and {Bird}, Jonathan C. and {Bizyaev}, Dmitry and {Blanc}, Guillermo A. and {Blomqvist}, Michael and {Bolton}, Adam S. and {Boquien}, M{\'e}d{\'e}ric and {Borissova}, Jura and {van den Bosch}, Remco and {Bovy}, Jo and {Brandt}, William N. and {Brinkmann}, Jonathan and {Brownstein}, Joel R. and {Bundy}, Kevin and {Burgasser}, Adam J. and {Burtin}, Etienne and {Busca}, Nicol{\'a}s G. and {Cappellari}, Michele and {Delgado Carigi}, Maria Leticia and {Carlberg}, Joleen K. and {Carnero Rosell}, Aurelio and {Carrera}, Ricardo and {Chanover}, Nancy J. and {Cherinka}, Brian and {Cheung}, Edmond and {G{\'o}mez Maqueo Chew}, Yilen and {Chiappini}, Cristina and {Choi}, Peter Doohyun and {Chojnowski}, Drew and {Chuang}, Chia-Hsun and {Chung}, Haeun and {Cirolini}, Rafael Fernando and {Clerc}, Nicolas and {Cohen}, Roger E. and {Comparat}, Johan and {da Costa}, Luiz and {Cousinou}, Marie-Claude and {Covey}, Kevin and {Crane}, Jeffrey D. and {Croft}, Rupert A.~C. and {Cruz-Gonzalez}, Irene and {Garrido Cuadra}, Daniel and {Cunha}, Katia and {Damke}, Guillermo J. and {Darling}, Jeremy and {Davies}, Roger and {Dawson}, Kyle and {de la Macorra}, Axel and {Dell'Agli}, Flavia and {De Lee}, Nathan and {Delubac}, Timoth{\'e}e and {Di Mille}, Francesco and {Diamond-Stanic}, Aleks and {Cano-D{\'\i}az}, Mariana and {Donor}, John and {Downes}, Juan Jos{\'e} and {Drory}, Niv and {du Mas des Bourboux}, H{\'e}lion and {Duckworth}, Christopher J. and {Dwelly}, Tom and {Dyer}, Jamie and {Ebelke}, Garrett and {Eigenbrot}, Arthur D. and {Eisenstein}, Daniel J. and {Emsellem}, Eric and {Eracleous}, Mike and {Escoffier}, Stephanie and {Evans}, Michael L. and {Fan}, Xiaohui and {Fern{\'a}ndez-Alvar}, Emma and {Fernandez-Trincado}, J.~G. and {Feuillet}, Diane K. and {Finoguenov}, Alexis and {Fleming}, Scott W. and {Font-Ribera}, Andreu and {Fredrickson}, Alexander and {Freischlad}, Gordon and {Frinchaboy}, Peter M. and {Fuentes}, Carla E. and {Galbany}, Llu{\'\i}s and {Garcia-Dias}, R. and {Garc{\'\i}a-Hern{\'a}ndez}, D.~A. and {Gaulme}, Patrick and {Geisler}, Doug and {Gelfand}, Joseph D. and {Gil-Mar{\'\i}n}, H{\'e}ctor and {Gillespie}, Bruce A. and {Goddard}, Daniel and {Gonzalez-Perez}, Violeta and {Grabowski}, Kathleen and {Green}, Paul J. and {Grier}, Catherine J. and {Gunn}, James E. and {Guo}, Hong and {Guy}, Julien and {Hagen}, Alex and {Hahn}, ChangHoon and {Hall}, Matthew and {Harding}, Paul and {Hasselquist}, Sten and {Hawley}, Suzanne L. and {Hearty}, Fred and {Gonzalez Hern{\'a}ndez}, Jonay I. and {Ho}, Shirley and {Hogg}, David W. and {Holley-Bockelmann}, Kelly and {Holtzman}, Jon A. and {Holzer}, Parker H. and {Huehnerhoff}, Joseph and {Hutchinson}, Timothy A. and {Hwang}, Ho Seong and {Ibarra-Medel}, H{\'e}ctor J. and {da Silva Ilha}, Gabriele and {Ivans}, Inese I. and {Ivory}, KeShawn and {Jackson}, Kelly and {Jensen}, Trey W. and {Johnson}, Jennifer A. and {Jones}, Amy and {J{\"o}nsson}, Henrik and {Jullo}, Eric and {Kamble}, Vikrant and {Kinemuchi}, Karen and {Kirkby}, David and {Kitaura}, Francisco-Shu and {Klaene}, Mark and {Knapp}, Gillian R. and {Kneib}, Jean-Paul and {Kollmeier}, Juna A. and {Lacerna}, Ivan and {Lane}, Richard R. and {Lang}, Dustin and {Law}, David R. and {Lazarz}, Daniel and {Lee}, Youngbae and {Le Goff}, Jean-Marc and {Liang}, Fu-Heng and {Li}, Cheng and {Li}, Hongyu and {Lian}, Jianhui and {Lima}, Marcos and {Lin}, Lihwai and {Lin}, Yen-Ting and {Bertran de Lis}, Sara and {Liu}, Chao and {de Icaza Lizaola}, Miguel Angel C. and {Long}, Dan and {Lucatello}, Sara and {Lundgren}, Britt and {MacDonald}, Nicholas K. and {Deconto Machado}, Alice and {MacLeod}, Chelsea L. and {Mahadevan}, Suvrath and {Geimba Maia}, Marcio Antonio and {Maiolino}, Roberto and {Majewski}, Steven R. and {Malanushenko}, Elena and {Malanushenko}, Viktor and {Manchado}, Arturo and {Mao}, Shude and {Maraston}, Claudia and {Marques-Chaves}, Rui and {Masseron}, Thomas and {Masters}, Karen L. and {McBride}, Cameron K. and {McDermid}, Richard M. and {McGrath}, Brianne and {McGreer}, Ian D. and {Medina Pe{\~n}a}, Nicol{\'a}s and {Melendez}, Matthew and {Merloni}, Andrea and {Merrifield}, Michael R. and {Meszaros}, Szabolcs and {Meza}, Andres and {Minchev}, Ivan and {Minniti}, Dante and {Miyaji}, Takamitsu and {More}, Surhud and {Mulchaey}, John and {M{\"u}ller-S{\'a}nchez}, Francisco and {Muna}, Demitri and {Munoz}, Ricardo R. and {Myers}, Adam D. and {Nair}, Preethi and {Nandra}, Kirpal and {Correa do Nascimento}, Janaina and {Negrete}, Alenka and {Ness}, Melissa and {Newman}, Jeffrey A. and {Nichol}, Robert C. and {Nidever}, David L. and {Nitschelm}, Christian and {Ntelis}, Pierros and {O'Connell}, Julia E. and {Oelkers}, Ryan J. and {Oravetz}, Audrey and {Oravetz}, Daniel and {Pace}, Zach and {Padilla}, Nelson and {Palanque-Delabrouille}, Nathalie and {Alonso Palicio}, Pedro and {Pan}, Kaike and {Parejko}, John K. and {Parikh}, Taniya and {P{\^a}ris}, Isabelle and {Park}, Changbom and {Patten}, Alim Y. and {Peirani}, Sebastien and {Pellejero-Ibanez}, Marcos and {Penny}, Samantha and {Percival}, Will J. and {Perez-Fournon}, Ismael and {Petitjean}, Patrick and {Pieri}, Matthew M. and {Pinsonneault}, Marc and {Pisani}, Alice and {Poleski}, Rados{\l}aw and {Prada}, Francisco and {Prakash}, Abhishek and {Queiroz}, Anna B{\'a}rbara de Andrade and {Raddick}, M. Jordan and {Raichoor}, Anand and {Barboza Rembold}, Sandro and {Richstein}, Hannah and {Riffel}, Rogemar A. and {Riffel}, Rog{\'e}rio and {Rix}, Hans-Walter and {Robin}, Annie C. and {Rockosi}, Constance M. and {Rodr{\'\i}guez-Torres}, Sergio and {Roman-Lopes}, A. and {Rom{\'a}n-Z{\'u}{\~n}iga}, Carlos and {Rosado}, Margarita and {Ross}, Ashley J. and {Rossi}, Graziano and {Ruan}, John and {Ruggeri}, Rossana and {Rykoff}, Eli S. and {Salazar-Albornoz}, Salvador and {Salvato}, Mara and {S{\'a}nchez}, Ariel G. and {Aguado}, D.~S. and {S{\'a}nchez-Gallego}, Jos{\'e} R. and {Santana}, Felipe A. and {Santiago}, Bas{\'\i}lio Xavier and {Sayres}, Conor and {Schiavon}, Ricardo P. and {da Silva Schimoia}, Jaderson and {Schlafly}, Edward F. and {Schlegel}, David J. and {Schneider}, Donald P. and {Schultheis}, Mathias and {Schuster}, William J. and {Schwope}, Axel and {Seo}, Hee-Jong and {Shao}, Zhengyi and {Shen}, Shiyin and {Shetrone}, Matthew and {Shull}, Michael and {Simon}, Joshua D. and {Skinner}, Danielle and {Skrutskie}, M.~F. and {Slosar}, An{\v{z}}e and {Smith}, Verne V. and {Sobeck}, Jennifer S. and {Sobreira}, Flavia and {Somers}, Garrett and {Souto}, Diogo and {Stark}, David V. and {Stassun}, Keivan and {Stauffer}, Fritz and {Steinmetz}, Matthias and {Storchi-Bergmann}, Thaisa and {Streblyanska}, Alina and {Stringfellow}, Guy S. and {Su{\'a}rez}, Genaro and {Sun}, Jing and {Suzuki}, Nao and {Szigeti}, Laszlo and {Taghizadeh-Popp}, Manuchehr and {Tang}, Baitian and {Tao}, Charling and {Tayar}, Jamie and {Tembe}, Mita and {Teske}, Johanna and {Thakar}, Aniruddha R. and {Thomas}, Daniel and {Thompson}, Benjamin A. and {Tinker}, Jeremy L. and {Tissera}, Patricia and {Tojeiro}, Rita and {Hernandez Toledo}, Hector and {de la Torre}, Sylvain and {Tremonti}, Christy and {Troup}, Nicholas W. and {Valenzuela}, Octavio and {Martinez Valpuesta}, Inma and {Vargas-Gonz{\'a}lez}, Jaime and {Vargas-Maga{\~n}a}, Mariana and {Vazquez}, Jose Alberto and {Villanova}, Sandro and {Vivek}, M. and {Vogt}, Nicole and {Wake}, David and {Walterbos}, Rene and {Wang}, Yuting and {Weaver}, Benjamin Alan and {Weijmans}, Anne-Marie and {Weinberg}, David H. and {Westfall}, Kyle B. and {Whelan}, David G. and {Wild}, Vivienne and {Wilson}, John and {Wood-Vasey}, W.~M. and {Wylezalek}, Dominika and {Xiao}, Ting and {Yan}, Renbin and {Yang}, Meng and {Ybarra}, Jason E. and {Y{\`e}che}, Christophe and {Zakamska}, Nadia and {Zamora}, Olga and {Zarrouk}, Pauline and {Zasowski}, Gail and {Zhang}, Kai and {Zhao}, Gong-Bo and {Zheng}, Zheng and {Zheng}, Zheng and {Zhou}, Xu and {Zhou}, Zhi-Min and {Zhu}, Guangtun B. and {Zoccali}, Manuela and {Zou}, Hu},
        title = "{Sloan Digital Sky Survey IV: Mapping the Milky Way, Nearby Galaxies, and the Distant Universe}",
      journal = {\aj},
         year = 2017,
        month = jul,
       volume = {154},
       number = {1},
          eid = {28},
        pages = {28},
          doi = {10.3847/1538-3881/aa7567},
archivePrefix = {arXiv},
       eprint = {1703.00052},
 primaryClass = {astro-ph.GA},
       adsurl = {https://ui.adsabs.harvard.edu/abs/2017AJ....154...28B}
}

@ARTICLE{2006A&A...460..339J,
       author = {{Jordi}, K. and {Grebel}, E.~K. and {Ammon}, K.},
        title = "{Empirical color transformations between SDSS photometry and other photometric systems}",
      journal = {\aap},
         year = 2006,
        month = dec,
       volume = {460},
       number = {1},
        pages = {339-347},
          doi = {10.1051/0004-6361:20066082},
archivePrefix = {arXiv},
       eprint = {astro-ph/0609121},
 primaryClass = {astro-ph},
       adsurl = {https://ui.adsabs.harvard.edu/abs/2006A&A...460..339J}
}

@ARTICLE{2013A&A...554A.131V,
       author = {{Vargas-Maga{\~n}a}, M. and {Bautista}, J.~E. and {Hamilton}, J. -Ch. and {Busca}, N.~G. and {Aubourg}, {\'E}. and {Labatie}, A. and {Le Goff}, J. -M. and {Escoffier}, S. and {Manera}, M. and {McBride}, C.~K. and {Schneider}, D.~P. and {Willmer}, Ch. N.~A.},
        title = "{An optimized correlation function estimator for galaxy surveys}",
      journal = {\aap},
         year = 2013,
        month = jun,
       volume = {554},
          eid = {A131},
        pages = {A131},
          doi = {10.1051/0004-6361/201220790},
archivePrefix = {arXiv},
       eprint = {1211.6211},
 primaryClass = {astro-ph.CO},
       adsurl = {https://ui.adsabs.harvard.edu/abs/2013A&A...554A.131V}
}

@ARTICLE{1993ApJ...412...64L,
       author = {{Landy}, Stephen D. and {Szalay}, Alexander S.},
        title = "{Bias and Variance of Angular Correlation Functions}",
      journal = {\apj},
         year = 1993,
        month = jul,
       volume = {412},
        pages = {64},
          doi = {10.1086/172900},
       adsurl = {https://ui.adsabs.harvard.edu/abs/1993ApJ...412...64L}
}

@article{1991ApJ...379..482K,
  title = {Power-spectrum analysis of one-dimensional redshift surveys},
  volume = {379},
  ISSN = {1538-4357},
  url = {http://dx.doi.org/10.1086/170523},
  DOI = {10.1086/170523},
  journal = {The Astrophysical Journal},
  publisher = {American Astronomical Society},
  author = {Kaiser,  N. and Peacock,  J. A.},
  year = {1991},
  month = oct,
  pages = {482}
}

@article{Wolz_2021,
   title={H <scp>i</scp> constraints from the cross-correlation of eBOSS galaxies and Green Bank Telescope intensity maps},
   volume={510},
   ISSN={1365-2966},
   url={http://dx.doi.org/10.1093/mnras/stab3621},
   DOI={10.1093/mnras/stab3621},
   number={3},
   journal={Monthly Notices of the Royal Astronomical Society},
   publisher={Oxford University Press (OUP)},
   author={Wolz, Laura and Pourtsidou, Alkistis and Masui, Kiyoshi W and Chang, Tzu-Ching and Bautista, Julian E and Müller, Eva-Maria and Avila, Santiago and Bacon, David and Percival, Will J and Cunnington, Steven and Anderson, Chris and Chen, Xuelei and Kneib, Jean-Paul and Li, Yi-Chao and Liao, Yu-Wei and Pen, Ue-Li and Peterson, Jeffrey B and Rossi, Graziano and Schneider, Donald P and Yadav, Jaswant and Zhao, Gong-Bo},
   year={2021},
   month=dec, pages={3495–3511} }

@article{Anderson_2018,
   title={Low-amplitude clustering in low-redshift 21-cm intensity maps cross-correlated with 2dF galaxy densities},
   volume={476},
   ISSN={1365-2966},
   url={http://dx.doi.org/10.1093/mnras/sty346},
   DOI={10.1093/mnras/sty346},
   number={3},
   journal={Monthly Notices of the Royal Astronomical Society},
   publisher={Oxford University Press (OUP)},
   author={Anderson, C J and Luciw, N J and Li, Y -C and Kuo, C Y and Yadav, J and Masui, K W and Chang, T-C and Chen, X and Oppermann, N and Liao, Y-W and Pen, U-L and Price, D C and Staveley-Smith, L and Switzer, E R and Timbie, P T and Wolz, L},
   year={2018},
   month=feb, pages={3382–3392} }

@dataset{2014yCat..35660001T,
       author = {{Tempel}, E. and {Tamm}, A. and {Gramann}, M. and {Tuvikene}, T. and {Liivamagi}, L.~J. and {Suhhonenko}, I. and {Kipper}, R. and {Einasto}, M. and {Saar}, E.},
        title = "{VizieR Online Data Catalog: Flux- and volume-limited groups for SDSS galaxies (Tempel+, 2014)}",
 howpublished = {VizieR On-line Data Catalog: J/A+A/566/A1. Originally published in: 2014A\&A...566A...1T},
         year = 2014,
        month = apr,
          eid = {J/A+A/566/A1},
          doi = {10.26093/cds/vizier.35660001},
       adsurl = {https://ui.adsabs.harvard.edu/abs/2014yCat..35660001T}
}

@ARTICLE{2016A&A...588A..14T,
       author = {{Tempel}, E. and {Kipper}, R. and {Tamm}, A. and {Gramann}, M. and {Einasto}, M. and {Sepp}, T. and {Tuvikene}, T.},
        title = "{Friends-of-friends galaxy group finder with membership refinement. Application to the local Universe}",
      journal = {\aap},
         year = 2016,
        month = apr,
       volume = {588},
          eid = {A14},
        pages = {A14},
          doi = {10.1051/0004-6361/201527755},
archivePrefix = {arXiv},
       eprint = {1601.01117},
 primaryClass = {astro-ph.CO},
       adsurl = {https://ui.adsabs.harvard.edu/abs/2016A&A...588A..14T}
}

@ARTICLE{1982ApJ...257..423H,
       author = {{Huchra}, J.~P. and {Geller}, M.~J.},
        title = "{Groups of Galaxies. I. Nearby groups}",
      journal = {\apj},
         year = 1982,
        month = jun,
       volume = {257},
        pages = {423-437},
          doi = {10.1086/160000},
       adsurl = {https://ui.adsabs.harvard.edu/abs/1982ApJ...257..423H}
}
